%% file: 00-paper_top.tex
\documentclass[conference,letterpaper,twocolumn,10pt]{IEEEtran}
\usepackage{graphicx}
\usepackage{tikz}
\usepackage{amsmath}
\usepackage{amssymb}
\usepackage{booktabs}  
\usepackage{paralist}
\newcommand{\para}[1]{\smallskip\noindent\textbf{{#1.}}}
\usepackage{minted}
\usepackage[utf8]{inputenc}    
\usepackage{enumitem}    
\usepackage{comment}
\usepackage{subcaption}
\usepackage{hyperref}      
\usepackage{cleveref}
\usepackage{tabularx}      
\usepackage{makecell}      
\usepackage{textcomp}      
\usepackage{ragged2e}      
\usepackage{lipsum}
\usepackage{threeparttable}
\usepackage[table]{xcolor}
\crefname{algocf}{algorithm}{algorithms}
\Crefname{algocf}{Algorithm}{Algorithms}
\usepackage[T1]{fontenc} 
\usepackage{iftex} 
\usepackage{caption} 
\usepackage[inkscapelatex=false]{svg} 

\newcolumntype{L}{>{\RaggedRight\arraybackslash}X}
\newcolumntype{C}{>{\centering\arraybackslash}X}

\definecolor{tableheadspec}{HTML}{446688}

\ifLuaTeX
\protected\def\pdfmapline {\pdfextension mapline}
\fi
\usepackage{fix-cm}    

\makeatletter
\newcommand\HUGE{\@setfontsize\Huge{37}{34}}
\makeatother  

\pdfmapline{+federation < federation.ttf <T1-WGL4.enc}

\newcommand\tngfont[1]{{\usefont{T1}{federation}{m}{n} #1 }}

\usepackage[backend=bibtex,style=ieee,citestyle=numeric-comp,sortcites=true,maxnames=3]{biblatex}
\DeclareFieldFormat[online]{title}{\mkbibquote{#1\addperiod}\addspace}

\AtBeginBibliography{%
  \footnotesize 

  \setlength{\bibitemsep}{0pt} 
  \setlength{\itemsep}{0pt} 

  \settowidth{\bibhang}{\printfield{labelprefix}99\printfield{labelsuffix}\hspace*{0.5em}}%
}

\PassOptionsToPackage{hyphens}{url}\usepackage{hyperref}
\usepackage[dvipsnames,svgnames]{xcolor} 
\hypersetup{
    colorlinks=true,
    linkcolor=ForestGreen,
    citecolor=ForestGreen,
    urlcolor=RoyalBlue,
    filecolor=IndianRed,
    pdftitle={SPEC CPU: The Next Generation}, 
    pdfauthor={TBD}, 
    pdfsubject={Benchmarks}, 
    pdfkeywords={Standardization, Benchmarks, CPU Performance}, 
    pdfnewwindow=true,
    pdfdisplaydoctitle=true,
    bookmarksopen=true
}

\begin{document}

\title{\HUGE{\tngfont{SPEC\,CPU:\,the\,next\,generation}}}

\author{
    \begin{tabular}{c}
        \parbox[t]{0.92\textwidth}{
            \centering
            Mahesh~Madhav\textsuperscript{1},
            Allen~Lee\textsuperscript{2},
            Andres~Mejia\textsuperscript{3},
            Branden~Moore\textsuperscript{4},
            Charan~Soppadandi\textsuperscript{5},
            Chris~Cambly\textsuperscript{6},
            Christoph~Müllner\textsuperscript{7},
            Daniel~Bowers\textsuperscript{8},
            David~Reiner\textsuperscript{4},
            Denis~Bakhvalov\textsuperscript{9},
            Di~Zhao\textsuperscript{1},
            Duane~Voth\textsuperscript{4},
            Feng~Xue\textsuperscript{1},
            Frédérique~Silber-Chaussumier\textsuperscript{10},
            James~Bucek\textsuperscript{8},
            James~Southern\textsuperscript{11},
            Jiangning~Liu\textsuperscript{1},
            Jim~Himer\textsuperscript{8},
            John~Henning\textsuperscript{8},
            Kevin~Smith\textsuperscript{1},
            Kristen~Yang\textsuperscript{4},
            Kunal~Kashyap\textsuperscript{4},
            Mason~Guy\textsuperscript{3},
            Mat~Colgrove\textsuperscript{12},
            Michael~Berg\textsuperscript{13},
            Prasad~Battini\textsuperscript{3},
            Prasad~Joshi\textsuperscript{3},
            Rohit~Prasad\textsuperscript{3},
            Shay~Bhattacharya\textsuperscript{4},
            Sriyash~Caculo\textsuperscript{1},
            Stefan~Reimbold\textsuperscript{6},
            Sundar~Iyengar\textsuperscript{3},
            Van~Smith\textsuperscript{4}, and
            Zarko~Todorovski\textsuperscript{6}
        }
        \\[15ex] 
        
        \parbox[t]{\textwidth}{
            \centering\small
            \textsuperscript{1}Ampere Computing,
            \textsuperscript{2}IEIT,
            \textsuperscript{3}Intel,
            \textsuperscript{4}AMD,
            \textsuperscript{5}Dell Technologies,
            \textsuperscript{6}IBM,
            \textsuperscript{7}VRULL,
            \textsuperscript{8}SPEC,
            \textsuperscript{9}Rivos,
            \textsuperscript{10}ARM,
            \textsuperscript{11}HPE,
            \textsuperscript{12}NVIDIA,
            \textsuperscript{13}SiFive
        }
    \end{tabular}
}

\maketitle
\input{commands} 
\input{10-abstract}
\input{20-intro}

\input{50-bm_table}

\input{30-changes}

\input{40-benchmarks}

\input{60-analysis}

\input{70-roundrobin}

\input{80-conclusion}

\printbibliography
\appendix

\input{99-appendix}

\end{document}

%% file: commands.tex
\newcommand{\Red}[1]{{\color{red} #1}}
\newcommand{\ignore}[1]{}
\newcommand{\asm}[1]{\texttt{#1}}
\newcommand{\sys}[1]{\texttt{#1}}
\newcommand{\kw}[1]{\textit{#1}}
\newcommand{\kwb}[1]{\textbf{#1}}
\newcommand{\type}[1]{\textit{#1}}
\newcommand{\Response}[1]{{\color{blue} #1}}
\newcommand{\XXX}[1]{\Red{\textbf{XXX[}#1\textbf{]}}}
\newcommand{\tocite}[1]{\Red{CITE:\cite{#1}}}
\newcommand{\toref}[1]{\Red{REF:\ref{#1}}}
\newcommand{\parasub}[1]{\smallskip\noindent\textit{{#1:}\xspace}}
\newcommand{\mahesh}[1]{\textcolor{purple}{#1}}
\newcommand{\anyone}[1]{\textcolor{blue}{#1}}
\newcommand{\niparagraph}[1]{\noindent\textbf{\textsf{#1}\hspace{0.5em}}}
\newcommand\TODO[1]{\textcolor{red}{TODO: #1}}

\newenvironment{CompactItemize}%
  {\begin{list}{$\blacktriangleright$}%
    {\leftmargin=\parindent \itemsep=2pt \topsep=2pt
     \parsep=0pt \partopsep=0pt}}%
  {\end{list}}
\renewcommand{\labelitemi}{$\blacktriangleright$}

%% file: 10-abstract.tex
\begin{abstract}

The march toward developing relevant and robust CPU benchmarks continues with the introduction of SPEC CPU$^\circledR$2026, the next generation suite for measuring processor performance. This paper details the methodology behind its creation, showcasing a process centered on community collaboration and principled development. The suite is built upon a foundation of modern, open-source applications, selected and hardened through a process that emphasizes workload diversity, portability, and software longevity. A key contribution is Rolling Round-Robin Rate, a novel and standardized approach to running heterogeneous, multiprogrammed workloads that addresses a long-standing gap in benchmarking practice. Additionally, the suite features an expanded set of multithreaded benchmarks and introduces workloads with distinct microarchitectural profiles, reflecting the demands of contemporary software. By detailing our principled approach to benchmark selection, adaptation, and validation, we demonstrate how the SPEC CPU$^\circledR$2026 suite sets the standard for performance evaluation in the next era of computer architecture research and development. 

\end{abstract}

%% file: 20-intro.tex
\section{Introduction}

David Patterson's principle that ``benchmarks shape a field'' posits that well-designed metrics accelerate progress by enabling fair and objective comparisons \cite{patterson_field}. For more than three decades, the SPEC CPU suites have served this role, establishing a trusted lineage from CPU89 to CPU2017 \cite{Dixit1993OverviewOT, cpu2000, cpu2006, cpu2017, mrob_spec_overview}. This paper introduces SPEC CPU$^\circledR$2026, the newest iteration in this lineage, born from a multi-year effort to reflect the evolution of general-purpose computing. In an era dominated by discourse on specialized AI accelerators, the performance of general-purpose CPUs remains fundamental. The enduring importance of CPU microarchitectural innovation, distinct from advances in process technology or software, has been quantitatively demonstrated through longitudinal studies that rely on SPEC CPU data to normalize performance across generations \cite{cpudb}. SPEC CPU$^\circledR$2026 reaffirms its relevance by providing a refreshed and essential tool to measure performance and energy on this bedrock of modern computation.

To fulfill this role effectively, SPEC CPU adheres to a specific set of principles that define its scope. The suite exclusively encompasses natively compiled C, C++, and Fortran code, measuring performance in both single-threaded and multi-threaded scenarios. It does not cover managed runtimes (e.g., Java, Python, Julia), as the complexities of Just-In-Time compilation introduce portability concerns and significant run-to-run variance that conflicts with SPEC's foundational requirement for reproducibility \cite{dacapo_variance, dacapo_variance2, ren_variance}. This commitment to determinism is inseparable from a commitment to correctness. While some modern benchmark suites have only recently added result validation \cite{ren_16}, SPEC CPU has, since its inception, not only measured how fast a workload runs, but how fast it runs \emph{correctly}. This rigorous result validation is a hallmark of an industrial- and research-grade benchmark suite and remains central to its design.

This adherence to principle does not imply stagnation. On the contrary, the development of each new suite is an introspective process informed by lessons from the past and the evolving needs of the industry. Past suites were critiqued for having a limited number of microarchitectural behaviors \cite{only_four_bottlenecks} or for including workloads vulnerable to non-portable optimizations \cite{specint_analysis}, feedback that SPEC has taken seriously.

SPEC CPU2026 is the direct result of this methodical, forward-looking process, embracing its role as a flexible research harness. It addresses past shortcomings and reflects contemporary software trends through several key enhancements. The suite is significantly larger, a deliberate strategy to increase application diversity and offer a variety of behaviors. Responding to the growing complexity of modern software, it introduces a new emphasis on front-end bound integer benchmarks characterized by large code footprints. A landmark improvement is the inclusion of multiple multithreaded integer benchmarks, addressing a critical gap in CPU2017. Finally, the suite introduces Rolling Round-Robin Rate (RRR), a new exhibition methodology provided to evaluate system performance with heterogeneous multiprogrammed benchmarks.

A fundamental role of SPEC CPU is to serve as a standardized tool for exploration, a principle occasionally misunderstood in external critiques. Some analyses \cite{evaluatology1} posit that SPEC CPU is arbitrarily vague or arbitrarily specific, based on incorrect assumptions that it mandates certain compiler flags or thread counts or operating systems. Such interpretations confuse the rules for formally compliant, \emph{published} results with the suite's broader purpose. The extensive configuration space is not forbidden; its exploration is simply required to be documented for transparency and reproducibility.

The most accurate analogy for SPEC CPU is that of a tape measure: it is a tool for measurement, not a dictate on what must be measured. While a small portion of the users submit official scores, the vast majority use the suite as a flexible harness for internal research on compilers, hardware, and software. Recognizing this primary use case, CPU2026 expands the suite's analytical capabilities. For example, the raw output files have been augmented to include detailed statistics, such as variance and standard deviations. This evolution enhances the rigor of the already flexible harness, empowering users with deeper and more reliable data for their own explorations.

This paper details the principles of benchmark selection, the engineering efforts required to ensure portability and software longevity, and the key changes that define the new suite. By providing this transparent account, it demonstrates how SPEC CPU continues its long-standing mission: to provide a fair, relevant, and trusted standard for performance evaluation in the next era of computer architecture.

%% file: 50-bm_table.tex
\begin{table*}[th!]
\centering
\setlength{\extrarowheight}{1.1ex} 
\caption{\large{\textbf{The SPEC CPU\textregistered{}2026 Benchmarks}}\\}
\label{tab:spec-benchmarks}

\begin{tabularx}{\textwidth}{@{} l l c c r r L c @{}}
\\
\arrayrulecolor{lightgray!40}
\rowcolor{tableheadspec}

\thead[l]{\textbf{\makecell[l]{\textcolor{white}{SPECrate\textregistered 2026} \\ \textcolor{white}{Integer}}}} &
\thead[l]{\textbf{\makecell[l]{\textcolor{white}{SPECspeed\textregistered 2026} \\ \textcolor{white}{Integer}}}} &
\thead[c]{\textbf{\makecell[c]{\textcolor{white}{speed} \\ \textcolor{white}{MT\textsuperscript{[A]}}}}} &
\thead[c]{\textbf{\textcolor{white}{Language}}} &
\thead[r]{\textbf{\makecell[r]{\textcolor{white}{KLOC}\\\textcolor{white}{files\textsuperscript{[B]}}}}} &
\thead[r]{\textbf{\makecell[r]{\textcolor{white}{KLOC}\\\textcolor{white}{hit\textsuperscript{[C]}}}}} &
\thead[l]{\textbf{\textcolor{white}{Application Domain}}} &
\thead[c]{\textbf{\textcolor{white}{ref}}} \\

& 801.xz\_s & MT & C++, C & 53 & 5 & Data compression & \cite{801-xz} \\
\hline
706.stockfish\_r & & & C++ & 13 & 5 & Game (chess) - A/B search, deep learning neural network & \cite{706-stockfish} \\
\hline
707.ntest\_r & 807.ntest\_s & MT & C++ & 16 & 5 & Game (othello) - A/B search with heuristic eval function & \cite{707-ntest} \\
\hline
708.sqlite\_r & & & C & 245 & 23 & SQL compiler/interpreter and database & \cite{708-sqlite} \\
\hline
710.omnetpp\_r & & & C++, C & 224 & 19 & Discrete event modeling - network and queuing simulations & \cite{710-omnetpp} \\
\hline
714.cpython\_r & & & C & 747 & 56 & Python interpreter & \cite{714-cpython} \\
\hline
& 817.flac\_s & MT & C++, C & 57 & 7 & Lossless audio codec & \cite{817-flac} \\
\hline
721.gcc\_r & 821.gcc\_s & MT & C++, C & 3,833 & 326 & C language optimizing compiler & \cite{721-gcc} \\
\hline
723.llvm\_r & 823.llvm\_s & MT & C++, C & 3,167 & 123 & C/C++ language optimizing compiler & \cite{723-llvm} \\
\hline
727.cppcheck\_r & 827.cppcheck\_s & MT & C++ & 287 & 111 & Static analysis of C/C++ code & \cite{727-cppcheck} \\
\hline
729.abc\_r & 829.abc\_s & & C++, C & 989 & 37 & Sequential logic synthesis and formal verification & \cite{729-abc} \\
\hline
734.vpr\_r & 834.vpr\_s & & C++, C & 210 & 30 & FPGA circuit place and route & \cite{734-vpr} \\
\hline
735.gem5\_r & 835.gem5\_s & & C++, C & 971 & 78 & Computer architecture simulation model & \cite{735-gem5} \\
\hline
& 838.diamond\_s & MT & C++, C & 239 & 12 & Bioinformatics - metagenomics and protein sequencing & \cite{838-diamond} \\
\hline
& 846.minizinc\_s & MT & C++, C & 372 & 33 & Constraint programming (solvers: gecode and chuffed) & \cite{846-minizinc} \\
\hline
750.sealcrypto\_r & & & C++, C & 39 & 5 & Security and privacy - Homomorphically Encrypted query & \cite{750-seal} \\
\hline
753.ns3\_r & 853.ns3\_s & & C++ & 942 & 71 & Discrete event network simulator for internet systems & \cite{753-ns3} \\
\hline
& 854.graph500\_s & MT & C & 10 & 1 & Graph analytics & \cite{854-graph500} \\
\hline
777.zstd\_r & & & C & 58 & 7 & Data compression/decompression & \cite{777-zstd} \\
\hline
\addlinespace 
\addlinespace
\rowcolor{tableheadspec}

\thead[l]{\textbf{\makecell[l]{\textcolor{white}{SPECrate\textregistered 2026} \\ \textcolor{white}{Floating Point}}}} &
\thead[l]{\textbf{\makecell[l]{\textcolor{white}{SPECspeed\textregistered 2026} \\ \textcolor{white}{Floating Point}}}} &
\thead[c]{\textbf{\makecell[c]{\textcolor{white}{speed} \\ \textcolor{white}{MT\textsuperscript{[A]}}}}} &
\thead[c]{\textbf{\textcolor{white}{Language}}} &
\thead[r]{\textbf{\makecell[r]{\textcolor{white}{KLOC}\\\textcolor{white}{files\textsuperscript{[B]}}}}} &
\thead[r]{\textbf{\makecell[r]{\textcolor{white}{KLOC}\\\textcolor{white}{hit\textsuperscript{[C]}}}}} &
\thead[l]{\textbf{\textcolor{white}{Application Domain}}} &
\thead[c]{\textbf{\textcolor{white}{ref}}} \\

& 800.pot3d\_s & MT & Fortran & 12 & 1 & Solar physics: finite diff method, conjugate gradient solver & \cite{800-pot3d} \\
\hline
& 803.sph\_exa\_s & MT & C++ & 3 & 1 & Astrophysics - Smoothed Particle Hydrodynamics (SPH) & \cite{803-sphexa} \\
\hline
709.cactus\_r &  809.cactus\_s & MT & C++, C & 187 & 19 & Astrophysics - relativity, finite difference, time integration & \cite{709-cactus} \\
\hline
& 811.tealeaf\_s & MT & C & 5 & 1 & High energy physics & \cite{811-tealeaf} \\
\hline
& 816.nab\_s & MT & C & 26 & 2 & Molecular modeling & \cite{816-nab} \\
\hline
& 820.cloverleaf\_s & MT & Fortran & 10 & 1 & Explicit hydrodynamics & \cite{820-cloverleaf} \\
\hline
722.palm\_r & 822.palm\_s & MT & Fortran & 218 & 6 & Atmospheric science & \cite{722-palm} \\
\hline
731.astcenc\_r & & & C++ & 43 & 8 & Computer vision - Adaptive Scalable Texture Compression & \cite{731-astcenc} \\
\hline
736.ocio\_r & & & C++ & 183 & 13 & Color management for visual effects and animation & \cite{736-ocio} \\
\hline
737.gmsh\_r & & & C++, C & 721 & 35 & Finite element mesh generation & \cite{737-gmsh} \\
\hline
748.flightdm\_r & & & C++ & 100 & 14 & Flight dynamics models for aeronautics & \cite{748-flightdm} \\
\hline
749.fotonik3d\_r & 849.fotonik3d\_s & MT & Fortran & 15 & 2 & Computational Electromagnetics (CEM) & \cite{749-fotonik3d} \\
\hline
& 857.namd\_s & MT & C++ & 9 & 2 & Classical molecular dynamics simulation & \cite{857-namd} \\
\hline
765.roms\_r & 865.roms\_s & MT & Fortran & 585 & 13 & Regional ocean modeling & \cite{765-roms} \\
\hline
766.femflow\_r & & & C++ & 2,505 & 23 & Fluid dynamics: high-order finite element method & \cite{766-femflow} \\
\hline
767.nest\_r & 867.nest\_s & MT & C++ & 208 & 17 & Neuroscience simulator for spiking neural network models & \cite{767-nest} \\
\hline
772.marian\_r & 872.marian\_s & MT & C++ & 219 & 15 & Neural machine translation for written language & \cite{772-marian} \\
\hline
782.lbm\_r & & & C & 1 & 1 & Computational fluid dynamics, Lattice Boltzmann Method & \cite{782-lbm} \\
\hline
& 881.neutron\_s & MT & C & 4 & 1 & Physics simulation of neutron transport in nuclear reactors & \cite{881-neutron} \\
\hline
\addlinespace
\end{tabularx}
\par
\begin{flushleft}
\small 
~\\
~\\
\textsuperscript{[A]} MT indicates the SPECspeed benchmark uses parallelism; either multi-threading or multi-tasking. \\
\textsuperscript{[B]} KLOC = line count in thousands. Counts all files in the source directory; includes comments and blank lines. \\
\textsuperscript{[C]} KLOC = line count in thousands. Only counts lines which are exercised based on benchmark code coverage metrics.
\end{flushleft}
\setlength{\extrarowheight}{0pt}
\end{table*}

%% file: 30-changes.tex
\section{Key Changes in SPEC CPU2026}

CPU2026 introduces a series of significant enhancements and strategic changes compared to its predecessor, CPU2017. These modifications are designed to reflect the evolution of modern hardware, contemporary software stacks, and principled benchmarking methodologies. The key changes are summarized below, with further details later in this paper.

\para{Enhanced Suite Diversity} The composition of the suite has evolved strategically. The new suite deliberately elicits a broader set of microarchitecture responses than before, covers a wider span of application domains, and makes more extensive use of multiple workloads per benchmark. All the benchmarks can be seen in Table~\ref{tab:spec-benchmarks}.

\para{Increased Scale and Resource Requirements} The suite is substantially larger, featuring more component benchmarks. To reflect the growing memory capacity of modern systems, the memory footprint of the SPECspeed$^\circledR$ large multi-threaded benchmark suite has increased from 16\,GB to 64\,GB. The SPECrate$^\circledR$ throughput benchmark suite's footprint remains the same at 2\,GB per copy. Each benchmark retains \texttt{test} and \texttt{train} sizes, which have much shorter runtimes than the measured and reported \texttt{ref} size.

\para{Updated Language Standards} The suite adopts modern ISO standards: C18 (from C99), C++17 (from C++03), and Fortran 2018 (from Fortran 2003). In addition to OpenMP, language parallelism is employed through C++'s \texttt{std::thread} and Fortran's \texttt{DO\_CONCURRENT}. Some floating point benchmarks require precise math for functionality and verification, and thus may not work correctly with fast/relaxed math. 

\para{New Analytical Capabilities} The harness and reporting tools have been enhanced. The raw output now includes additional population statistics for SPECrate such as coefficient of variation, quartiles, min/max/average copy times, and standard deviation, enabling more rigorous analysis of multi-copy runs. Additionally, a new exhibition run style, Rolling Round-Robin Rate (RRR), has been introduced to facilitate systems research with heterogeneous, multiprogrammed workloads.

\para{Expanded Reporting Categories} To better represent real-world deployment scenarios, two major reporting categories have been added. First, official, compliant scores can now be submitted from bare-metal instances on public cloud platforms, moving beyond the ``estimated'' status of the past. Second, a new category distinguishes between results obtained using vendor-supported and community-supported open-source compilers. This is motivated by the widespread use of community compilers like GCC and LLVM, and their distinct performance characteristics compared to vendor compilers  \cite{comparing_c_compilers, gcc_vs_icc, compilers_on_amd, cpu2017_energy_characterization}, providing a more representative view of the performance users experience on different software stacks.

%% file: 40-benchmarks.tex
\section{Benchmark Development}

The benchmark development pipeline commenced by soliciting candidates from the open-source community, academic researchers, and industry practitioners. This stage was conducted through the CPUv8 search program \cite{cpuv8_search}, which ran from February 2020 to March 2023. Candidates then progressed through the steps of adaptation into a formal benchmark, workload selection, and performance characterization, before being considered for final selection. It is important to note that this pipeline is not strictly linear, as these stages ran concurrently: new candidates are continually introduced while others mature or are culled from the process. 

\subsection{Search Program}
\label{sec:search_program}
The CPUv8 benchmark search program proved remarkably successful, drawing 33 benchmark candidates into consideration. An impressive 29 of these completed initial porting and workload definition \cite{cpuv8_step3} to advance into the benchmark selection stage, with 24 of those external candidates ultimately integrated into the final suite (Table~\ref{tab:spec-benchmarks}). A significant portion of these submissions originated from projects with strong open-source community backing. This offered collaborations with authors and their communities, which became essential for cross-system porting and for addressing issues discovered during the development and testing process. The resulting benchmarks are diverse and meaningful, including prize-winning drug discovery programs vital to COVID-19 vaccine research \cite{namd_bell_prize}, flight simulators used by government agencies \cite{jsbsim_cpuv8}, brain modeling tools \cite{nest_cpuv8}, and even a media application that won an Academy Award \cite{ocio_oscars}. The intensive evaluation process didn't just fortify these applications into SPEC CPU benchmarks; it also led to valuable fixes and enhancements that were upstreamed back to the original projects. The development process stands as a powerful testament to the symbiotic relationship between SPEC and the open-source world.

\subsection{Adaptation}
\label{sec:development_process}

The development of the SPEC CPU suite has always been guided by a core philosophy that prioritizes the selection of real-world applications with significant user bases. While purpose-built microbenchmarks can be valuable for exercising specific CPU features, the committee intentionally biases towards production software used in the field. This approach ensures that the performance characteristics measured are representative of genuine computational workloads, and that optimization of the hardware systems running these benchmarks does indeed help improve performance in the community. Transitioning a real-world application into a trusted benchmark requires an adaptation process to satisfy the non-negotiable principles of determinism, reproducibility, and portability.

This adaptation can be analogized to studying an organism in a controlled environment versus its native habitat. To enable systematic and reproducible study, the \emph{organism} of the application must be carefully adapted to the \emph{laboratory} of the SPEC CPU harness. This process involves a series of modifications designed to eliminate external sources of variance and ensure that the benchmark clearly measures the performance of the System Under Test (SUT), and not the surrounding environment. The fundamental goal is to ensure that the benchmark executes an identical amount of user-space work across any compliant system, and produces a identical result on every run within a given tolerance. To achieve this level of rigor, each candidate benchmark undergoes a series of modifications:

\para{Elimination of Non-Determinism} Sources of high entropy like reads from \texttt{/dev/random}, or calls to hardware-based true random number generators, are replaced with deterministic pseudo-random number generators like the Mersenne Twister \texttt{std::mt19937}. Additionally, the C++17 standard does not define strict results for \texttt{std::sort}, so we convert those usages to \texttt{std::stable\_sort} which is guaranteed to produce the same results across library implementations; and likewise for other unstable standard algorithms that have stable equivalents. These compromises provide reproducibility in both application control logic and data results, from run to run and across hardware systems and compiler libraries.

\para{Maximization of Portability} All platform-specific code, including hand-coded assembly and compiler intrinsics, is removed and replaced with portable C, C++, or Fortran equivalents (\S\ref{sec:portability}). This is often the most significant deviation from the original application but is essential for ensuring forward and backward compatibility of the suite across decades of architectures. Consequently, benchmark candidates with a heavy reliance on non-portable, hand-optimized code are generally disfavored and do not make it far in the selection process.

\para{Isolation from the Environment} The benchmark is decoupled from its external execution environment. This includes removing calls that query or modify the environment (\texttt{getenv}, \texttt{setenv}), eliminating internal measurement and control (\texttt{gettimeofday}, \texttt{getrusage}, \texttt{setrlimit}), and excising any debugging hooks that could alter program behavior (\texttt{signal}, \texttt{sigaction}, \texttt{dlopen}).

\para{Focus on User-Space Execution} To ensure the benchmark primarily measures CPU and memory subsystem performance rather than OS efficiency, system calls are minimized. The target is for at least 95\% of the execution time to be spent within the user-space code provided by the benchmark. Exceptions are consciously made for ubiquitous library functions (\texttt{malloc}, \texttt{strcpy}, standard math functions), as their performance is of broad interest to the community.

\para{Suppression of Threading Artifacts} For single-threaded SPECrate benchmarks derived from multithreaded applications, synchronization primitives like locks and mutexes introduce a ``threading tax'' that is irrelevant to measuring single-threaded performance. This overhead is systematically identified and suppressed for SPECrate builds, while the necessary hooks are preserved for use in the multi-threaded SPECspeed version of the benchmark.

\para{Validation Across Multiple Systems} The committee runs a continuous integration process through kit build and testing which involves extensive validation across a matrix of hardware, operating systems, and compilers. The committee membership consists of representatives from companies that implement the x86, ARM, POWER, and RISC-V architectures; their products are used to exercise the candidate benchmarks under Linux, Windows, macOS, and other operating systems. We build the codes using the latest open-source compilers, GCC and LLVM, as well as vendor compilers from Intel, AMD, IBM, NVidia, HP/Cray, and Microsoft. Each of these systems must produce the same answer, within tolerance, of the golden reference results to be considered a successful run. This exhaustive testing at multiple optimization levels (\texttt{-O2}, \texttt{-O3}, LTO, PGO) validates the benchmark's portability fitness. 

\para{Legal} Finally, a thorough legal review is conducted on all included source code and data inputs. Every file must have its provenance cited and be licensed correctly, to ensure the suite is commercially distributable on solid legal foundations \cite{cpu26_license}.

\subsection{Floating-Point versus Integer}

Traditionally, SPEC CPU is split into two suites: integer (INT) and floating-point (FP). Although some may consider this as a throwback from an era when floating-point units (FPUs) were optional co-processors, the bifurcation actually began with SPEC CPU2000. So, although CPUs from that era already integrated powerful FPUs, the suites were created to distinguish workload domains. In the modern era, this can be seen where INT versus FP tends to correlate loosely with cloud computing versus high-performance computing (HPC). Since 2000, SPEC has employed a quantitative criterion for this classification: applications with over 10\% of their dynamic instructions being floating-point were designated FP, while those with less than 1\% were designated INT.

This clear delineation, effective through the CPU 2017 suite, has become increasingly blurred by architectural evolution. In contemporary processors, integer and floating-point SIMD execution units are often co-located or share scheduling resources. Moreover, it is now common for predominantly integer-based SIMD applications to leverage floating-point load/store instructions for vector memory accesses, which complicates instruction-based accounting; the counts can also differ significantly based on compiler optimization levels \cite{cpu2000}. This architectural and programmatic convergence has created a significant ``gray zone'' for applications that exhibit an FP instruction composition between 1\% and 10\%.

A substantial number of open-source candidates for the CPU2026 suite fell squarely into this ambiguous category, possessing characteristics of integer workloads while incorporating a non-trivial amount of floating-point operations. To resolve this, the SPEC committee adopted a qualitative methodology, applied on a case-by-case basis. For each application within the gray zone, the final classification was determined by a pragmatic assessment based on the application's primary computational purpose and its established reputation within its user community as either an integer or floating-point workload.

\subsection{Workload Selection}
\label{sec:workload_selection}

The performance profile of an application is highly sensitive to its input workloads, which encompass command-line arguments, configuration files, and datasets. The SPEC CPU suite has consistently biased its selection towards real-world datasets that are relevant and representative of common usage in the field. The challenge of selecting a single, representative input from a near-infinite space is well-documented \cite{picking_inputs, best_practices}. To address this, SPEC heavily relies on the guidance of benchmark authors and their respective user communities, a principle institutionalized through the search program (\S\ref{sec:search_program}). This collaborative approach ensures that the selected workloads are authentic and accurately reflect the benchmark's intended application domain.

One objective for CPU2026 was to enhance the suite's resilience to targeted, non-generalizable optimizations. This goal was pursued by expanding the use of multiple, distinct workloads for a single benchmark—a capability present in prior suites but more systematically employed in this version. This multi-workload approach serves several purposes: it increases the diversity of exercised code paths, captures a wider range of application behaviors, and makes the benchmark more difficult to crack, in that a compiler or hardware optimization might yield a significant speedup on one specific input but fail to generalize across the others. The diversity among these sub-workloads is qualitatively visualized in the Basic Block Vector (BBV) plots explained in Figure \ref{fig:bbv_perf_853} and presented in the Appendix for all single-threaded benchmarks (Figs.~\ref{fig:bbv_int_rate} and \ref{fig:bbv_perf_fp}), where one can compare the execution profiles across inputs.

This methodological enhancement, however, introduces a notable trade-off for performance analysis. By design, the aggregate score for a multi-workload benchmark represents a composite of behaviors, which could obscure and dilute the characteristics of its individual components. For instance, 729.abc and 727.cppcheck contain a mix of core-bound, high-IPC workloads and memory-bound, low-IPC workloads (as shown in Fig.~\ref{fig:bbv_int_rate}). Consequently, the aggregate benchmarks cannot be singularly categorized as high-bandwidth or high-IPC, as their subcomponents exhibit a multitude of behaviors. To support deeper, more granular analysis and to empower the research community, each benchmark's documentation includes instructions for creating custom input sets. This enables further investigation, in the spirit of research projects such as the Alberta Workloads \cite{alberta_workloads}.

\section{Principles of Benchmark Selection}
\label{sec:selection_principles}

The methodology behind the SPEC CPU suite has been recognized as a best practice in performance evaluation, with adjacent benchmarking communities either adopting similar principles or validating the approach \cite{evaluating_the_evaluations, better_bench}. This section details the foundational tenets of crafting the suite.

\subsection{Culling}
\label{sec:culling_principles}

The selection of benchmarks for a SPEC CPU suite is a rigorous, multi-faceted process designed to curate a balanced, relevant, and scientifically sound collection of workloads. While many promising applications are proposed as candidates, a process of principled attrition is necessary to ensure the final suite meets stringent quality standards. The primary criteria that led to excluding candidates are given here (with concrete examples offered in \S\ref{sec:culled}). The first two rationales listed below are fundamental, and the remaining are based on suite-level composition, logistical, and technical considerations.

\para{Determinism} A foundational requirement for any SPEC benchmark is the execution of a deterministic and reproducible quantum of work. This principle was a critical filter for applications based on heuristic search algorithms, such as linear solvers, constraint programming, or those employing gradient descent. These programs often exhibit non-deterministic execution paths, where minor architectural or compiler differences can lead to ``short-cuts'' to the solution. As this violates the core tenet of equal work across all test platforms, such candidates, while valuable in their own right, are unsuitable for comparative CPU benchmarking. If deterministic execution was not possible, the candidate was excluded.

\para{Development Divergence} A second essential factor is the benchmark's representativeness of its real-world counterpart after undergoing necessary modifications for portability. To ensure broad compatibility, all platform-specific code, such as intrinsics and hand-tuned assembly, must be removed. For certain domains, particularly modern AI and media encoding applications, this ``defanging'' process caused the benchmark's performance profile to diverge significantly from the original highly optimized application. When the resulting portable code no longer reflected the computational characteristics of the software used in the field, its value as a representative benchmark was diminished, leading to its removal.

\para{Domain Redundancy and Scope} To ensure broad coverage, the suite cannot be over-represented by a single application domain. In fields with multiple high-quality candidates (e.g., file compression, scripting languages), a ``horse race'' ensued, with only the most prominent or relevant candidate being selected. Conversely, applications with an exceptionally narrow user base that did not address broad industry challenges were deemed too specialized for inclusion.

\para{Codebase Health and Maintainability} Preference was given to modern, actively maintained codebases. Candidates based on decades-old, unmaintained code were rejected as they are poor indicators of future computational trends. All code must be clear, well-structured, and maintainable by the committee for the suite's lifespan. Obfuscated or unwieldy code presents an unacceptable maintenance burden.

\para{Insufficient Maturity} Some candidates, though promising, were not sufficiently developed to meet the production timeline and were deferred for future consideration. Usually this was due to lack of portability across systems, which required more development and testing effort.

\para{Excessive I/O or ``Peaky'' Profiles} The suite is designed to measure CPU and memory performance. Candidates with significant file I/O were rejected as their performance would be unduly influenced by storage subsystems and OS calls \cite{cpu2006_fileio}. Similarly, benchmarks with ``peaky'' profiles \cite{cpu2006_hotspots}, where runtime is dominated by a few functions, were disfavored due to their vulnerability to narrow or non-portable compiler optimizations that would not benefit general-purpose computing.

\para{Potential Bias} To maintain objectivity, candidates perceived as being pre-tuned for a specific committee member's architecture are heavily scrutinized and often excluded to prevent inherent bias. The development process itself is conducted transparently, ensuring there is no intentional attempt to mislead or obfuscate the benchmark's behavior.

\subsection{Selection}

Despite SPEC's historical foundation of selecting benchmarks from real-world, portable source code, a persistent misconception is that the suite is "synthetic" \cite{ampere_benchmarks_youtube}, or not composed of "real workloads" \cite{amazon_benchmarks_youtube}. This paper directly addresses that perception. The final selection process is guided by  principles designed to curate a balanced, diverse, and forward-looking suite, rooted in a transparent and open development culture where proprietary interests are subordinated to the creation of technically credible and vendor-neutral benchmarks \cite{john_suite_growth}.

The selection process was also driven by a goal to enhance the suite's behavioral diversity and address known gaps from previous versions \cite{lizy_2017}. A notable outcome in the integer suite is the inclusion of benchmarks bottlenecked by the CPU's front-end, characterized by heavy instruction delivery pressure, high ITLB miss rates, or frequent branch mispredictions. This shift reflects the changing landscape of modern software, which increasingly features large code footprints and complex control flow, moving beyond just the predominantly back-end-bound workloads of previous suites \cite{dcperf}.

\subsection{Number of Benchmarks}

A defining characteristic of CPU2026 is an expanded benchmark count of 52 (up from 43 in CPU2017), made possible by the prolific outcome of the search program (\S\ref{sec:search_program}). The ability to craft a larger suite provided a chance to address key challenges observed in the academic and industrial use of previous suites. Historically, concerns have been raised about researchers creating subsets of benchmarks from inside and outside of the suite without a clear justification \cite{MisSPECulation}, a practice which confounds reviewers and severely hinders the direct comparison of results across studies \cite{abuse_of_spec_subsetting}. Furthermore, past suites have been critiqued for potential redundancy in workload behavior \cite{lizy_subsetting2, superfluous_spec, designing_workloads}. By increasing the number of benchmarks, we aimed for a greater diversity of microarchitectural behaviors, programming styles, and application domains. A larger, more diverse suite offers more optimization problems to solve, diminishes the impact of ``cracking'' a single benchmark via the 1/N rule, and disincentivizes non-portable, benchmark-specific optimizations.

The expansion introduces trade-offs, namely analysis complexity. While individual benchmarks are shorter to keep the total runtime of the suite comparable to that of CPU2017, the increased analysis volume is an intentional feature. It steers hardware and compiler systems designers away from narrow tuning and toward developing more general-purpose optimizations that ultimately deliver greater uplift to end-users.

\subsection{Culled Benchmark Candidates}
\label{sec:culled}

In response to reviewer feedback requesting concrete examples, this section details the rationale behind the exclusion of several notable candidates and application domains. These stories illustrate the practical application of the selection principles outlined in Section~\S\ref{sec:culling_principles}.

\para{Modern AI Workloads} We evaluated portable CPU inference engines from the transformer/LLM era, including llama.cpp \cite{llama-cpp} and whisper.cpp \cite{whisper-cpp}. These candidates advanced deep into the evaluation process due to the domain's importance. However, restricting them to portable C++ codepaths (with intrinsics removed) caused a fundamental divergence from their real-world behavior. The resulting benchmarks became uncharacteristically compute-bound, with longer instruction path lengths between transactions and significantly lower MPKI than their field-deployed counterparts. Both devolved into workloads where 95\% of the runtime was spent in a single, inefficient hot loop. Optimizations on such a narrow, unrepresentative loop would offer no value to the industry. Furthermore, llama.cpp faced significant challenges with result verification across systems, requiring us to ``put inference on rails''—forcing determinism by capturing the sequence of generated tokens offline and then forward-feeding them back into the model during runtime—a process which further separated the candidate from realistic use.

\para{Cryptography} While a critical CPU workload, production cryptography (e.g., AES, RSA) is dominated by hand-tuned assembly and ISA intrinsics. Removing these architecture-specific hooks to create a portable benchmark results in code that is not representative of real-world deployments. A generic, unoptimized crypto workload would be a poor proxy for what is run in the field. We did retain 750.sealcrypto (homomorphic encryption) because its core exercises finite-field mathematics, which is an algorithmic phase that will remain relevant to general-purpose CPUs even as other parts of the HE stack migrate to specialized hardware.

\para{Media Codecs} We considered AV1/AOM \cite{av1oam} and Opus \cite{opus_codec}, two codecs used widely for internet video and audio. Both make such heavy use of architecture-specific assembly that removing these implementations rendered their performance profiles unrepresentative. And in the case of Opus, the program processed audio so quickly that the workload shifted from CPU-bound to I/O-bound, with multi-copy SPECrate runs stalling on disk activity while the CPU remained idle. 817.flac doesn't have this issue since the benchmark is a multi-threaded program writing into just a single file.

\para{Key-Value Stores} A high performance key-value store library was adapted as a benchmark and reached the final stages of selection. After extensive experimentation, the committee concluded that the benchmark's workload did not reflect real-world deployments of that application, as it exhibited pathological behavior such as abnormal memory bandwidth usage and redundant data decompression. Various different configurations were attempted to alleviate these issues, but it became clear that none of the options could represent field usage meaningfully. As we ran out of time for more development, this database candidate was dropped.

\para{Duplicate Application Domains and Behaviors} We evaluated four high-quality data compression candidates (xz, brotli \cite{brotli}, 7-zip \cite{seven-zip}, zstd). While compression is an important domain, the performance behaviors of these four were highly similar, creating redundancy. We selected 777.zstd for the intrate suite due to its widespread adoption (Linux kernel, cloud services, databases). We retained 801.xz for the intspeed suite because it was the only compression candidate that offered a multithreaded implementation.

\para{Non-Determinism and the ``Equal Work'' Principle} Some candidates used search or non-linear optimization, where different platforms can complete different amounts of work while still arriving at the same solution. For example, 737.gmsh originally included an adaptive mesh refinement phase whose iteration count could vary by 30\% depending on FP numerics, compiler flags, or ISA. By disabling just the adaptive phase of meshing, we successfully created a fully deterministic workload while still retaining field behavior. In contrast, the candidate HiGHS \cite{highs-dev} is a linear programming solver that could not be similarly adapted. After consultation with its community \cite{highs_cpuv8}, it was determined that guaranteeing equal work would require using a trivial input problem with a single solution path, resulting in a workload which is not representative of field behavior. A different approach proved successful for the 846.minizinc constraint solver benchmark, which avoids this pitfall by using unsatisfiable problem inputs. When no solutions exist, the solver must exhaustively search all paths, naturally ensuring equal work on all systems.

\para{Legacy Kernels and Bespoke Microbenchmarks} A proposal to include the NASA Parallel Benchmarks \cite{nasa-npb} was declined. While historically significant, these kernels are now over three decades old, and SPEC prioritizes modern, actively maintained codebases. Similarly, a microbenchmark inspired by the financial services industry was rejected. Although the domain is of interest, the small code did not represent modern FSI algorithms, and SPEC disfavors member-authored bespoke benchmarks to avoid any perception of bias.

\para{Virtualization and containers} These environments are important and growing, but SPEC CPU measures natively compiled, application-level CPU and memory behavior under a single-workload harness. Virtualization adds system-level effects and policies that fit better under other SPEC suites and methodologies, such as SPECvirt$^\circledR$2021 \cite{specvirt}.

\para{Looking Forward} These experiences highlight a central challenge: balancing portability with representativeness for workloads that rely on ISA-specific optimizations. For future suites, the committee will evaluate whether allowing optional, architecture-specific libraries alongside a generic reference implementation could enable the inclusion of these important modern applications without compromising the suite’s core principles of equal work, portability, and vendor neutrality.

\section{Longevity and Portability}
\label{sec:portability}

After the benchmarks were selected, the focus turned to another mainstay philosophy of SPEC CPU, namely portability. This principle is upheld through two key practices: methodical adherence to ISO language standards and a commitment to high-quality, warning-free code. This approach is the primary reason for the suite's exceptional longevity, as evidenced by suites like SPEC CPU2000 remaining relevant a quarter-century after their release, particularly in embedded systems.

By ensuring both forward compatibility (allowing older suites on new systems) and backward compatibility (enabling new suites on legacy hardware), this standards-based design preserves SPEC CPU's value across decades. The CPU2026 suite continues this tradition with its baseline of C++17, C18, and Fortran 2018. The following subsections detail the specific development practices and validation efforts essential to upholding these principles.

\subsection{Code Hardening and Standards Compliance}
\label{sec:code_hardening}

This commitment to standards is enforced through a multi-faceted code hardening process. The goal is not merely to compile the code, but to ensure it is robust, warning-free under pedantic mode (\texttt{-Wpedantic}), and free of ambiguous behavior. This ensures that a standards-compliant compiler developed decades from now will be able to build the suite. This process addressed several common categories of issues:

\para{Elimination of Undefined Behavior} A primary focus was the elimination of undefined, unspecified, and implementation-defined behavior, which can lead to non-deterministic results or outright failures across different compilers and platforms. This was achieved through both static analysis via compiler warnings and dynamic analysis using runtime sanitizers. These efforts rectified critical issues such as incompatible types which resulted in dangerous pointer conversions \cite{jsbsim_ptrconv} and data overflows \cite{cppc_sani}; uninitialized member variables that led to divergent behavior and segmentation faults \cite{diamond_uninit, ntest_uninit, nest_sanitizer}; and subtle object construction races like the ``initialization-order fiasco'' detected by Address Sanitizer \cite{diamond_fiasco}.

\para{Ensuring Data Model Portability} A second critical task was to ensure the code was agnostic to platform-specific data models. This involved correcting a class of warnings related to mismatched integer types and signed/unsigned comparisons, often by standardizing on consistent types for object sizes and indices \cite{chuffed_types}. Mismatches in type usage were also corrected to prevent portability issues, for example on platforms where fundamental types have different sizes \cite{gmsh_mac_ce}.

\para{Modernization to C++17 Standards} The process also involved modernizing legacy codebases to comply with the selected C++17 standard, which required addressing a wide range of issues identified by modern, standards-compliant compilers \cite{gmsh_warnings, vpr_cxx20}. Specific modernization efforts included replacing deprecated features such as \texttt{std::bind2nd} with modern lambda functions, converting \texttt{std::random\_shuffle} to the newer \texttt{std::shuffle} with a deterministic random engine, removing the now-obsolete \texttt{register} keyword \cite{gmsh_cxx17}, and standardizing the use of \texttt{std::nan} in place of custom NaN implementations \cite{ns3_nan}.

\para{Removal of Non-Standard Language Extensions} Finally, to guarantee maximal portability across all compliant compilers, non-standard language features and compiler-specific extensions were spliced out. This included replacing compiler-specific attributes like `always\_inline' with the standard `inline' keyword, and removing uses of the non-standard `restrict' keyword. This adherence to the ISO standard ensures that the performance of the benchmarks is not dependent on proprietary features that favor a particular compiler.

\subsection{Endianness}
\label{sec:big_endian}

To ensure the broad applicability and architectural neutrality of the CPU2026 suite, a dedicated validation effort was undertaken to guarantee portability to big-endian systems. The primary platform for this validation was IBM AIX running on the POWER architecture. This process uncovered and led to the resolution of several classes of portability issues, with many of the resulting patches being upstreamed to the originating open-source communities \cite{jsbsim_be, chuffed_be, abc_be, gem5_be}.

\para{Endian-Dependent Input Data Formats} Several benchmarks assumed a little-endian format for their on-disk input files, requiring modifications to ensure data could be correctly interpreted on big-endian systems. For instance, the 731.astcenc benchmark initially failed because its input textures were encoded using a third-party library lacking big-endian support. The resolution involved transitioning the input data to a new, endian-agnostic format. Similarly, 772.marian presumed a little-endian layout for its model files, which was rectified by integrating byte-swapping routines into the data loading process for big-endian systems.

\para{Memory Layout and Type-Punning Assumptions} A common class of errors stemmed from C/C++ code that made implicit assumptions about the in-memory byte order of data structures, often through pointer casting and dereferencing. A verification failure in 748.flightdm was traced to an unsafe type-cast that violated memory layout assumptions on big-endian systems. Another in 846.minizinc manifested due to layout of bit fields in union types where the union held either pointers or numbers, and these fields fell out of alignment. Collaboration with the upstream developers resulted in patches that implemented more robust, endian-neutral data handling \cite{jsbsim_be, chuffed_be}. Code in 729.abc and 735.gem5 also contained pointer dereferences and data operations that implicitly assumed a little-endian memory model. These sections were refactored to use endian-agnostic methods and explicit endianness checks, with fixes contributed back to the respective projects \cite{abc_be, gem5_be}.

\para{Exposure of Latent Software Bugs} The porting process of 721.gcc to AIX uncovered a latent bug in the GCC compiler's \texttt{tree-vrp} optimization pass specific to big-endian targets \cite{gcc_be}. As backporting the upstream fix was infeasible for the software version used by the benchmark, the committee implemented a targeted workaround by disabling the problematic pass (\texttt{-fno-tree-vrp}) from one specific workload that was affected. This ensured consistent benchmark execution across all platforms without altering its fundamental behavior.

\subsection{Operating Systems}

The committee prioritized enabling CPU2026 to run across multiple operating systems, including Microsoft Windows running on both x86-64 and aarch64 platforms. Most of the applications considered for CPU2026 were developed for Unix-like systems, which means they had either limited or no prior support for running on Windows. A key aid to supporting Windows was to use MinGW (Minimalized GNU for Windows) \cite{mingw} with gcc and gfortran compilers.  Using MinGW helped diagnose whether issues were attributable to the Windows OS, to Windows compilers, or to a code dependency on GNU or POSIX functions.
 
Some applications needed little to no modification; others required substantial investigation and patches. For example, 721.gcc required changes in around 6000 distinct lines in a code base of over 4 million total lines. Bringing 735.gem5 to Windows/MSVC included splicing out vast chunks of the code base that were unexercised by the chosen workloads, an approach which greatly reduced the total porting effort.
 
The source code modifications fell into several common categories. A significant portion of the work involved refactoring file I/O and path handling to accommodate Windows-specific file system conventions. Another common task was resolving data model discrepancies, most notably by addressing differences in the size of the \texttt{long} data type (4 bytes on Windows vs. 8 bytes on most 64-bit Unix-like systems). Extensive changes were needed to resolve platform-specific dependencies, typically by substituting non-portable GNU/POSIX library functions with Windows-native equivalents and including the correct header files. In addition, some changes were needed to handle OS-specific constraints, such as maximum file path length and executable file size.

This validation process uncovered real computation bugs in the software, one which required consulting an astrodynamics textbook written in 1971! \cite{jsbsim_bugfix}. Other issues that caused Windows-only errors were proactively upstreamed and accepted by the community \cite{gem5_port, nest_bugfix}. A code sequence with multiple virtual base pointers in 734.vpr exposed a memory size calculation error in LLVM, which was only seen in the Microsoft ABI \cite{vpr_llvm}.

These initiatives underscore a key benefit of cross-platform testing. Porting to less common architectures (big-endian), or operating systems (Windows, macOS, Android), rigorously tests the implicit assumptions made during development on more homogeneous platforms. The process not only hardens the benchmark code, making it more robust and portable, but also provides tangible benefits back to the open-source communities through upstreamed bug fixes. This feedback loop enhances the quality of both the SPEC CPU suite and the foundational applications upon which it is built.

\subsection{IO Analysis and Reduction} 

The SPEC CPU benchmark suite is designed primarily to assess a processor’s computational performance and its interaction with the hierarchical memory subsystem. Introducing disk or network I/O into such benchmarks can lead to unpredictable delays that vary significantly across platforms, thereby obscuring the true scalability of the processor. Sources of variability include differences in storage device performance (e.g., NVMe, SSD, HDD), file system overhead, and network latency. When the CPU is forced to wait for I/O operations, overall utilization decreases, resulting in an inaccurate representation of computational capability. This issue is particularly pronounced in multi-threaded workloads, where threads can block waiting on I/O rather than fully exercising the processor’s computational resources \cite{cpu2006_fileio}.

A variety of Linux-based tools were employed to quantify I/O interactions and assess their impact on benchmark accuracy including \texttt{strace} \cite{strace_project} to monitor system call interactions with the kernel and \texttt{sar} \cite{godard_sysstat} to capture historical I/O activity. Additionally, \texttt{emon} \cite{intel_emon} was used to compare I/O bandwidth between 1-copy and 256-copy configurations. The analysis focused on quantifying the frequency and volume of I/O-related system calls to ensure that workloads remained fundamentally compute-bound, and that many-copy scaling was not impacted by increases in I/O activity.

Consistent with the suite's long-standing character, several candidate workloads required optimization to bring their I/O behavior in line with (or better than) prior suites. Several mitigation techniques were applied across the benchmarks. To reduce I/O volume, input and output files were trimmed, and early or periodic writes were removed. At the code level, unbuffered output operations (e.g., \texttt{fprintf}) were replaced with buffered equivalents like \texttt{std::stringstream} to decimate the number of write system calls \cite{ntest_io, vpr_io}. Additionally, extraneous stream flushes were removed by replacing \texttt{std::endl} with the newline character '\texttt{\textbackslash n}' \cite{gem5_io}, and inefficient, repeated \texttt{open}/\texttt{close} call sequences were consolidated \cite{ns3_io}. These changes leverage the operating system’s ability to buffer writes within a memory page, thereby minimizing the frequency of system calls and keeping the performance focus on the CPU and memory subsystems.

\subsection{Memory Safety and Code Sanitization}
\label{sec:memory_safety}

As part of a broader commitment to delivering high-quality, robust software, a specific focus was placed on ensuring memory safety. Recent guidance from expert practitioners \cite{acm_safety, safety_for_skeptics, practical_security} as well as government agencies \cite{lord2023urgent, house2024back}, has made it an important aspect of modern software development. This industry-wide imperative is particularly relevant for the SPEC CPU suite due to its long-term, archival nature; in fact, modern analysis tools have identified out-of-bounds violations in CPU2006 \cite{AddressSanitizer} and CPU2017 \cite{heapcheck_safety, floatzone}. Thus, as benchmarks are frozen upon release and used for decades, there is a heightened responsibility to ensure their codebases are free from latent defects, particularly memory safety vulnerabilities.

To meet this responsibility, a comprehensive validation process was employed, utilizing a suite of software and hardware-assisted sanitization techniques. Each benchmark was tested with the Address Sanitizer (ASan) from both GCC and LLVM \cite{ASAN_gcc, ASAN_llvm} to detect issues like buffer overruns and use-after-free errors. The multithreaded SPECspeed benchmarks were then tested with Thread Sanitizer (TSan) \cite{ThreadSanitizer} to identify data races. This process was augmented by hardware-accelerated validation using the ARM Memory Tagging Extension (MTE) available on AmpereOne$^\circledR$ processors \cite{ampere_mte}.

This multi-faceted approach proved effective, successfully identifying and enabling the correction of several previously unknown issues. For instance, MTE was instrumental in discovering memory safety defects in 767.nest \cite{nest_safety} and 735.gem5 \cite{gem5_sani}. TSan then uncovered thread data races in 867.nest \cite{nest_threadrace} and 837.gmsh \cite{gmsh_threadrace}. All identified issues were patched with help from the community, with fixes contributed back to the respective upstream open-source projects. This exhaustive sanitization process provides high confidence for memory safety, at least within the scope of the code paths exercised by each benchmark's workloads.

\section{Reference system}

SPEC chooses a reference machine to normalize the performance and energy metrics used in the CPU benchmark suites. Each benchmark is run and measured on this machine to establish a reference time and energy for that benchmark. These values are then used in the SPEC ratio calculations to establish reportable scores.

\begin{table}[h]
\footnotesize
\centering
\setlength{\aboverulesep}{0pt}
\setlength{\belowrulesep}{0pt}
\setlength{\extrarowheight}{.90ex}

\caption{SPEC CPU Reference Machines}
\label{tab:ref-machine}

\begin{tabularx}{\columnwidth}{@{} 
  >{\hsize=0.50\hsize}L
  >{\hsize=1.12\hsize}L
  >{\hsize=1.38\hsize}L
  @{}} 
\arrayrulecolor{lightgray!40}

\rowcolor{tableheadspec}
\thead[l]{\textbf{\textcolor{white}{CPU suite}}} &
\thead[l]{\textbf{\textcolor{white}{System}}} &
\thead[l]{\textbf{\textcolor{white}{CPU}}} \\

CPU 89/92 &  DEC VAX-11/780 & 5 MHz DEC KA780  \\
\bottomrule
CPU 95 &  Sun SPARCstation 10/40 & 40 MHz SuperSPARC SM40 \\
\bottomrule
CPU 2000 &  Sun Ultra5\_10 & 300 MHz UltraSPARC IIi \\
\bottomrule
CPU 2006 &  Sun Ultra Enterprise 2 & 296 MHz UltraSPARC II \\
\bottomrule
CPU 2017 &  Sun Fire V490 & 2.1 GHz UltraSPARC-IV+ \\
\bottomrule
CPU 2026 &  Lenovo TS HR330A & 3.0 GHz Ampere eMAG 8180 \\
\bottomrule
\end{tabularx}
\par
\setlength{\aboverulesep}{0.6ex} 
\setlength{\belowrulesep}{0.9ex} 
\setlength{\extrarowheight}{0pt}
\end{table}

The reference machine for CPU2026 is a historical Lenovo ThinkSystem HR330A \cite{ampere_lenovo} which uses the Ampere eMAG™ 8180 64-bit processor \cite{ampere_emag}. The eMAG processor, introduced in 2018, used the ARMv8 aarch64 ISA and supported up to 32 cores. Table~\ref{tab:ref-machine} shows the reference machines used over time. One motivation for choosing older hardware is to ensure that resulting scores for modern machines will be above 1.0.

Note that when comparing any two systems measured with SPEC CPU (within the same suite version), their performance relative to each other would remain the same even if a different reference machine were used. This is a consequence of the math involved in calculating the individual and overall geomean metrics.

%% file: 60-analysis.tex
\section{Analysis}

Benchmark characterization has been used for decades to guide architectural design \cite{benchmark_characterization_1991}. Many experiments were conducted on previous generations of SPEC CPU to correlate candidate behavior to contemporary application trends. These include studies conducted during development to discover hot function routines \cite{cpu2006_hotspots}, to highlight challenges and insights derived from event-based analysis \cite{cpu2006_perfmon}, to compare suite versions \cite{intel_cpu2000_06, cpu2017_characterization, cpu2006_17_memsystem}, or to perform statistical analyses \cite{lizy_subsetting}. Academic studies provide a memory-centric characterization of SPEC CPU2017, detailing memory footprints and bandwidth patterns \cite{cpu2017_memory_centric}, or evaluating memory hierarchy response \cite{cpu2017_another_mem_hier_charz}. Some even combined top-down analysis with energy metrics \cite{cpu2017_energy_characterization}.

Prior SPEC CPU suites have served as a cornerstone for academic and industry research, enabling countless studies on workload characterization, architectural innovation, and performance modeling; the expectation is that CPU2026 will carry this legacy forward. Therefore, the analyses presented below are merely introductory, to showcase the kinds of data used by the committee for the process of benchmark selection. 

\subsection{PMC Characterization}
\label{pmc_characterization}
Performance Monitoring Counters (PMCs) are registers built into modern CPUs that record low-level microarchitectural events during program execution, such as instructions commits, cache misses, and branch mispredictions. PMC characterization and Top-down Microarchitectural Analysis \cite{yasin_tma} is used to better understand the general behaviors and performance bottlenecks of the benchmark candidates. Detailed breakdowns of the bottlenecks can be found in Appendix~\ref{specrate_characterization}.

Taken together, instructions-per-cycle (IPC) and stall distributions offer a high-level perspective on the SPEC CPU2026 suites. The integer suites tend to exhibit more balanced frontend and backend bottlenecks, whereas floating-point suites are more consistently backend-bound. Some exceptions stand out, for example, 709.cactus shows notable frontend pressure despite being a floating-point workload, and compression oriented applications such as 777.zstd and 731.astcenc exhibit the highest fraction of cycles lost to speculation, consistent with the control-flow irregularity typical in data compression. These observations reflect the behavior of one particular system; other microarchitectures will showcase different bottlenecks.

\subsection{BBV Recurrence Plots}
\label{bbv_description}
\begin{figure}[t]
    \centering 

    \begin{subfigure}{\columnwidth}
        \includegraphics[width=\linewidth]{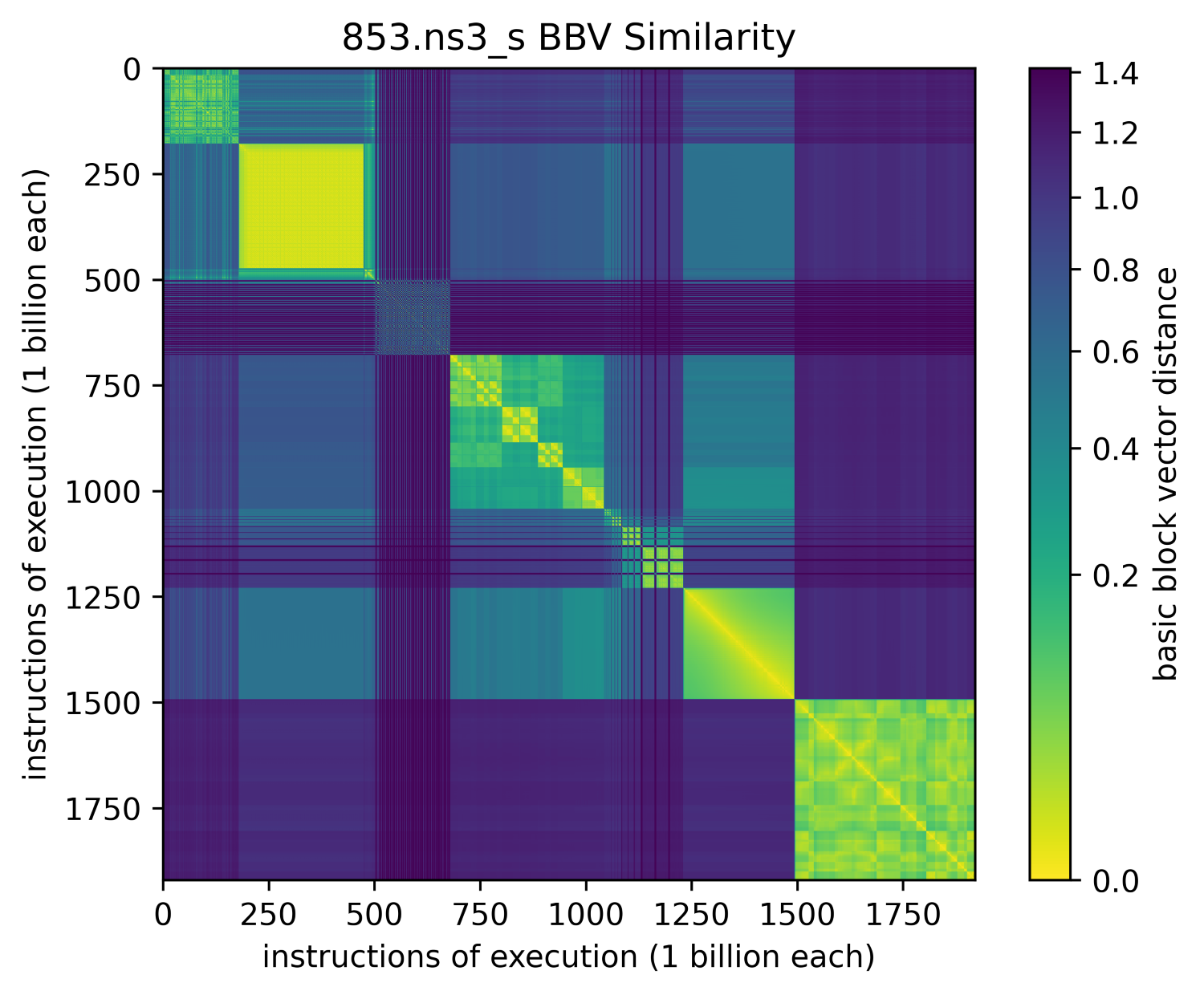}
    \end{subfigure}

    \vspace{-12pt} 

    \begin{subfigure}{\columnwidth}
        \centering
        \hspace*{-14pt} 
        \includegraphics[width=0.82\linewidth]{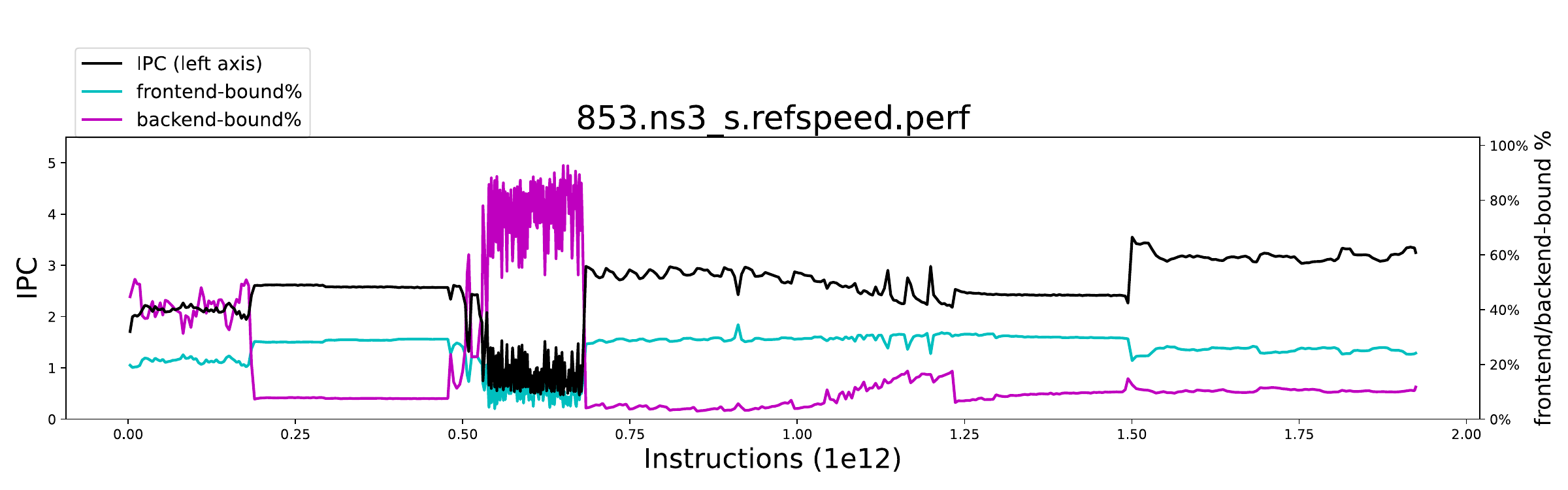}
    \end{subfigure}

    \caption{Self-similarity recurrence plot alongside a performance plot collected from an AMD EPYC™ 9755. 853.ns3 is one of the few single-threaded SPECspeed benchmarks, which means it can accommodate this study. It consists of seven workloads which can be seen as seven major squares; the correlation is visually apparent between BBV recurrence regions and high level performance behaviors of the perf plot. The third square, starting at 500B instructions, is a workload that exhibits a large degree of DTLB misses, hence the spike in backend bottlenecks and lower IPC in that region. It is also the most dissimilar workload from the other six, as evidenced by the dark regions when comparing it to other workloads. The seventh workload, which starts at 1500B, is also dissimilar to the others.}
    \label{fig:bbv_perf_853}

\end{figure}

One way to observe the internal behavior of a program is the analysis of its functional execution phases through Basic Block Vector (BBV) analysis, a method pioneered by the SimPoints toolkit \cite{simpoints} and available in Valgrind \cite{bbv_valgrind}. A basic block is a sequence of instructions with a single entry and exit point. A BBV captures the execution frequency of each basic block, indexed by the program counter of its entry instruction, over a fixed interval of execution (e.g., 10 million instructions). Running this over the entire benchmark results in a series of very high-dimensional vectors, each representing a snapshot of the program's behavior. The similarity between any two execution intervals can be quantified by calculating the Euclidean distance between their corresponding BBVs; a smaller distance implies more similar behavior. Every BBV can be compared to every other BBV, which creates an NxN matrix. This matrix of distance can be visualized to characterize the program's phase behavior across its entire run \cite{mav_perf}, resulting in the self-similarity plot as seen in Figure~\ref{fig:bbv_perf_853}.

In the recurrence plot, each point (\textit{i}, \textit{j}) is colored based on the distance between the BBVs in interval \textit{i} and interval \textit{j}. This plot provides a qualitative assessment of a benchmark's execution diversity. For workload selection, it helps identify redundancy; if two different inputs for a benchmark produced nearly identical regions in the recurrence plot, it signaled an opportunity to prune one in favor of a workload that exercised different code paths. Benchmarks with ``peaky'' function profiles tend to produce monotone yellow recurrence plots which can be spotted quickly. This style of visualization provides more insight than a textual function-level analysis alone, and offers another tool to assist in benchmark selection.

\subsection{Perf Plots}

To complement the BBV plots, performance plots can be generated from PMCs sampled over time. This helps examine how behavior evolves during execution. For each benchmark, these plots monitor IPC, frontend-bound percentage, and backend-bound percentage over time, allowing correlations among the three metrics, as well as fluctuations and phase changes within the individual workloads that constitute each benchmark. These time-series charts reveal finer-grained dynamics, highlighting periods of pipeline efficiency or stall dominance that vary across different workloads. Normalizing time-series to instructions allows overlaying with the BBV plots, offering a deeper analysis as seen in Figure~\ref{fig:bbv_perf_853}. These pairings are offered in Appendix~\ref{sec:bbv_appendix} for all the single-threaded benchmarks.

\subsection{Parallelism }
In SPECspeed, all 13 of the floating point benchmarks use parallelism, and 9 out of 13 integer benchmarks use parallelism. These are marked as MT in Table~\ref{tab:spec-benchmarks}. In total, these 22 parallel benchmarks use one of the following techniques: OpenMP 3.0, C++'s \texttt{std::thread}, Fortran's \texttt {DO CONCURRENT}, or task-based process spawning. All of the SPECspeed threaded benchmarks are classified as strong scaling scenarios, as described in Table~\ref{tab:scaling}. The SPECrate benchmarks are considered weak scaling, as the amount of work grows as the number of copies is increased.

\begin{table}[h!]
\footnotesize 
\centering 

\caption{Types of parallel software performance scaling in SPEC CPU}
\label{tab:scaling}
\setlength{\extrarowheight}{.75ex}

\begin{tabularx}{\columnwidth}{@{} X X @{}}
\arrayrulecolor{lightgray!40}
\rowcolor{tableheadspec}
\textbf{\textcolor{white}{SPECrate\textregistered~2026}} &
\textbf{\textcolor{white}{SPECspeed\textregistered~2026}} \\
\hline 
\textbf{Weak Scaling} & \textbf{Strong Scaling} \\
\hline
Workload size increases with copies & Workload size remains constant \\
\hline
Goal is to maintain a constant time to complete the tasks as the workload size grows & Goal is to decrease the total time to complete the fixed-size workload by splitting it amongst processors \\
\hline
Gustafson's Law: speedup based on the workload size scaling up to match the number of processors \cite{gustafson} & Amdahl's Law: speedup is limited by the portions of the program that cannot be parallelized \cite{amdahl} \\
\hline
\end{tabularx}
\end{table}

In addition to language based parallelism, for the first time SPEC CPU offers two benchmarks with task-based parallelism: 821.gcc and 823.llvm. These are based on the two most popular open source compilers, and in both of these benchmarks, thousands of unique command lines are invoked to build multiple input source files. These benchmarks mimic the way `make~-j~N' runs in the field; each command line spawns a new compiler process, which keeps N cores active until the large pile of work is completed.

%% file: 70-roundrobin.tex
\section{RRR - heterogeneous schedule}

\begin{figure*}[!th]
    \centering 

    \begin{tabular}{@{}c@{\hfill}c@{\hfill}c@{}}

        \begin{subfigure}[b]{0.313\textwidth}
            \includegraphics[width=\linewidth]{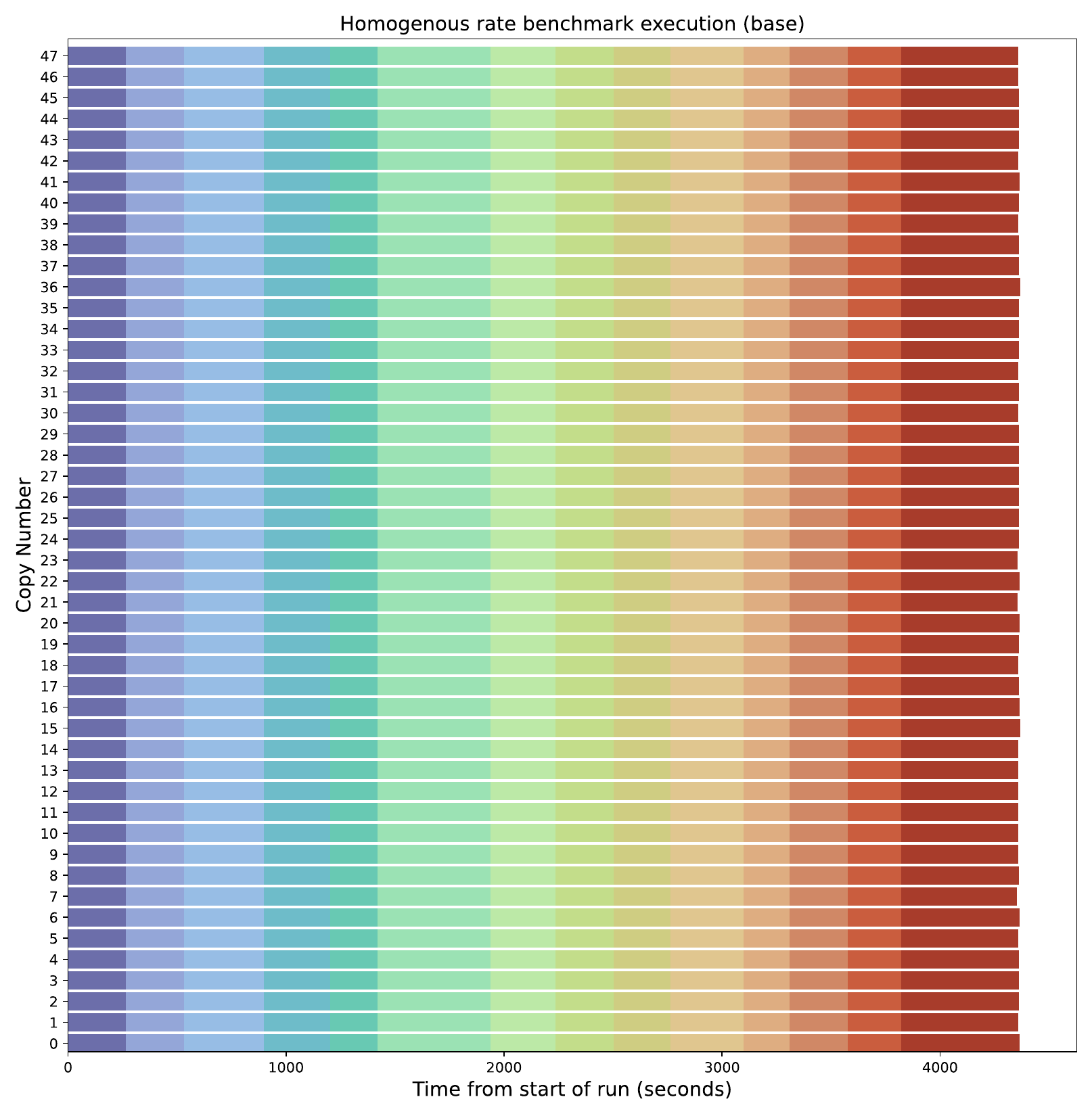}
            \caption{Standard refrate}
            \label{fig:sub_a}
        \end{subfigure}
        & 
        
        \begin{subfigure}[b]{0.313\textwidth}
            \includegraphics[width=\linewidth]{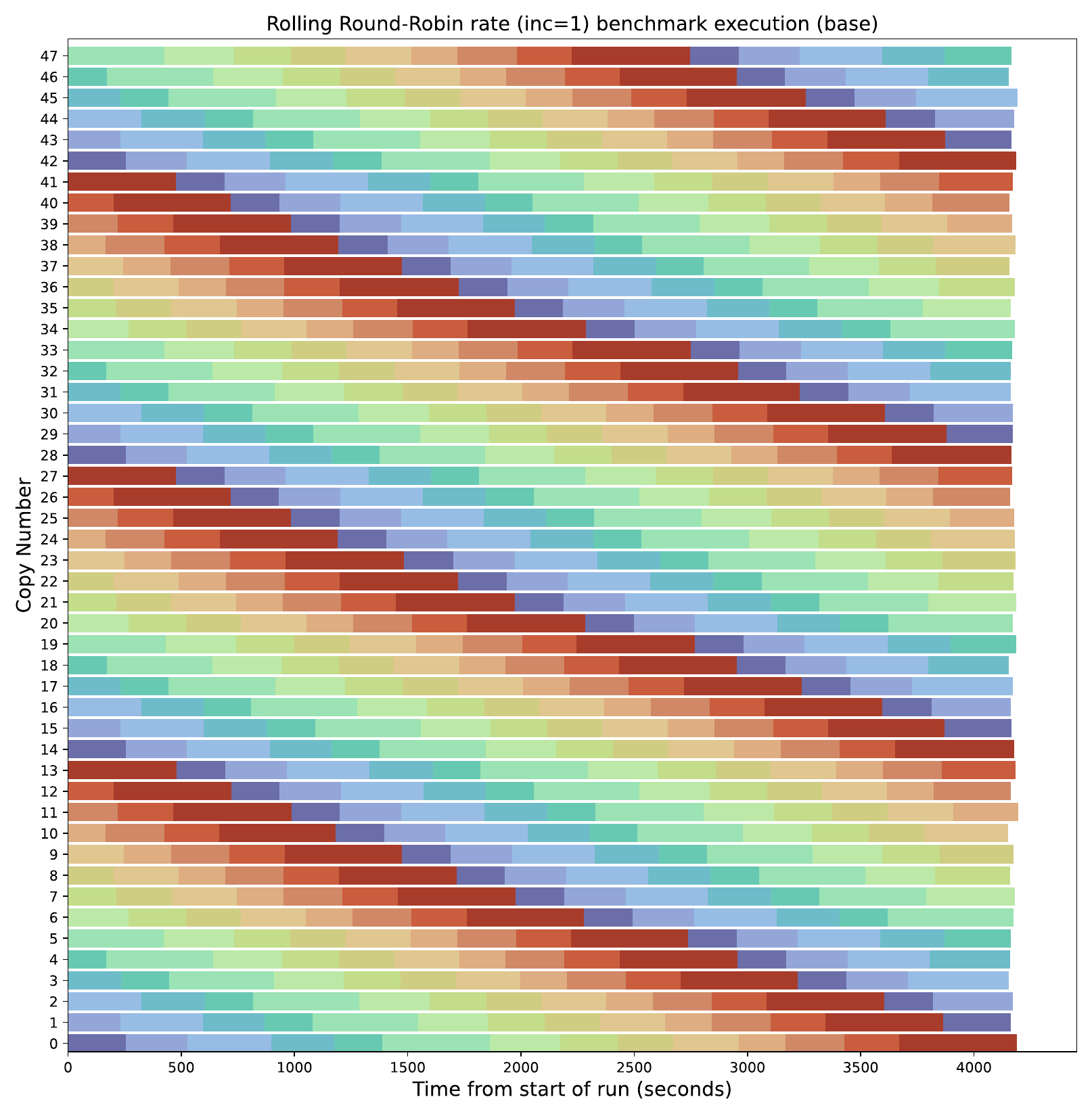}
            \caption{Round-robin rate, inc=1}
            \label{fig:sub_b}
        \end{subfigure}
        & 
        
        \begin{subfigure}[b]{0.3775\textwidth}
            \includegraphics[width=\linewidth]{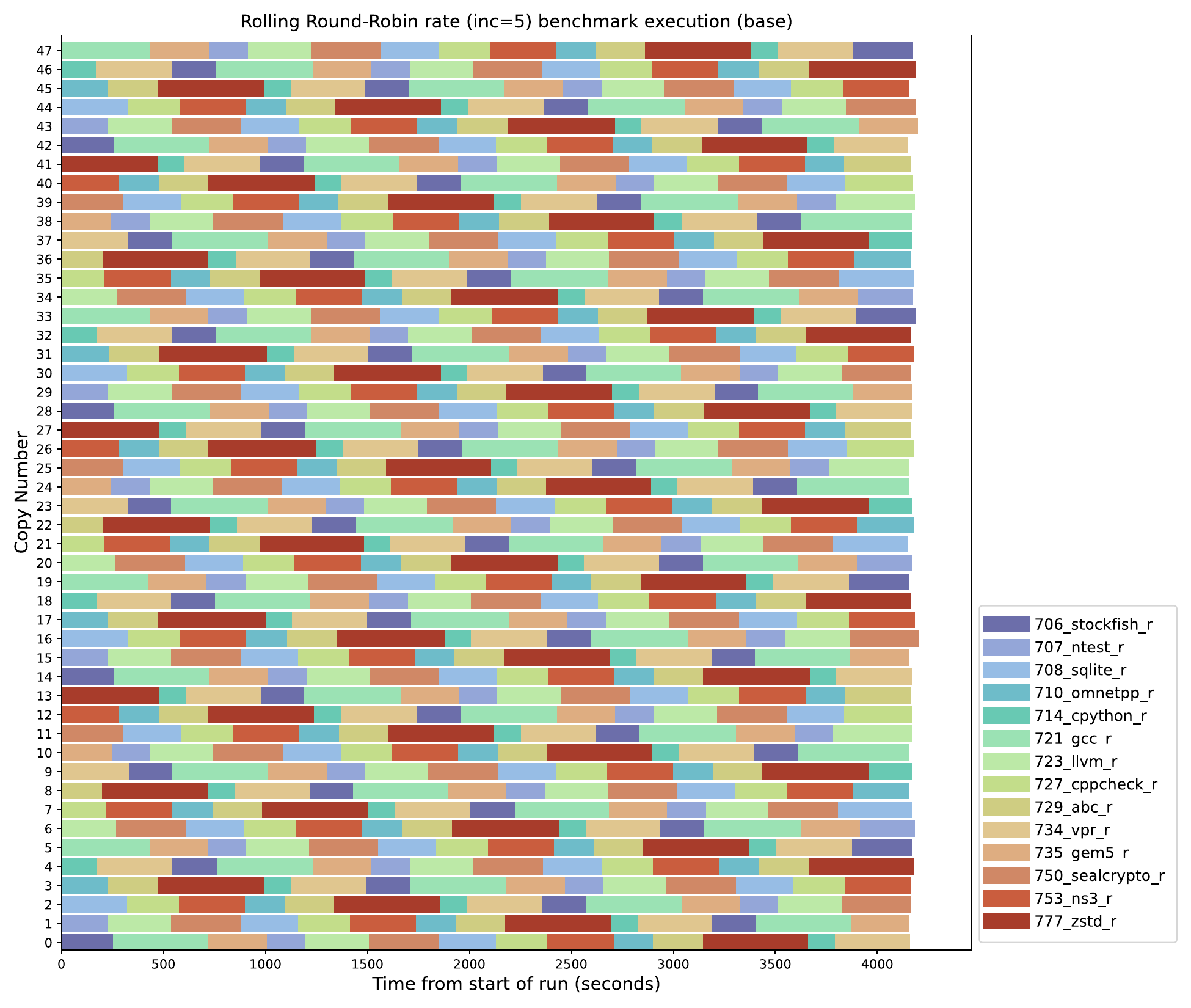}
            \caption{Round-robin rate, inc=5}
            \label{fig:sub_c}
        \end{subfigure}
        
    \end{tabular} 

    \caption{Time plots showing all the integer rate benchmarks executing on schedules for homogenous (refrate) and heterogeneous (rrrrate) on 48 copies. In all cases, each copy runs every benchmark once. The harness interface allows using a subset of benchmarks as well, allowing the user to craft their own scenario to run using the round-robin scheduler. The ``inc'' parameter tells the scheduler how to increment through the roster of benchmarks, allowing for some customization in the heterogeneous scheduling. Plots were generated with the \texttt{rate\_timeplot.py} script found in the \texttt{scripts.misc} directory bundled with the CPU2026 distribution.}
    \label{fig:rrrplots}
\end{figure*}

CPU2026 introduces a new heterogeneous style of running benchmarks in multi-copy mode called Rolling Round-Robin, or RRR. Here we explain the motivation for this run style of multi-programmed workloads and a description of the methodology. RRR is in exhibition as the scoring methodology is not well-established; this is an open call to assist with its continuing evaluation.

SPEC CPU's primary multi-copy benchmark, SPECrate, stresses systems with a homogeneous load, with each copy running the same component benchmark simultaneously. This single-program, homogeneous capacity methodology has been foundational to SPEC CPU since 1992 \cite{spec_homogenous_capacity}. In today's CPUs with hundreds of cores, this method can expose system corner cases, yet modern server systems do not usually operate under homogeneous loads. Multi-tenant systems operate with VMs running all manners of workloads simultaneously, hence the motivation for a heterogeneous style of benchmarking.

The increasing complexity and heterogeneity of modern multicore SoCs, particularly those deployed in cloud and data center environments for AI agentic workflows, necessitate robust methodologies for multiprogrammed workload characterization and performance evaluation. Single-program homogeneous runs cannot expose intricate cross-process interactions, hetergeneous resource contention, and scheduling dynamics that are prevalent in contemporary multiprogrammed systems. A challenge identified in the literature is the lack of a standardized approach to generate multiprogrammed benchmarks. Researchers frequently resort to custom-crafted benchmark mixtures and ad-hoc scheduling policies \cite{selecting_benchmark_combos, predicting_concurrent_execution_time, multicore_interference, SMT_pairs, dynamic_adaptive_power_multicore, phase_aware_cache_partitioning}, hindering direct comparison across studies and limiting the generalization of architectural and software optimizations. This problem extends from defining benchmark composition and workload sampling methods \cite{hetero_bench4, hetero_bench3} to the selection of appropriate performance and fairness metrics \cite{hetero_bench5, hetero_bench6, survey_of_multiprogram, four_metrics_to_evaluate_multicore}. Similar work \cite{multiprogram_workload} highlights the importance of distinguishing between sample imbalance (differences in standalone runtimes) and schedule imbalance (asymmetric contention), yet many ad-hoc methodologies still introduce both. This fragmentation underscores a need for a standardized and reproducible methodology for evaluating multiprogrammed performance on heterogeneous multicore systems.

To address this gap, CPU2026 introduces the Rolling Round-Robin benchmarking mode \cite{rrrrate}. RRR offers a standardized, deterministic schedule and repeatable method for utilizing the existing intrate and fprate suites as multiprogrammed benchmarks. For a suite comprising N benchmarks (e.g., 14 for intrate) running on M cores, the RRR methodology operates as follows: Each of M cores executes all N benchmarks sequentially, in a predetermined fixed order, rotating through the benchmarks. For example, if Core 0 starts with Benchmark A, Core 1 would start with Benchmark B, and so on, cyclically through the entire benchmark roster. Each core continues to run its assigned schedule of benchmarks until all N benchmarks have completed on all M cores. This design ensures that every benchmark runs on every core, providing a uniform exposure while allowing for the exploration of diverse contention scenarios in a controlled and repeatable fashion. This methodical approach inherently eliminates sample imbalance, as each benchmark is guaranteed to run to completion an equal number of iterations across the system, ensuring identical instruction counts for each benchmark and each core, from the perspective of user-level CPU execution. Figure~\ref{fig:rrrplots} compares three different styles of schedule methods, from runs on an AmpereOne$^\circledR$ system executing 48 copies. 

The RRR methodology provides a much-needed common ground for rigorous research into multicore performance. Since the benchmarks have been hardened through portability (\S\ref{sec:portability}), there is no worry of syscalls or IO or similar wrenches impacting the schedule. This allows the focus to remain on user level CPU performance under heterogenous load. By standardizing the workload construction, RRR enables direct and fair comparisons of architectural innovations, OS scheduling policies \cite{hetero_aware_os_scheduler_for_asymmetric_multicore, efficient_multicore_proc_scheduling}, and resource partitioning schemes \cite{resource_partitioning_in_colocation, contention_aware_job_colocation}. It allows multiprogramming studies to move beyond methods that substitute manual kernel isolation for full complex applications \cite{hetero_bench1, hetero_speccast}. RRR enables benchmarks to interact with the microarchitecture and each other in ways more representative of their complete execution profiles, preserving ``unwanted cross-effects'' that might otherwise be overlooked \cite{hetero_bench1}. While RRR standardizes the generation of one style of multiprogrammed workloads, the research community continues to debate the optimal metrics for evaluating such systems (e.g., cumulative IPC, average throughput, harmonic mean, fairness indices) \cite{hetero_bench5, hetero_bench6, survey_of_multiprogram, four_metrics_to_evaluate_multicore}. RRR provides the raw, standardized execution data upon which various metrics can be proposed and compared, fostering further discussion on the most appropriate performance and fairness indicators for complex heterogeneous systems. By offering a robust and deterministic framework to control heterogenous load, the SPEC CPU committee believes RRR will significantly enhance the rigor and comparability of future multicore research.

%% file: 80-conclusion.tex
\section{Conclusion}

The physicist Lord Kelvin famously stated, "If you can not measure it, you can not improve it." This maxim has served as the unspoken charter for the field of computer architecture, where progress is inextricably linked to the ability to perform fair, repeatable, and relevant measurements. For over three decades, the SPEC CPU suite has been the industry's primary instrument for this purpose. This paper has detailed the creation of CPU2026, the seventh version of this essential tool, reaffirming the enduring importance of general-purpose compute performance in an era of increasing specialization. By providing a transparent account of the methodology, we have demonstrated that the suite is not a synthetic construct, but a carefully curated collection of real-world applications, hardened into a robust and portable benchmark.

This paper has chronicled the principled and methodical process behind the development of CPU2026. We have shown how a symbiotic relationship with the open-source community, formalized through a multi-year search program, yielded a suite with an expanded set of benchmarks exhibiting novel microarchitectural profiles not seen in prior suites. Key contributions were detailed, including the landmark introduction of multithreaded integer workloads and the new Rolling Round-Robin Rate (RRR) methodology for characterizing heterogeneous throughput on multicore processors. Furthermore, we outlined the extensive engineering discipline required to create a long-lived suite, from meticulous code hardening for ISO compliance and portability across diverse platforms like big-endian systems and Microsoft Windows, to a modern commitment to memory safety validated by a battery of software and hardware sanitizers.

In his lecture \emph{Technology and Courage}, Ivan Sutherland spoke of the courage required to embark on a new and uncertain endeavor \cite{tech_courage}. In that light, this work honors the courage of our SPEC predecessors who, despite being commercial competitors, came together to pursue a shared vision of microprocessor benchmarks that are fair, comparable, and representative \cite{Dixit1993OverviewOT}. The current generation of stewards has continued to manifest this vision with the release of SPEC CPU2026. The suite is now entrusted to the community—to the architects, compiler developers, software experts, and researchers who will use it to test their own courageous ideas. The baton is now passed to you, the assiduous reader, to build upon this foundation and continue the march toward ever stronger computing benchmarks. You are the Next Generation.

\section*{Acknowledgement}

The creation of SPEC CPU$^\circledR$2026 was a significant collaborative effort, and the authors wish to express their profound gratitude to the individuals whose leadership and contributions were essential. Special thanks are extended to the four individuals who served as committee chair over the course of the development process: Jeff Reilly, James Bucek, Van Smith, and Frédérique Silber-Chaussumier. Their guidance was instrumental in steering a diverse committee toward a common goal and successfully delivering this benchmark suite. Deepest gratitude is extended to John Henning, the committee secretary, for his overall mentorship drawn from his encyclopedic knowledge of SPEC CPU history, and for scrutinizing the suite with an eye for licensing.  We are also grateful to Cloyce Spradling, the release manager, for his meticulous work in integrating the components and managing the final publication. We thank Ronen Zohar for his expertise on C/C++ language standards and complex compiler behaviors, and Sunil Vijay Sathe for sharing his specialized knowledge of HPC workloads.  Finally, we acknowledge Ruihao Li, Neeraja Yadwadkar, and Lizy John for providing the dendrogram analysis that served as a vital tool in the benchmark selection process.

This suite would not have been possible without the collective dedication and expertise of the entire SPEC CPU committee, supporting contributors, and the participants of the CPUv8 benchmark search program.

%% file: 99-appendix.tex
\subsection{Testimonials}
The following testimonials are provided by the authors and maintainers of open source projects who collaborated with SPEC during benchmark development and integration of their applications. Their reflections highlight the technical benefits and improvements in portability, performance, and code quality that resulted from this partnership.

\para{735.gem5} \emph{Jason Lowe-Power}: ``For decades, computer architecture simulators have relied on SPEC CPU as the gold standard for evaluating new hardware designs. It is gratifying to come full circle with the inclusion of gem5 in SPEC CPU2026, where the rigorous adaptation process has already resulted in multiple upstreamed fixes. I am confident this partnership will continue to foster a virtuous cycle of innovation for the entire computer architecture community.''

\para{767.nest} \emph{Hans Ekkehard Plesser}: ``Working with the SPEC committee to prepare the NEST simulator for SPEC CPU2026 has been a great experience. The SPEC team tested NEST on a much wider range of operating systems, compilers, and hardware than we have available in our regular test and benchmark setup. The small number of incompatibilities revealed in this process provided good learning opportunities for us. Stress-testing of multithreading in NEST by the SPEC CPU team complemented our own efforts and contributed to further ascertain the thread-safety of NEST. As an added bonus for us and NEST users, the SPEC CPU team even contributed some small optimizations to the simulator. We are excited that NEST, having served as a reference for neuromorphic computing systems over the past decade,  will help to drive CPU performance as part of SPEC CPU2026.''

\para{753.ns3} \emph{Gabriel de Carvalho Ferreira}: ``The ns-3 community is delighted to have participated in the selection process for the new SPEC CPU benchmark suite. During the porting and testing phase, we identified several issues that impacted the reproducibility of results, including platform related precision issues, and hindered compilation and execution across diverse platforms. The fixes for these were promptly integrated upstream. As a result, ns-3's platform support has been greatly expanded to a wider array of systems. Additionally, the committee's analysis of our different workloads provided valuable insights and opportunities to improve the ns-3 simulator code.''

\para{729.abc} \emph{Alan Mishchenko}: ``Working with you on adapting ABC into a benchmark has been a genuinely rewarding collaboration. I took the feedback and portability requirements seriously, updating the code multiple times to meet the benchmarking needs. While I wouldn't call any single bug fix absolutely critical on its own, the process as a whole sharpened my understanding of how to write more portable, robust code - lessons that have directly benefited ABC's development going forward. I've enjoyed the collaboration and appreciate the care you puts into making this a two-way relationship between SPEC and the open-source community.''


\para{707.ntest} \emph{Vlad Petric}: ``The Othello player community, including but not limited to legends of the game and myself, is grateful to the SPEC.org members that brought many fixes and improvements to the NTest engine. These include code cleanups, bug fixes, portability enhancements, performance speedups, and scalability from mobile processors to high core-count systems. We intend to make NTest the premier engine in the Othello world again, and also the basis for a strong, player-usable computational solution to the game (God's algorithm). The collaboration greatly helped us with these goals.''

\para{748.flightdm} \emph{Sean McLeod}: ``The JSBSim community had a great time working with the representatives from SPEC. The process of transforming our application into a benchmark resulted in the discovery and fixing of many issues, ranging from astronautics algorithm corner cases to incorrect usage of C++ language constructs. The SPEC engineers exercised our codes on more systems/compilers than what we usually test; this enabled validation of these issues and subsequently gave us confidence that JSBSim is now more portable than ever.''

\emph{Bertrand Coconnier}: ``It is an honor for JSBSim to have been chosen for SPEC CPU2026, and I see it as great recognition of our work. Working with you has undoubtedly made JSBSim better software. A number of subtle bugs and errors have been uncovered and fixed thanks to your contribution, which is greatly appreciated. Your rigor, but also your goodwill, have been key to this result.''

\para{706.stockfish} \emph{Gian-Carlo Pascutto}: ``Personally, one of the great benefits of contributing to SPEC is that it makes sure compiler optimizer authors and CPU designers are aware and to some extent focus on these workloads. I am sure that because of the work we did on SPEC with Stockfish, compilers will get better at optimizing and autovectorizing integer neural network inference kernels, which will benefit a much wider set of software than chess engines only. Same for dealing with rather branchy code with unpredictable memory access.''

\para{734.vpr} \emph{Vaughn Betz}: ``Working with SPEC has helped ensure the large VPR code base is fully compliant with recent C++ standards and achieves consistent results across platforms. The collaboration with SPEC identified several instances of platform/compiler dependent code and result differences, which were fixed collaboratively and upstreamed to ensure the community could use VPR on the widest range of platforms with confidence. We also believe the hardware design community will benefit from having CAD tools represented in the SPEC benchmark suite, as that will help drive future CPU improvements to speed up these compute-intensive tools.''

\para{838.diamond} \emph{Benjamin J. Buchfink}: ``DIAMOND solves a computationally hard problem fundamental to biology and contains a lot of performance-critical and carefully optimized code. Working with the committee helped improve and harden my code with respect to microarchitectural subtleties and compiler peculiarities on different platforms. Due to the needs of scientific computation and the limited resources that most scientists have, both reproducibility and correctness as well as performance are important points for the user community.''

\para{727.cppcheck} \emph{Daniel Marjam{\"a}ki}: ``Our aim is to write truly portable code, and I remember you uncovering — and resolving — a few platform-specific issues while porting Cppcheck. That work was greatly appreciated. You even identified a case of undefined behavior along the way, which was a real bonus. It was a pleasure to be part of this program, and I’d be happy to take part again. I hope our participation will help inspire valuable optimizations in the future.''

\para{737.gmsh} \emph{Christophe Guizaine}: ``Working with the SPEC committee has been a very constructive and technically rewarding experience. The process of adapting Gmsh into a benchmark---defining representative workloads, improving portability, and hardening the code for diverse platforms---has led to concrete improvements that directly benefit the project and the Gmsh user community. Overall, beyond visibility, the collaboration has helped us better understand performance-critical paths and portability constraints, with benefits that extend well beyond the benchmark itself.'' 

\para{811.tealeaf/820.cloverleaf} \emph{Simon McIntosh-Smith}: ``The process for proposing candidate benchmarks for the new SPEC CPU suite was remarkably straightforward, and once we’d made it through the first few stages, we got very useful feedback on our two candidate codes. The SPEC team were easy to work with, and provided high-quality feedback on the performance, accuracy, and portability of the codes. Several important improvements have been fed back into the codes as a result, benefitting the HPC community. The new SPEC CPU suite looks to be a significant step forwards over previous iterations, and we’re excited to have contributed to what will be an invaluable tool in comp-arch research and development.''

\para{800.pot3d} \emph{Ronald Caplan}: ``The process of submitting, adapting, and creating workloads for our POT3D code to be included in SPEC benchmarks has been a rewarding experience for Predictive Science Inc. in several ways.  It launched our first major open-source code release, which has paved the way for several more since.  It helped our upstream code through the testing process, finding compatibility issues and work-arounds for cutting edge features and hardware.  The validation requirements have helped us craft new test suites for several of our codes.  By having POT3D in SPEC, we can view the submitted results to preview how our codes will perform across new architectures, and helps guide optimizations.  This also helps other groups with similar memory-bound stencil codes.''

\para{854.graph500} \emph{David Bader}: ``As graph-based workloads become increasingly central to machine learning and AI--from graph neural networks to knowledge graphs powering modern LLMs--the inclusion of 854.graph500 in SPEC CPU2026 reflects a critical shift in what `general-purpose' computing must handle. Working with the SPEC committee to adapt Graph500 was a genuinely symbiotic process: their rigorous hardening for portability and determinism produced fixes we upstreamed to the community, while SPEC gained a benchmark that captures the irregular memory access patterns and data-dependent parallelism that define this growing class of applications.'' 

\para{846.minizinc} \emph{Guido Tack}: ``Working with SPEC on integrating MiniZinc, Gecode, and Chuffed into the benchmark suite was a collaborative and highly constructive experience that we greatly appreciated. The process uncovered approximately a dozen portability, correctness and performance issues, many of which would have been difficult to identify outside SPEC’s rigorous cross-platform environment, and the resulting fixes and improvements were incorporated upstream. We are grateful for the care and technical depth of the feedback from the committee, which strengthened the robustness and maturity of our software systems and will provide lasting benefits to our user community, while also helping new users discover constraint programming and adopt our solutions.''


\subsection{SPECrate Characterization}
\label{specrate_characterization}
The performance characterization presented in this section is based on data collected on a system built around an AMD EPYC™ 9005 Series processor featuring the ``Zen 5'' core. This microarchitecture supports an 8-wide dispatch in the frontend, yielding a maximum theoretical IPC of eight instructions per cycle. Table~\ref{tab:sys-config} summarizes the full hardware and software operating environment. All SPECrate benchmarks were executed using a single copy. This initial analysis reports the IPC and a top-level breakdown of each benchmark, categorizing execution time into frontend bound, backend bound, bad speculation or lost cycles, and retiring. These results provide a high-level view of how the benchmarks spend their cycles and how efficiently each benchmark utilizes the core in practice.

\begin{table}[h]
\footnotesize
\centering
\setlength{\aboverulesep}{0pt}
\setlength{\belowrulesep}{0pt}
\setlength{\extrarowheight}{.75ex}

\caption{Experimental System Configuration}
\label{tab:sys-config}

\begin{tabularx}{\dimexpr\columnwidth - 8pt\relax}{@{}
  >{\hsize=0.8\hsize}L
  >{\hsize=1.20\hsize}L
  @{}}
\arrayrulecolor{lightgray!40}

\rowcolor{tableheadspec}
\thead[l]{\textbf{\textcolor{white}{Component}}} &
\thead[l]{\textbf{\textcolor{white}{Configuration}}} \\

\bottomrule
Processor &  AMD EPYC™ 9755 \cite{epyc9755} \\
\bottomrule
Frequency & 2.7 GHz (Max. Boost to 4.1 GHz) \\
\bottomrule
Memory &  2.3 TiB DDR5-6400 \\
\bottomrule
L1 Cache &  32 KiB I + 48 KiB D \\
\bottomrule
L2 Cache &  1 MiB \\
\bottomrule
L3 Cache &  512 MiB \\
\bottomrule
Compiler and Flags &  GCC 15.2 -O3 \\
\bottomrule
Operating Environment &  Ubuntu 24.04 LTS \\& Linux kernel: 6.8.0-44-generic \\
\bottomrule
\end{tabularx}
\par
\setlength{\aboverulesep}{0.6ex} 
\setlength{\belowrulesep}{0.9ex} 
\setlength{\extrarowheight}{0pt}
\end{table}

\begin{table}[tbp] 
\arrayrulecolor{lightgray!40}

\footnotesize
\centering
\caption{INT RATE benchmarks: IPC and Stall Ratio Breakdown}
\label{tab:intrate_stalls}
\setlength{\tabcolsep}{3pt} 
\setlength{\extrarowheight}{.75ex}
\setlength{\aboverulesep}{0pt}
\setlength{\belowrulesep}{0pt}
\begin{tabularx}{\columnwidth}{l *{5}{C}}
\rowcolor{tableheadspec}
\textbf{\textcolor{white}{Benchmark}} & 
\textbf{\textcolor{white}{IPC}} & 
\textbf{\textcolor{white}{Frontend}} & 
\textbf{\textcolor{white}{Backend}} & 
\textbf{\textcolor{white}{Lost}} & 
\textbf{\textcolor{white}{Retiring}} \\
\bottomrule

706.stockfish\_r & \cellcolor[HTML]{DEDEDE}3.12 & \cellcolor[HTML]{CDF3F3}0.34 & \cellcolor[HTML]{F6DBF6}0.24 & \cellcolor[HTML]{FCFCF0}0.05 & \cellcolor[HTML]{E9F4EE}0.37 \\
\bottomrule
707.ntest\_r     & \cellcolor[HTML]{D9D9D9}3.56 & \cellcolor[HTML]{E9FAFA}0.15 & \cellcolor[HTML]{F4D1F4}0.31 & \cellcolor[HTML]{F8F8E1}0.10 & \cellcolor[HTML]{E2F2E9}0.44 \\
\bottomrule
708.sqlite\_r    & \cellcolor[HTML]{E1E1E1}2.82 & \cellcolor[HTML]{E4F9F9}0.18 & \cellcolor[HTML]{EFC0EF}0.43 & \cellcolor[HTML]{FBFBEC}0.07 & \cellcolor[HTML]{EDF6F2}0.32 \\
\bottomrule
710.omnetpp\_r   & \cellcolor[HTML]{DEDEDE}3.17 & \cellcolor[HTML]{BCEFEF}0.45 & \cellcolor[HTML]{FBEDFB}0.12 & \cellcolor[HTML]{FDFDF5}0.04 & \cellcolor[HTML]{E6F3EC}0.39 \\
\bottomrule
714.cpython\_r   & \cellcolor[HTML]{DADADA}3.55 & \cellcolor[HTML]{B0EBEB}0.53 & \cellcolor[HTML]{FEF8FE}0.05 & \cellcolor[HTML]{FFFFFC}0.01 & \cellcolor[HTML]{E4F3EB}0.41 \\
\bottomrule
721.gcc\_r       & \cellcolor[HTML]{EFEFEF}1.57 & \cellcolor[HTML]{C7F1F1}0.38 & \cellcolor[HTML]{F1C5F1}0.39 & \cellcolor[HTML]{FBFBEE}0.06 & \cellcolor[HTML]{FCFCFF}0.18 \\
\bottomrule
723.llvm\_r      & \cellcolor[HTML]{F0F0F0}1.47 & \cellcolor[HTML]{B6EDED}0.49 & \cellcolor[HTML]{F5D5F5}0.28 & \cellcolor[HTML]{FAFAEA}0.07 & \cellcolor[HTML]{FCFCFF}0.17 \\
\bottomrule
727.cppcheck\_r  & \cellcolor[HTML]{E0E0E0}2.92 & \cellcolor[HTML]{AAEAEA}0.57 & \cellcolor[HTML]{FBEFFB}0.11 & \cellcolor[HTML]{FEFEF8}0.02 & \cellcolor[HTML]{EFF7F4}0.30 \\
\bottomrule
729.abc\_r       & \cellcolor[HTML]{E3E3E3}2.68 & \cellcolor[HTML]{DDF7F7}0.23 & \cellcolor[HTML]{F3CCF3}0.34 & \cellcolor[HTML]{F6F6DA}0.13 & \cellcolor[HTML]{F0F7F5}0.30 \\
\bottomrule
734.vpr\_r       & \cellcolor[HTML]{E4E4E4}2.55 & \cellcolor[HTML]{D1F4F4}0.31 & \cellcolor[HTML]{F5D6F5}0.28 & \cellcolor[HTML]{F8F8E3}0.09 & \cellcolor[HTML]{EFF7F4}0.31 \\
\bottomrule
735.gem5\_r      & \cellcolor[HTML]{E3E3E3}2.70 & \cellcolor[HTML]{BCEEEE}0.45 & \cellcolor[HTML]{F9E5F9}0.17 & \cellcolor[HTML]{FBFBF0}0.05 & \cellcolor[HTML]{EEF6F3}0.32 \\
\bottomrule
750.sealcrypto\_r& \cellcolor[HTML]{C8C8C8}5.23 & \cellcolor[HTML]{FEFFFF}0.01 & \cellcolor[HTML]{F2CBF2}0.35 & \cellcolor[HTML]{FFFFFF}0.00 & \cellcolor[HTML]{CEEAD8}0.63 \\
\bottomrule
753.ns3\_r       & \cellcolor[HTML]{E3E3E3}2.69 & \cellcolor[HTML]{AFEBEB}0.54 & \cellcolor[HTML]{FCF0FC}0.11 & \cellcolor[HTML]{FDFDF5}0.04 & \cellcolor[HTML]{EDF6F2}0.32 \\
\bottomrule
777.zstd\_r      & \cellcolor[HTML]{E4E4E4}2.57 & \cellcolor[HTML]{E0F8F8}0.21 & \cellcolor[HTML]{F2C8F2}0.37 & \cellcolor[HTML]{F5F5D6}0.14 & \cellcolor[HTML]{F1F8F6}0.28 \\
\bottomrule
\end{tabularx}
\par
\setlength{\aboverulesep}{0.6ex} 
\setlength{\belowrulesep}{0.9ex} 
\setlength{\extrarowheight}{0pt}
\end{table}

\begin{table}[tbp] 
\arrayrulecolor{lightgray!40}

\footnotesize
\centering
\caption{FP RATE Benchmarks: IPC and Stall Ratio Breakdown}
\label{tab:fprate_stalls}
\setlength{\tabcolsep}{3pt} 
\setlength{\extrarowheight}{.75ex}
\setlength{\aboverulesep}{0pt}
\setlength{\belowrulesep}{0pt}
\begin{tabularx}{\columnwidth}{l *{5}{C}}
\rowcolor{tableheadspec}
\textbf{\textcolor{white}{Benchmark}} & 
\textbf{\textcolor{white}{IPC}} & 
\textbf{\textcolor{white}{Frontend}} & 
\textbf{\textcolor{white}{Backend}} & 
\textbf{\textcolor{white}{Lost}} & 
\textbf{\textcolor{white}{Retiring}} \\
\bottomrule

709.cactus\_r    & \cellcolor[HTML]{E7E7E7}2.32 & \cellcolor[HTML]{B3ECEC}0.51 & \cellcolor[HTML]{F8E1F8}0.20 & \cellcolor[HTML]{FFFFFF}0.00 & \cellcolor[HTML]{F6FAFA}0.29 \\
\bottomrule
722.palm\_r       & \cellcolor[HTML]{D5D5D5}3.95 & \cellcolor[HTML]{EBFAFA}0.13 & \cellcolor[HTML]{F2C8F2}0.37 & \cellcolor[HTML]{FFFFFC}0.01 & \cellcolor[HTML]{E3F2E9}0.48 \\
\bottomrule
731.astcenc\_r   & \cellcolor[HTML]{E2E2E2}2.74 & \cellcolor[HTML]{CDF3F3}0.34 & \cellcolor[HTML]{FAEBFA}0.14 & \cellcolor[HTML]{F3F3CC}0.17 & \cellcolor[HTML]{F0F7F5}0.35 \\
\bottomrule
736.ocio\_r       & \cellcolor[HTML]{D4D4D4}4.04 & \cellcolor[HTML]{FCFFFF}0.02 & \cellcolor[HTML]{EFBEEF}0.43 & \cellcolor[HTML]{FFFFFD}0.01 & \cellcolor[HTML]{DDF0E4}0.53 \\
\bottomrule
737.gmsh\_r       & \cellcolor[HTML]{ECECEC}1.82 & \cellcolor[HTML]{DCF7F7}0.24 & \cellcolor[HTML]{EFBFEF}0.43 & \cellcolor[HTML]{F8F8E0}0.11 & \cellcolor[HTML]{FCFCFF}0.23 \\
\bottomrule
748.flightdm\_r   & \cellcolor[HTML]{DEDEDE}3.12 & \cellcolor[HTML]{CFF3F3}0.33 & \cellcolor[HTML]{F5D6F5}0.28 & \cellcolor[HTML]{FFFFFC}0.01 & \cellcolor[HTML]{ECF6F2}0.39 \\
\bottomrule
749.fotonik3d\_r  & \cellcolor[HTML]{DFDFDF}3.05 & \cellcolor[HTML]{FAFEFE}0.04 & \cellcolor[HTML]{EAABEA}0.56 & \cellcolor[HTML]{FEFEF9}0.02 & \cellcolor[HTML]{EDF6F2}0.38 \\
\bottomrule
765.roms\_r       & \cellcolor[HTML]{E1E1E1}2.86 & \cellcolor[HTML]{F8FEFE}0.05 & \cellcolor[HTML]{EAA8EA}0.58 & \cellcolor[HTML]{FEFEFA}0.02 & \cellcolor[HTML]{F0F8F5}0.35 \\
\bottomrule
766.femflow\_r    & \cellcolor[HTML]{D6D6D6}3.90 & \cellcolor[HTML]{F4FDFD}0.08 & \cellcolor[HTML]{F0C0F0}0.42 & \cellcolor[HTML]{FFFFFE}0.00 & \cellcolor[HTML]{E1F1E8}0.49 \\
\bottomrule
767.nest\_r       & \cellcolor[HTML]{D9D9D9}3.62 & \cellcolor[HTML]{F2FCFC}0.09 & \cellcolor[HTML]{F1C7F1}0.38 & \cellcolor[HTML]{FAFAE8}0.08 & \cellcolor[HTML]{E5F3EB}0.45 \\
\bottomrule
772.marian\_r     & \cellcolor[HTML]{C7C7C7}5.33 & \cellcolor[HTML]{F8FEFE}0.05 & \cellcolor[HTML]{F5D4F5}0.29 & \cellcolor[HTML]{FFFFFD}0.01 & \cellcolor[HTML]{D0EBD9}0.66 \\
\bottomrule
782.lbm\_r        & \cellcolor[HTML]{D6D6D6}3.89 & \cellcolor[HTML]{F8FEFE}0.05 & \cellcolor[HTML]{EFBDEF}0.44 & \cellcolor[HTML]{FFFFFC}0.01 & \cellcolor[HTML]{E1F1E8}0.49 \\
\bottomrule
\end{tabularx}
\par
\setlength{\aboverulesep}{0.6ex}
\setlength{\belowrulesep}{0.9ex}
\setlength{\extrarowheight}{0pt}
\end{table}

Tables~\ref{tab:intrate_stalls} and~\ref{tab:fprate_stalls} present the IPC and top-level stall distributions for the CPU2026 intrate and fprate benchmarks, respectively. Beyond the expected correlation between high IPC and a high fraction of retiring cycles, several broader behavioral categories emerge from the data. A subset of workloads is predominantly frontend-bound, characterized by instruction-delivery stalls exceeding other components; representative examples include 727.cppcheck and 753.ns3 in the intrate suite and 709.cactus in the fprate suite. In contrast, another group exhibits backend-bound behavior, with stalls dominated by memory latency or execution resource constraints, as seen in benchmarks such as 708.sqlite, 749.fotonik3d and 765.roms. Several others, such as 750.sealcrypto and 766.femflow, exhibit negligible lost cycles, indicating highly predictable control flow with minimal speculative penalties. At the suite level, the fprate benchmarks show a stronger tendency toward backend bottlenecks, whereas the intrate suite displays a more balanced mix of frontend and backend limited behavior. Lost cycle ratios in fprate are also generally lower, consistent with the more regular control flow of floating-point applications. These trends provide a structural view of the workload diversity across the two suites.


\subsection{SPECspeed Characterization}

Continuing the analysis from Section~\ref{specrate_characterization}, we offer the results from SPECspeed here. These benchmarks were run with 128 threads, noting that not all speed workloads fully scale to this thread count.

Tables~\ref{tab:intspeed_stalls} and~\ref{tab:fpspeed_stalls} present the IPC and stall distributions for the intspeed and fpspeed suites, respectively. As with the rate benchmarks, higher IPC values generally coincide with higher retiring fractions, while lower-IPC workloads tend to be dominated by stall behavior. The intspeed suite shows a mix of frontend-bound and backend-bound behavior, whereas the fpspeed suite is more uniformly backend limited, reflecting the memory intensive nature of many floating-point kernels.

With so many new multithreaded benchmarks, the community has an opportunity to study these to characterize highly contended locks, sharing of dirty lines between cores, and cache coherency issues that may stress snoop filters and other CPU features related to shared memory. A cursory analysis suggests 800.pot3d and 801.xz exhibit the most contention; the committee encourages deeper analysis on thread scalabilty, data sharing, and other performance metrics for MT.

\begin{table}[tbp] 
\arrayrulecolor{lightgray!40}

\footnotesize
\centering
\caption{INT SPEED Benchmarks: IPC and Stall Ratio Breakdown}
\label{tab:intspeed_stalls}
\setlength{\tabcolsep}{3pt} 
\setlength{\extrarowheight}{.75ex}
\setlength{\aboverulesep}{0pt}
\setlength{\belowrulesep}{0pt}
\begin{tabularx}{\columnwidth}{l *{5}{C}}
\rowcolor{tableheadspec}
\textbf{\textcolor{white}{Benchmark}} & 
\textbf{\textcolor{white}{IPC}} & 
\textbf{\textcolor{white}{Frontend}} & 
\textbf{\textcolor{white}{Backend}} & 
\textbf{\textcolor{white}{Lost}} & 
\textbf{\textcolor{white}{Retiring}} \\
\bottomrule

801.xz\_s       & \cellcolor[HTML]{EBEBEB}1.92 & \cellcolor[HTML]{EAFAFA}0.15 & \cellcolor[HTML]{EBADEB}0.55 & \cellcolor[HTML]{FBFBEC}0.07 & \cellcolor[HTML]{EFF7F4}0.23 \\
\bottomrule
807.ntest\_s     & \cellcolor[HTML]{D9D9D9}3.57 & \cellcolor[HTML]{EFFBFB}0.11 & \cellcolor[HTML]{F0C3F0}0.40 & \cellcolor[HTML]{FBFBED}0.06 & \cellcolor[HTML]{DCEFE3}0.42 \\
\bottomrule
817.flac\_s      & \cellcolor[HTML]{D1D1D1}4.38 & \cellcolor[HTML]{F3FCFC}0.08 & \cellcolor[HTML]{F2C9F2}0.36 & \cellcolor[HTML]{FFFFFD}0.01 & \cellcolor[HTML]{D3ECDC}0.52 \\
\bottomrule
821.gcc\_s       & \cellcolor[HTML]{E9E9E9}2.07 & \cellcolor[HTML]{B8EEEE}0.48 & \cellcolor[HTML]{F8E0F8}0.21 & \cellcolor[HTML]{FAFAE8}0.08 & \cellcolor[HTML]{EEF7F3}0.23 \\
\bottomrule
823.llvm\_s      & \cellcolor[HTML]{EBEBEB}1.92 & \cellcolor[HTML]{C2F0F0}0.41 & \cellcolor[HTML]{F3CFF3}0.32 & \cellcolor[HTML]{FCFCF2}0.04 & \cellcolor[HTML]{EFF7F4}0.22 \\
\bottomrule
827.cppcheck\_s  & \cellcolor[HTML]{E4E4E4}2.54 & \cellcolor[HTML]{A6E9E9}0.60 & \cellcolor[HTML]{FCF0FC}0.10 & \cellcolor[HTML]{FDFDF6}0.03 & \cellcolor[HTML]{EBF5F0}0.27 \\
\bottomrule
829.abc\_s       & \cellcolor[HTML]{F8F8F8}0.72 & \cellcolor[HTML]{E1F8F8}0.20 & \cellcolor[HTML]{E79CE7}0.66 & \cellcolor[HTML]{FCFCF3}0.04 & \cellcolor[HTML]{FCFCFF}0.08 \\
\bottomrule
834.vpr\_s       & \cellcolor[HTML]{E7E7E7}2.26 & \cellcolor[HTML]{CFF3F3}0.32 & \cellcolor[HTML]{F3CEF3}0.33 & \cellcolor[HTML]{FAFAE8}0.08 & \cellcolor[HTML]{ECF6F1}0.26 \\
\bottomrule
835.gem5\_s      & \cellcolor[HTML]{ECECEC}1.81 & \cellcolor[HTML]{BFEFEF}0.43 & \cellcolor[HTML]{F4D1F4}0.31 & \cellcolor[HTML]{FCFCF3}0.04 & \cellcolor[HTML]{F0F7F5}0.21 \\
\bottomrule
838.diamond\_s   & \cellcolor[HTML]{E3E3E3}2.71 & \cellcolor[HTML]{F3FCFC}0.08 & \cellcolor[HTML]{ECB0EC}0.53 & \cellcolor[HTML]{FCFCF4}0.04 & \cellcolor[HTML]{E3F2E9}0.36 \\
\bottomrule
846.minizinc\_s  & \cellcolor[HTML]{F1F1F1}1.38 & \cellcolor[HTML]{DBF6F6}0.25 & \cellcolor[HTML]{EAACEA}0.56 & \cellcolor[HTML]{FDFDF7}0.03 & \cellcolor[HTML]{F5F9F9}0.16 \\
\bottomrule
853.ns3\_s       & \cellcolor[HTML]{EAEAEA}2.05 & \cellcolor[HTML]{BDEFEF}0.45 & \cellcolor[HTML]{F5D5F5}0.29 & \cellcolor[HTML]{FEFEF8}0.02 & \cellcolor[HTML]{EDF6F2}0.24 \\
\bottomrule
854.graph500\_s  & \cellcolor[HTML]{F4F4F4}1.12 & \cellcolor[HTML]{DAF6F6}0.25 & \cellcolor[HTML]{E8A0E8}0.64 & \cellcolor[HTML]{FFFFFD}0.01 & \cellcolor[HTML]{FAFCFD}0.11 \\ 
\bottomrule

\end{tabularx}
\par
\setlength{\aboverulesep}{0.6ex}
\setlength{\belowrulesep}{0.9ex}
\setlength{\extrarowheight}{0pt}
\end{table}

\begin{table}[tbp] 
\arrayrulecolor{lightgray!40}
\footnotesize
\centering
\caption{FP SPEED Benchmarks: IPC and Stall Ratio Breakdown}
\label{tab:fpspeed_stalls}
\setlength{\tabcolsep}{3pt} 
\setlength{\extrarowheight}{.75ex}
\setlength{\aboverulesep}{0pt}
\setlength{\belowrulesep}{0pt}
\begin{tabularx}{\columnwidth}{l *{5}{C}}
\rowcolor{tableheadspec}
\textbf{\textcolor{white}{Benchmark}} & 
\textbf{\textcolor{white}{IPC}} & 
\textbf{\textcolor{white}{Frontend}} & 
\textbf{\textcolor{white}{Backend}} & 
\textbf{\textcolor{white}{Lost}} & 
\textbf{\textcolor{white}{Retiring}} \\
\bottomrule

800.pot3d\_s      & \cellcolor[HTML]{F6F6F6}0.90 & \cellcolor[HTML]{DAF6F6}0.25 & \cellcolor[HTML]{E7A0E7}0.64 & \cellcolor[HTML]{FFFFFF}0.00 & \cellcolor[HTML]{F7FAFB}0.11 \\
\bottomrule
803.sph\_exa\_s    & \cellcolor[HTML]{E6E6E6}2.36 & \cellcolor[HTML]{F3FCFC}0.08 & \cellcolor[HTML]{E9A7E9}0.59 & \cellcolor[HTML]{FDFDF7}0.03 & \cellcolor[HTML]{E5F3EB}0.30 \\
\bottomrule
809.cactus\_s    & \cellcolor[HTML]{F0F0F0}1.45 & \cellcolor[HTML]{BBEEEE}0.46 & \cellcolor[HTML]{F2CAF2}0.35 & \cellcolor[HTML]{FFFFFF}0.00 & \cellcolor[HTML]{F0F7F4}0.19 \\
\bottomrule
811.tealeaf\_s    & \cellcolor[HTML]{E9E9E9}2.12 & \cellcolor[HTML]{DBF6F6}0.24 & \cellcolor[HTML]{EEBAEE}0.46 & \cellcolor[HTML]{FEFEFB}0.01 & \cellcolor[HTML]{E7F4ED}0.28 \\
\bottomrule
816.nab\_s        & \cellcolor[HTML]{E9E9E9}2.15 & \cellcolor[HTML]{E7F9F9}0.17 & \cellcolor[HTML]{EEBAEE}0.47 & \cellcolor[HTML]{F8F8E2}0.10 & \cellcolor[HTML]{E8F4EE}0.27 \\
\bottomrule
820.cloverleaf\_s & \cellcolor[HTML]{FBFBFB}0.42 & \cellcolor[HTML]{FBFEFE}0.03 & \cellcolor[HTML]{DD76DD}0.92 & \cellcolor[HTML]{FFFFFF}0.00 & \cellcolor[HTML]{FCFCFF}0.05 \\
\bottomrule
822.palm\_s       & \cellcolor[HTML]{E8E8E8}2.17 & \cellcolor[HTML]{D8F5F5}0.27 & \cellcolor[HTML]{ECB3EC}0.51 & \cellcolor[HTML]{FFFFFF}0.00 & \cellcolor[HTML]{EEF7F3}0.20 \\
\bottomrule
849.fotonik3d\_s  & \cellcolor[HTML]{FBFBFB}0.38 & \cellcolor[HTML]{F1FCFC}0.09 & \cellcolor[HTML]{DF7FDF}0.85 & \cellcolor[HTML]{FFFFFF}0.00 & \cellcolor[HTML]{FCFCFF}0.05 \\
\bottomrule
857.namd\_s       & \cellcolor[HTML]{D8D8D8}3.67 & \cellcolor[HTML]{EBFAFA}0.14 & \cellcolor[HTML]{F3CDF3}0.34 & \cellcolor[HTML]{FAFAEA}0.07 & \cellcolor[HTML]{D7EDDF}0.46 \\
\bottomrule
865.roms\_s       & \cellcolor[HTML]{F8F8F8}0.67 & \cellcolor[HTML]{F1FCFC}0.10 & \cellcolor[HTML]{E184E1}0.82 & \cellcolor[HTML]{FFFFFE}0.00 & \cellcolor[HTML]{F9FBFD}0.08 \\
\bottomrule
867.nest\_s       & \cellcolor[HTML]{ECECEC}1.83 & \cellcolor[HTML]{DEF7F7}0.22 & \cellcolor[HTML]{EFBCEF}0.45 & \cellcolor[HTML]{F8F8E2}0.10 & \cellcolor[HTML]{ECF6F2}0.22 \\
\bottomrule
872.marian\_s     & \cellcolor[HTML]{D4D4D4}4.10 & \cellcolor[HTML]{F8FEFE}0.05 & \cellcolor[HTML]{EFBFEF}0.43 & \cellcolor[HTML]{FFFFFE}0.00 & \cellcolor[HTML]{D2EBDB}0.50 \\
\bottomrule
881.neutron\_s    & \cellcolor[HTML]{F3F3F3}1.20 & \cellcolor[HTML]{D4F4F4}0.29 & \cellcolor[HTML]{EFBCEF}0.45 & \cellcolor[HTML]{F8F8E1}0.10 & \cellcolor[HTML]{F3F9F7}0.15 \\
\bottomrule
\end{tabularx}
\par
\setlength{\aboverulesep}{0.6ex}
\setlength{\belowrulesep}{0.9ex}
\setlength{\extrarowheight}{0pt}
\end{table}

\subsection{Energy}

The SPEC CPU2017 benchmark suite introduced an optional metric to report energy consumption (in Joules) during CPU-intensive benchmarks. It requires a power measurement device and compliance with SPEC rules. The results help evaluate performance per watt, which is critical for energy efficiency in data centers and HPC environments. Reports typically include energy used and efficiency ratios, enabling comparisons beyond raw performance. This feature is retained as-is in CPU2026, and the methodology and measurement approach remain the same for the new set of benchmarks.

\subsection{Upstreamed Performance Improvements}

While the primary goal of the benchmark development process is to ensure portability and correctness, the work is inherently conducted by engineers with an expertise in performance analysis. As a result, opportunities are found for algorithmic optimizations which are beyond the scope of automated compiler technology. In the spirit of the symbiotic relationship with the open-source community, SPEC offers these enhancements back to the upstream projects. This section details some examples of benchmark code improvements which were accepted by their respective maintainers.

\para{767.nest} An analysis of hot functions in the modeling code revealed opportunities for memoization and strength reduction. A key constant involving a square root was being recalculated on every function call; this was refactored to be computed only once. Additionally, several loop-invariant divisions were moved outside a critical loop, and remaining divisions inside the loop were replaced with multiplications by their reciprocals. These optimizations resulted in a 10\% reduction in total application runtime upstream \cite{nest_memoization}.

\para{846.minizinc} An analysis of the source code identified a performance-critical section where a \texttt{std::vector} was being copied element-by-element using an explicit loop. This was refactored to use the more idiomatic and efficient assignment operator of \texttt{std::vector}. This change not only simplified and modernized the codebase but also resulted in a 7\% reduction in total application runtime upstream \cite{mzn_vector}.

\para{707.ntest} Several micro-optimizations were applied to the benchmark's scoring module. Floating-point overhead was reduced by moving variable declarations into narrower scopes to avoid unnecessary duplicated calculations and type conversions. Additionally, a strength-reduction optimization was applied, replacing a division with a multiplication by its pre-calculated reciprocal \cite{ntest_perf}. 

\para{Enrichment} Similar patches were upstreamed to other candidate applications \cite{tess_perf, brotli_perf, highs_perf, h3geo_perf}, even though these projects were ultimately culled for reasons detailed in Section~\S\ref{sec:culled}. 

\subsection{BBV Recurrence and Performance Plots}
\label{sec:bbv_appendix}
BBV plots were described in detail in Section~\S\ref{bbv_description}. Here we offer the self-similarity plots for all of the single-threaded benchmarks in CPU2026, alongside their corresponding performance bottleneck plots, in Figures \ref{fig:bbv_int_rate} and \ref{fig:bbv_perf_fp}. These results are captured from the same machine cited in Table \ref{tab:sys-config}, running a single copy of each benchmark. Since the BBV self-similarity analysis only makes sense for single-threaded runs, for the multi-threaded refspeed benchmarks we only offer performance plots, in Figures \ref{fig:perf_refspeed_int} and \ref{fig:perf_refspeed_fp}.

Here we can see how the benchmarks evoke a variety of microarchitectural behaviors from the underlying hardware, even within the workloads themselves. While the HPC centric benchmarks show high self-similarity and only exhibit one or two phases (i.e. 709.cactus, 722.palm, 749.fotonik3d, 765.roms, 782.lbm), the majority of the remainder have multiple phases and diversity in both code and hardware response.

\include{99.1-bbv_int_4x4}
\include{99.2-bbv_fp_perf_plot}

\clearpage
\include{99.5-intspeed-2wide}

\include{99.6-fpspeed-2wide}

%% file: 99.1-bbv_int_4x4.tex
\newcolumntype{C}{>{\centering\arraybackslash}X} 

\begin{figure*}[t]
    \centering 

    \begin{tabularx}{\textwidth}{@{} *{4}{C} @{}}

        \begin{subfigure}[b]{\linewidth}
            \includegraphics[width=\linewidth]{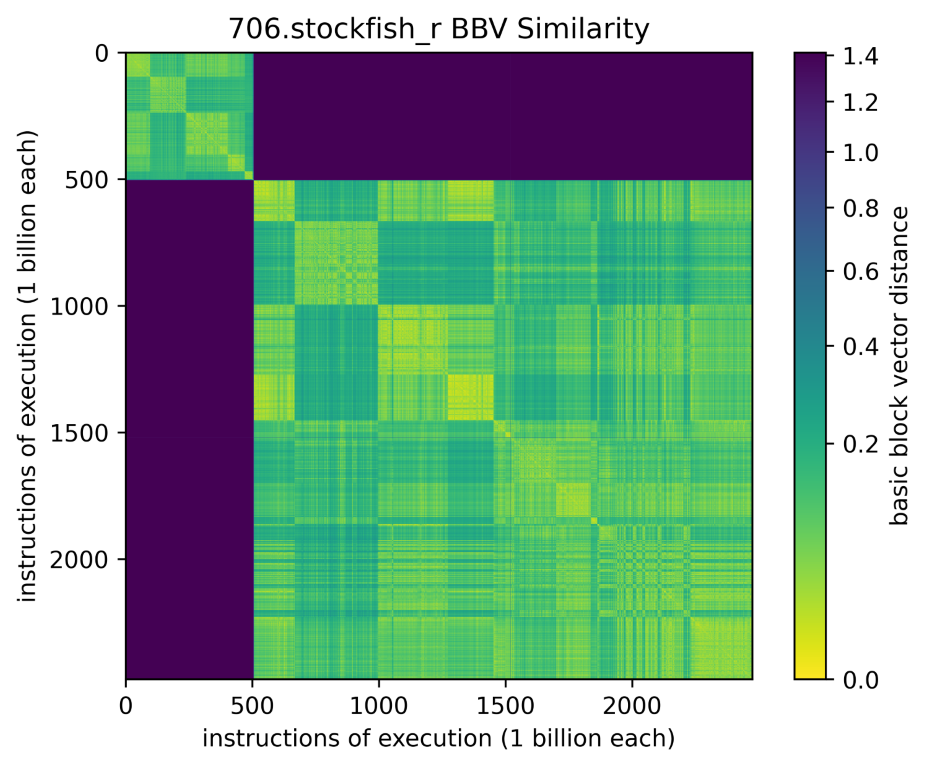}
            \label{fig:bbv_706}
        \end{subfigure}
        &
        \begin{subfigure}[b]{\linewidth}
            \includegraphics[width=\linewidth]{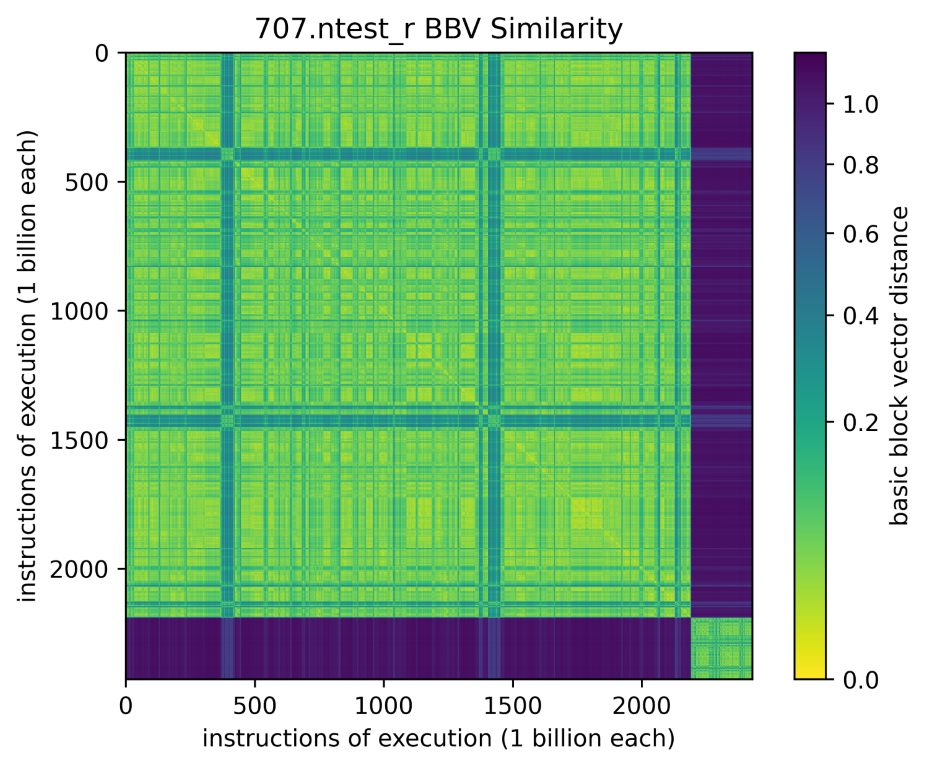}
            \label{fig:bbv_707}
        \end{subfigure}
        &
        \begin{subfigure}[b]{\linewidth}
            \includegraphics[width=\linewidth]{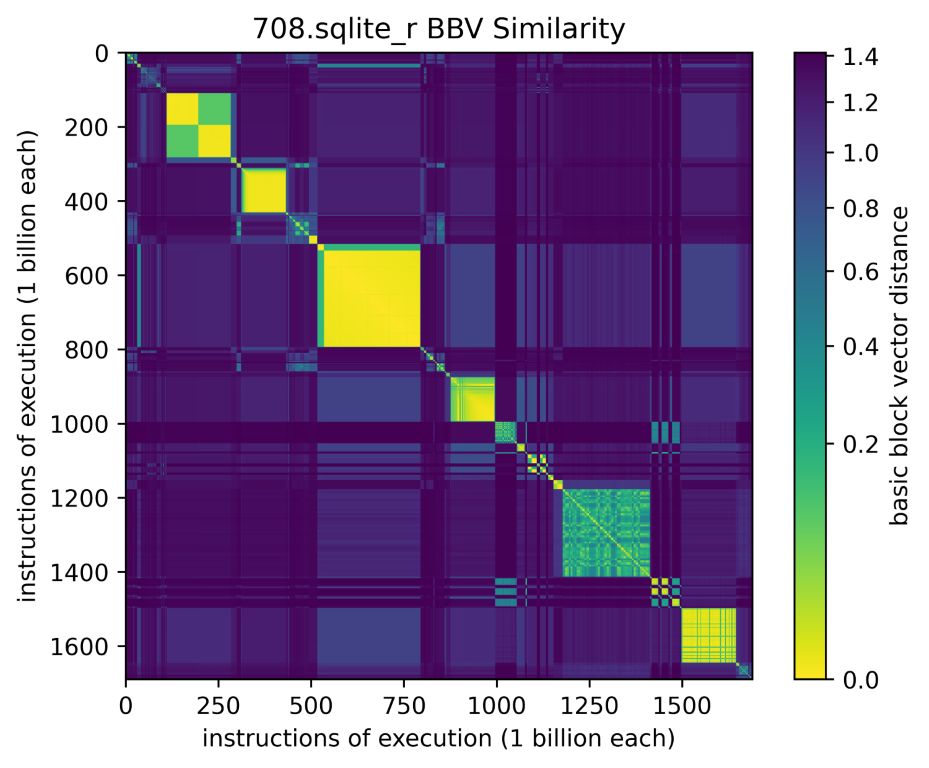}
            \label{fig:bbv_708}
        \end{subfigure}
        &
        \begin{subfigure}[b]{\linewidth}
            \includegraphics[width=\linewidth]{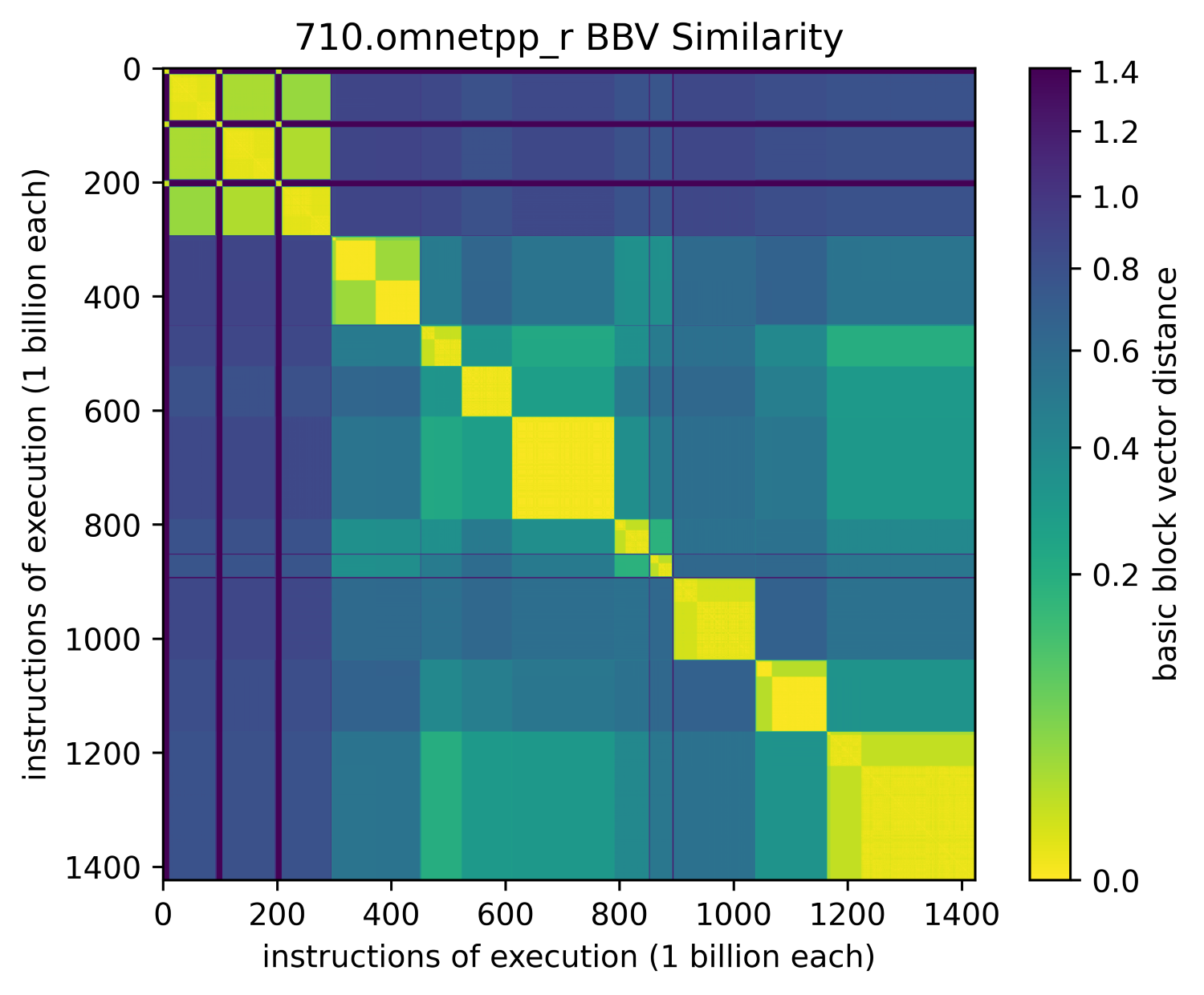}
            \label{fig:bbv_710}
        \end{subfigure}
        \\[-26pt] 

        \begin{subfigure}[b]{\linewidth}
            \centering
            \hspace*{-9pt} 
            \includegraphics[width=0.82\linewidth]{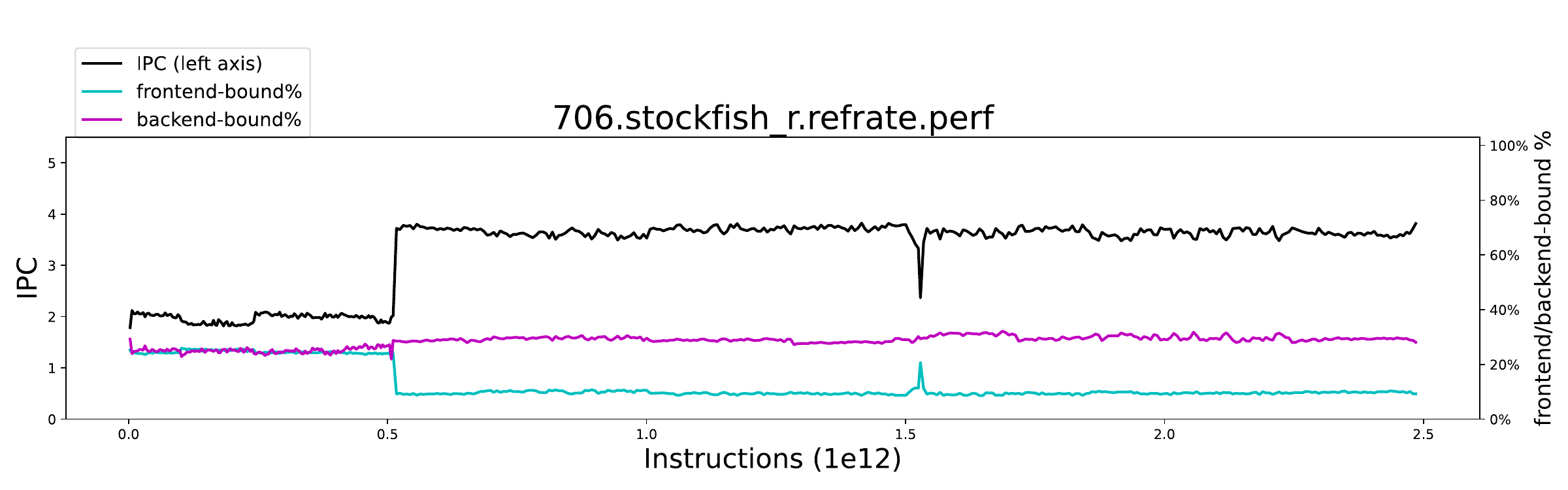}
        \end{subfigure}
        &
        \begin{subfigure}[b]{\linewidth}
            \centering
            \hspace*{-9pt} 
            \includegraphics[width=0.82\linewidth]{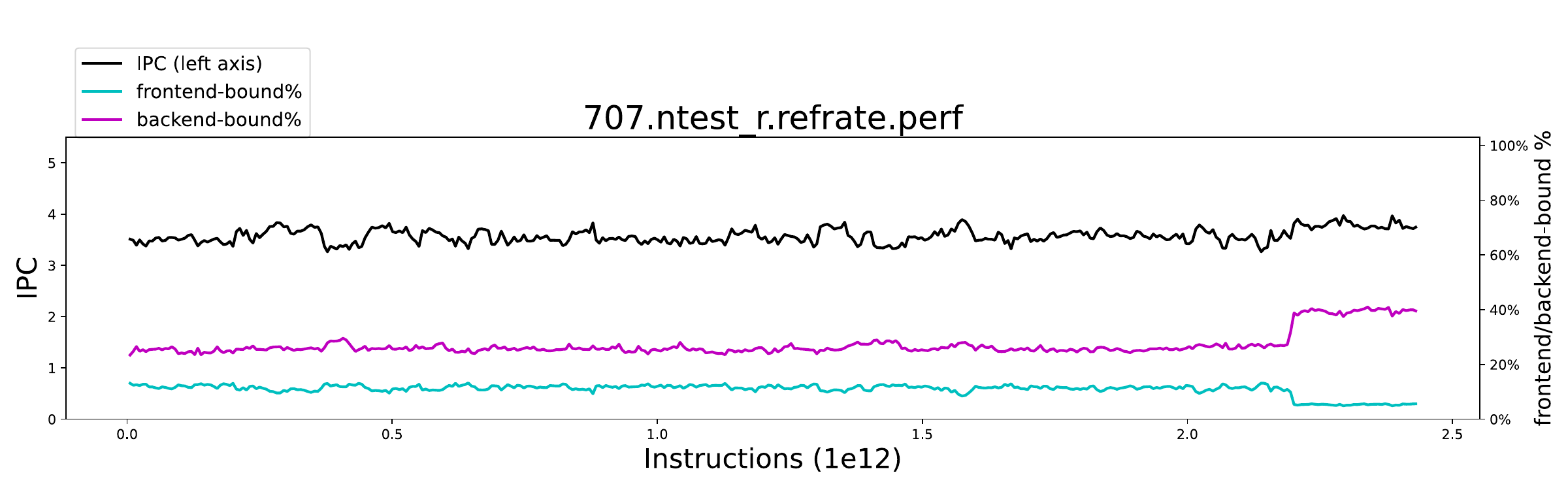}
        \end{subfigure}
        &
        \begin{subfigure}[b]{\linewidth}
            \centering
            \hspace*{-9pt} 
            \includegraphics[width=0.82\linewidth]{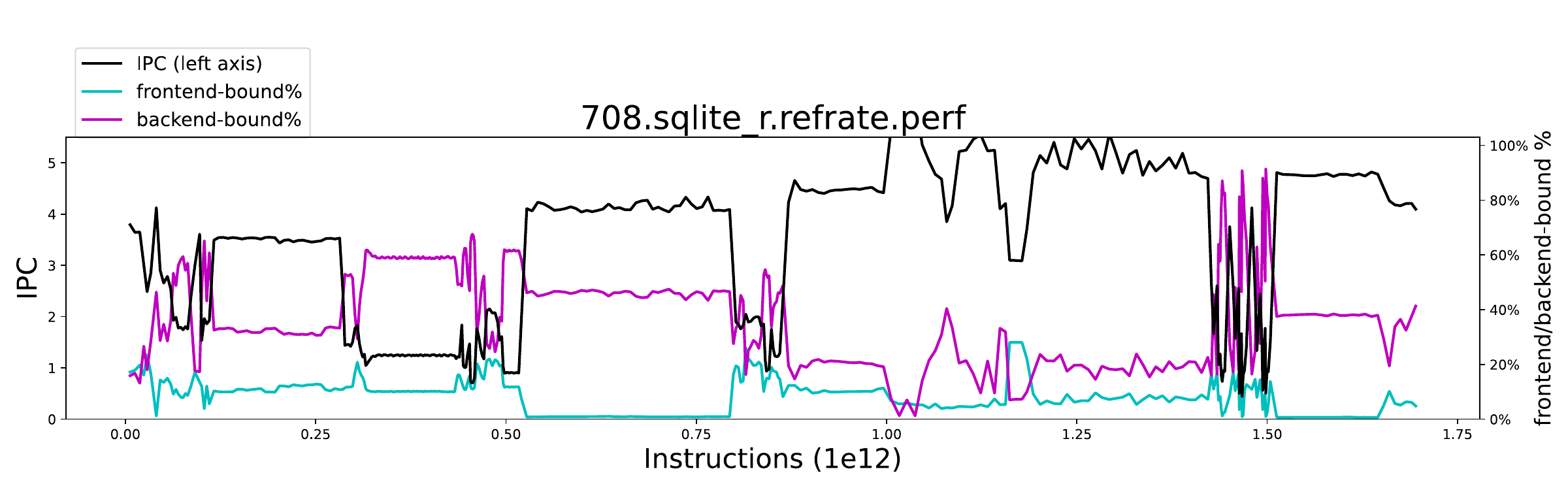}
        \end{subfigure}
        &
        \begin{subfigure}[b]{\linewidth}
            \centering
            \hspace*{-9pt} 
            \includegraphics[width=0.82\linewidth]{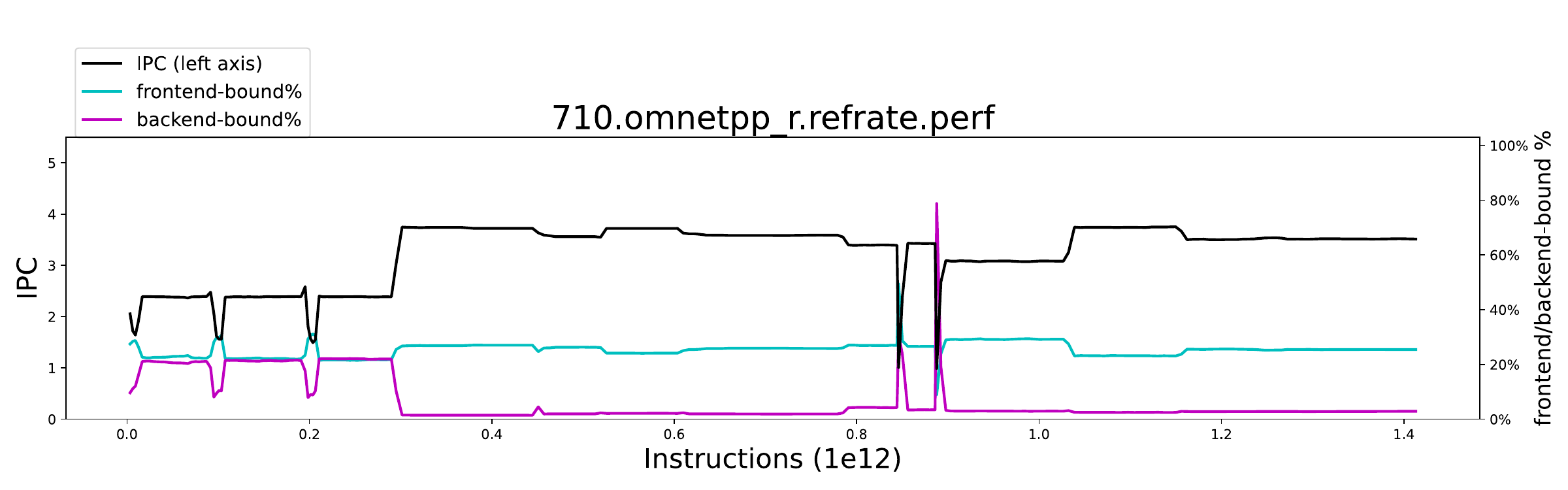}
            
        \end{subfigure}
        \\[+4pt]

        \begin{subfigure}[b]{\linewidth}
            \includegraphics[width=\linewidth]{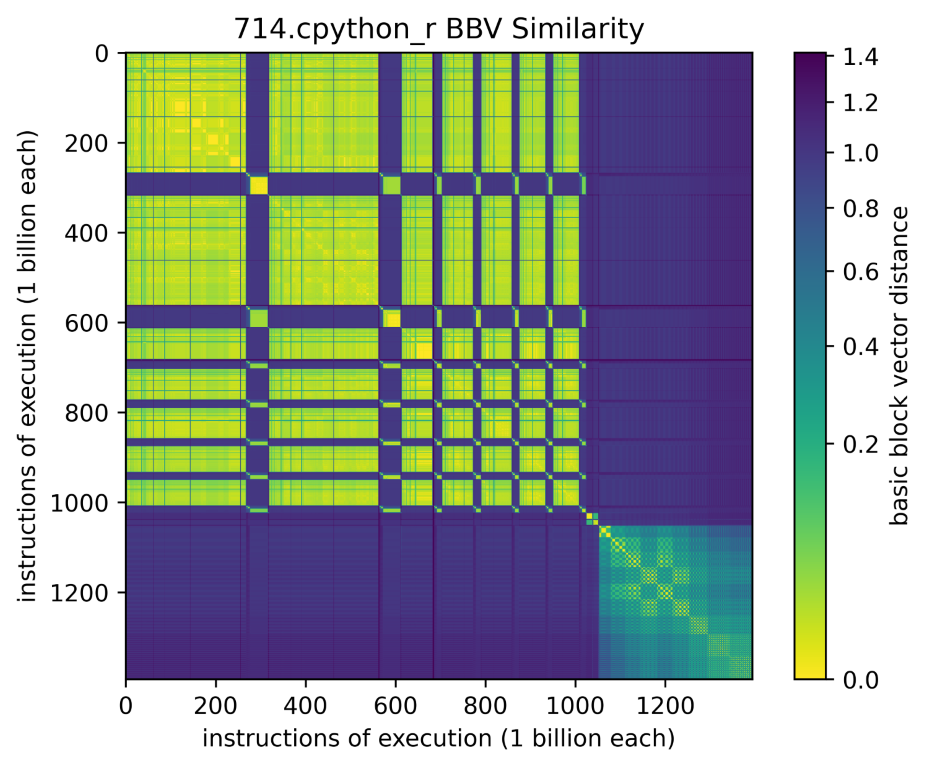}
            \label{fig:bbv_714}
        \end{subfigure}
        &
        \begin{subfigure}[b]{\linewidth}
            \includegraphics[width=\linewidth]{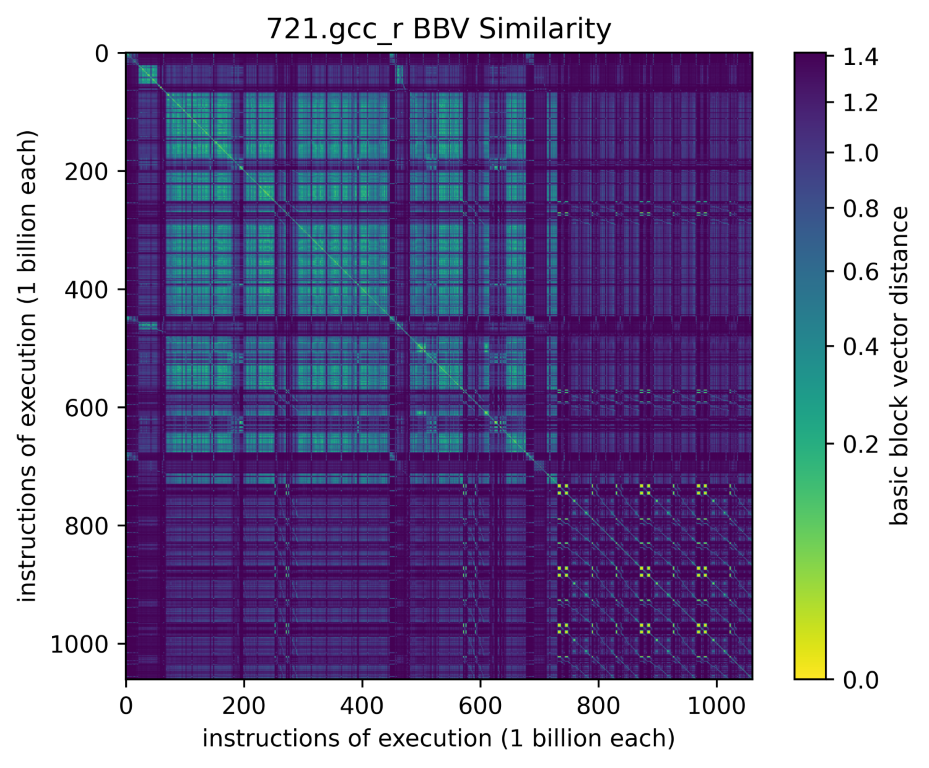}
            \label{fig:bbv_721}
        \end{subfigure}
        &
        \begin{subfigure}[b]{\linewidth}
            \includegraphics[width=\linewidth]{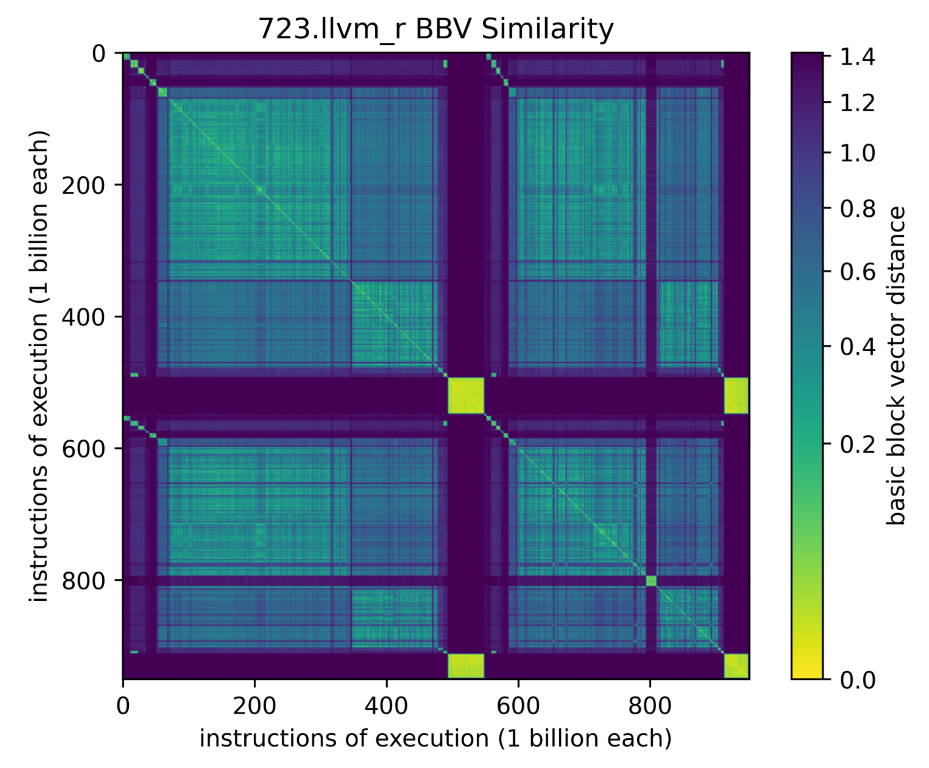}
            \label{fig:bbv_723}
        \end{subfigure}
        &
        \begin{subfigure}[b]{\linewidth}
            \includegraphics[width=\linewidth]{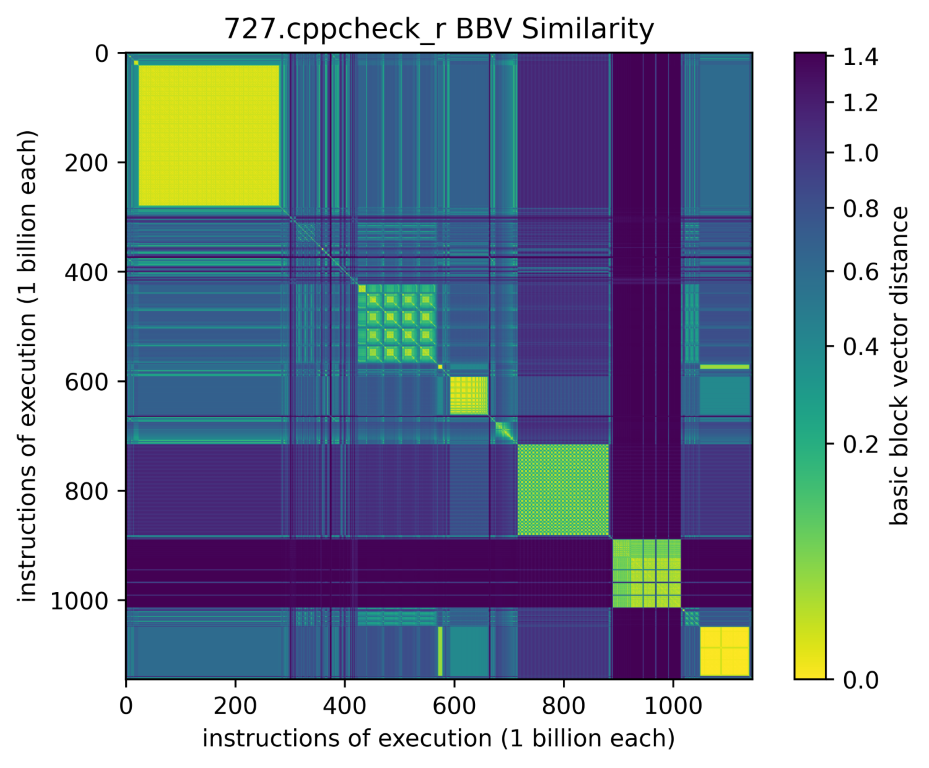}
            \label{fig:bbv_727}
        \end{subfigure}
        \\[-26pt] 

        \begin{subfigure}[b]{\linewidth}
            \centering
            \hspace*{-9pt} 
            \includegraphics[width=0.82\linewidth]{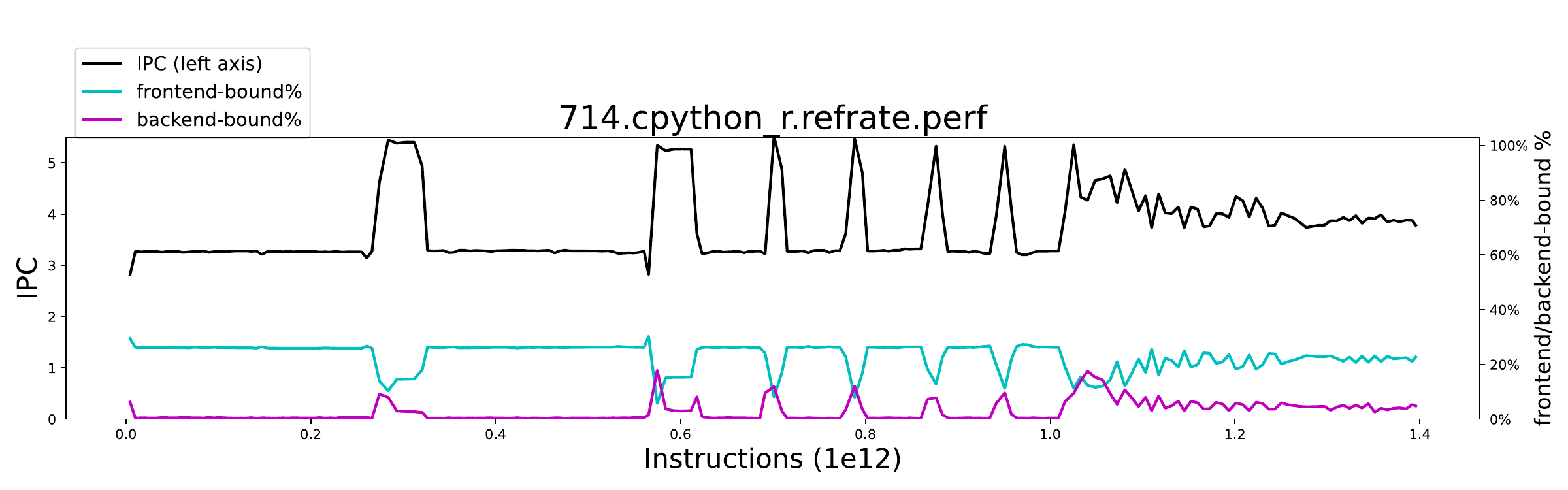}
        \end{subfigure}
        &
        \begin{subfigure}[b]{\linewidth}
            \centering
            \hspace*{-9pt} 
            \includegraphics[width=0.82\linewidth]{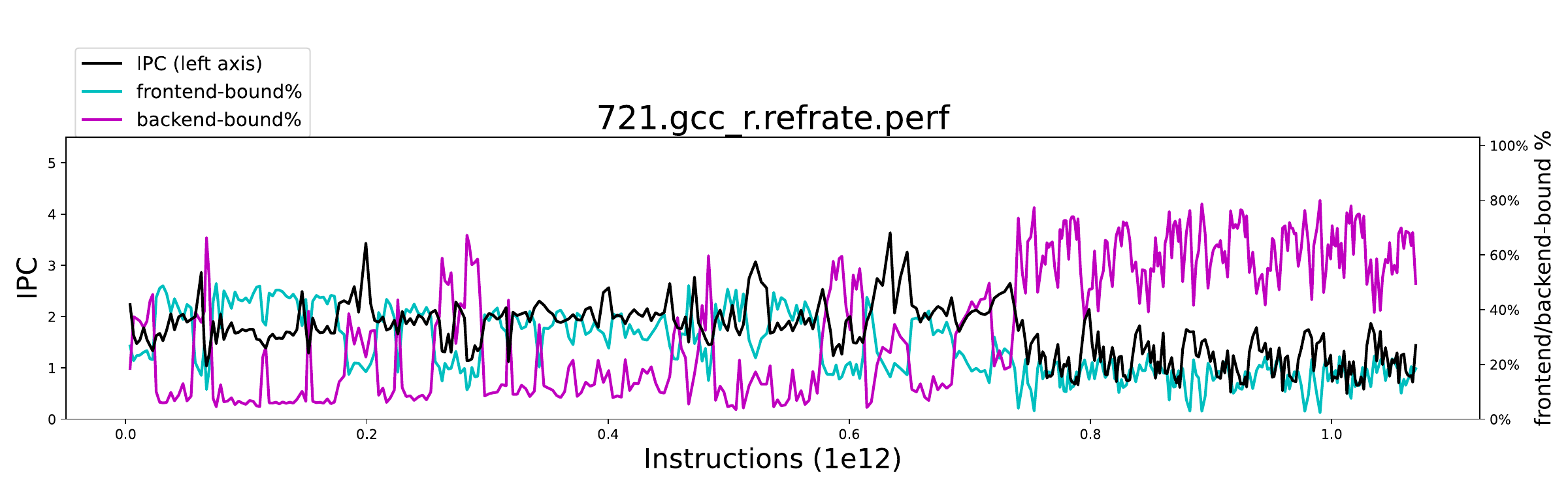}
        \end{subfigure}
        &
        \begin{subfigure}[b]{\linewidth}
            \centering
            \hspace*{-9pt} 
            \includegraphics[width=0.82\linewidth]{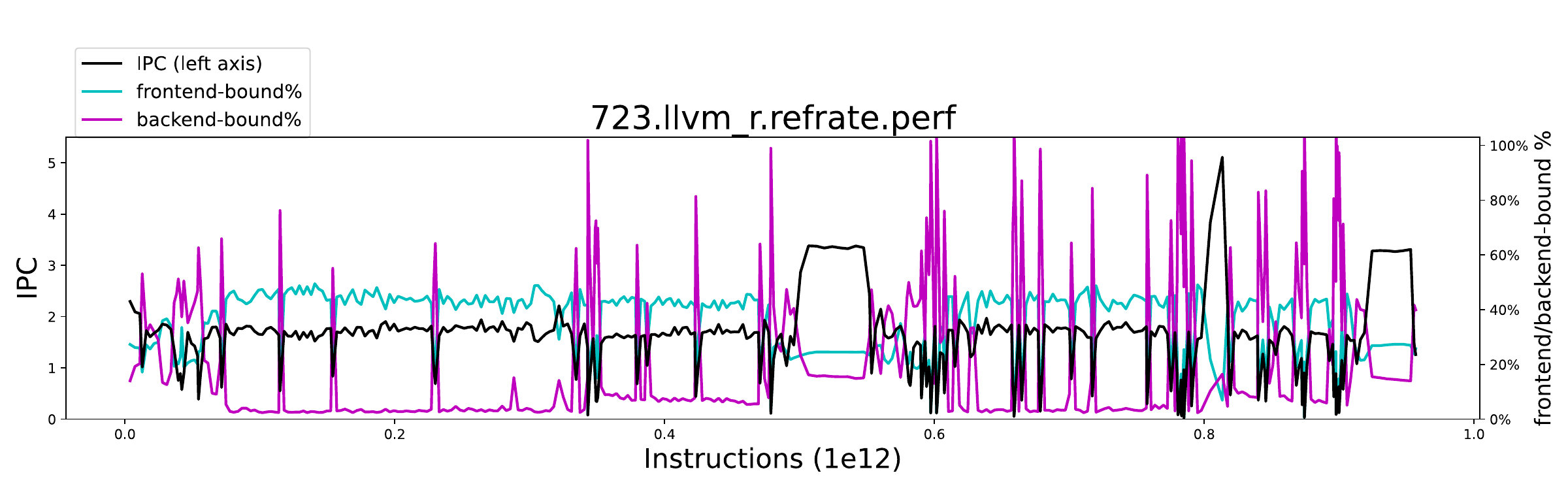}
        \end{subfigure}
        &
        \begin{subfigure}[b]{\linewidth}
            \centering
            \hspace*{-9pt} 
            \includegraphics[width=0.82\linewidth]{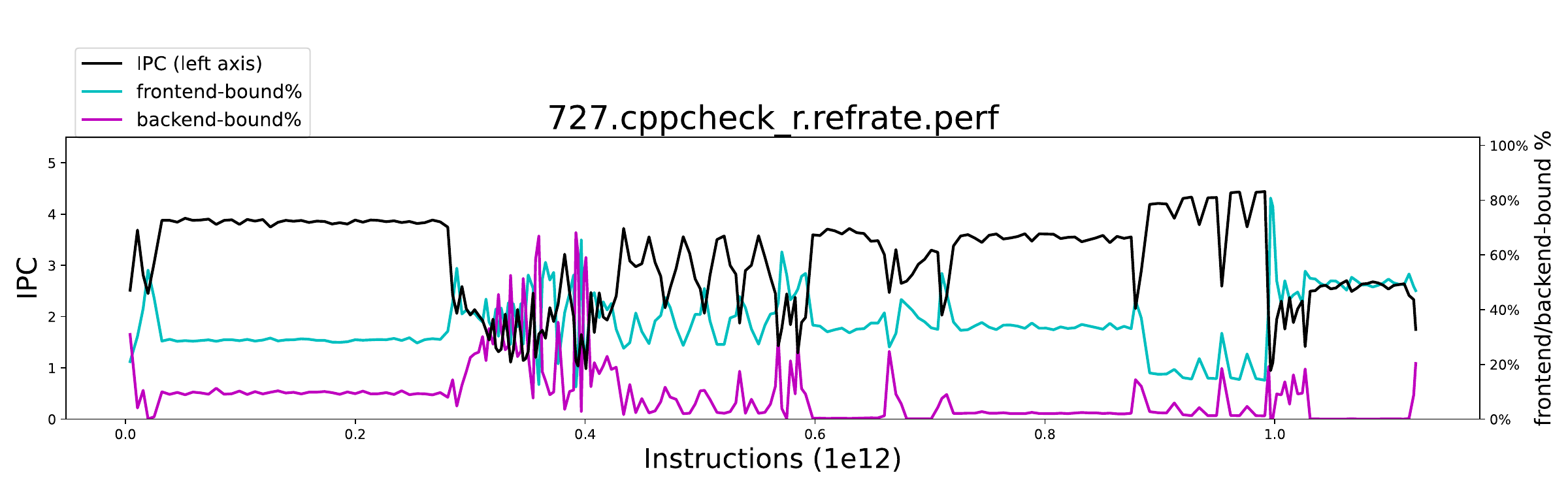}
            
        \end{subfigure}
        \\[+4pt]

        \begin{subfigure}[b]{\linewidth}
            \includegraphics[width=\linewidth]{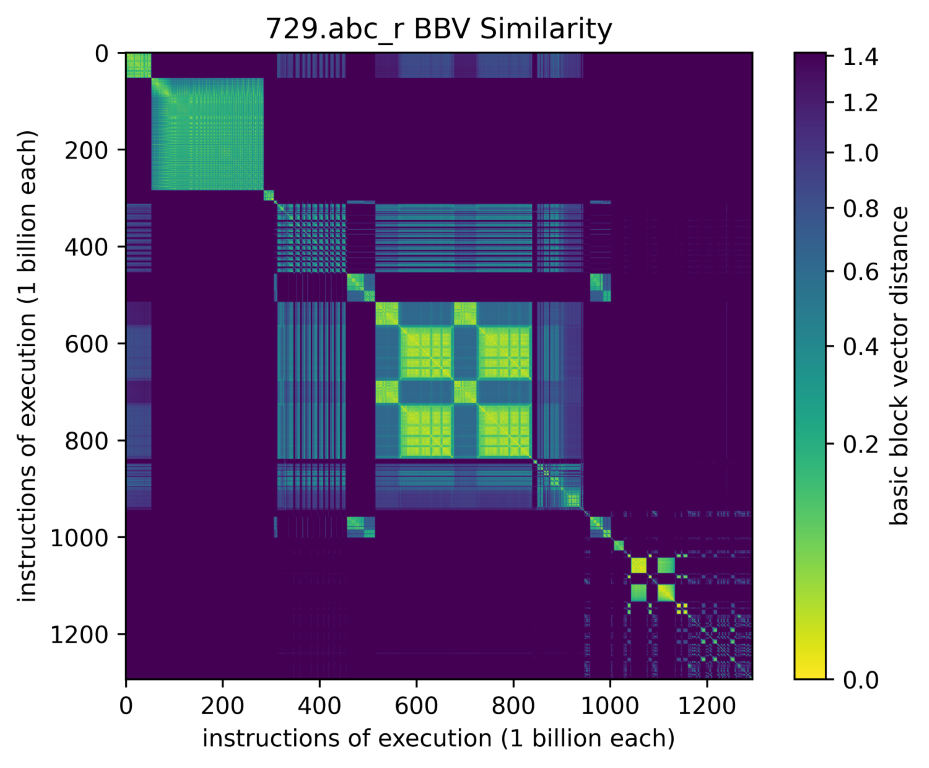}
            \label{fig:bbv_729}
        \end{subfigure}
        &
        \begin{subfigure}[b]{\linewidth}
            \includegraphics[width=\linewidth]{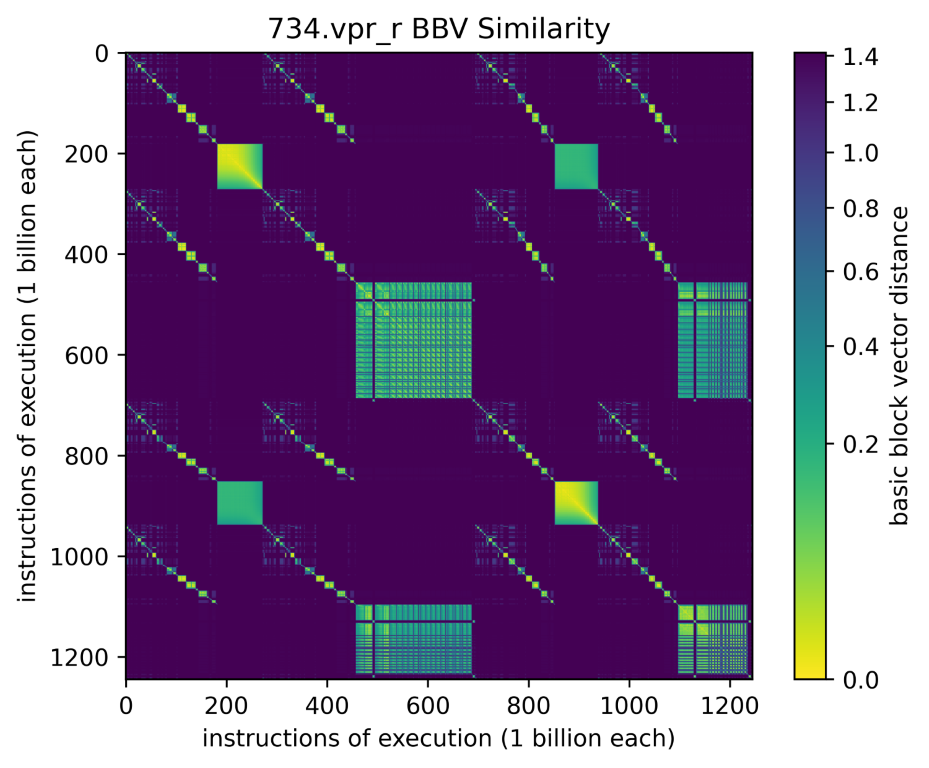}
            \label{fig:bbv_734}
        \end{subfigure}
        &
        \begin{subfigure}[b]{\linewidth}
            \includegraphics[width=\linewidth]{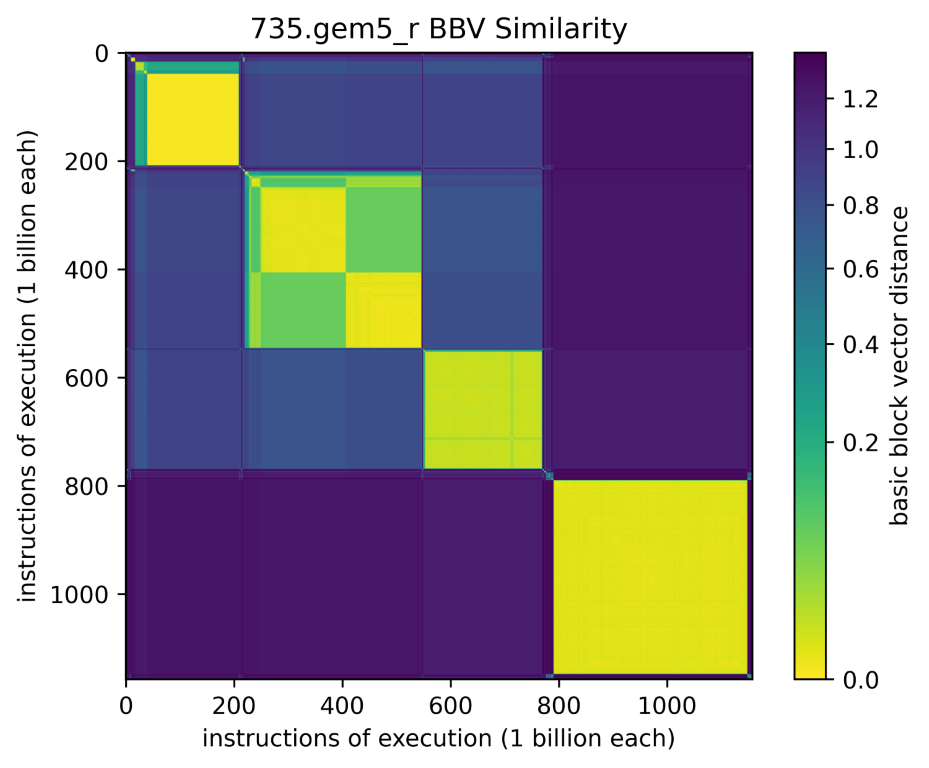}
            \label{fig:bbv_735}
        \end{subfigure}
        &
        \begin{subfigure}[b]{\linewidth}
            \includegraphics[width=\linewidth]{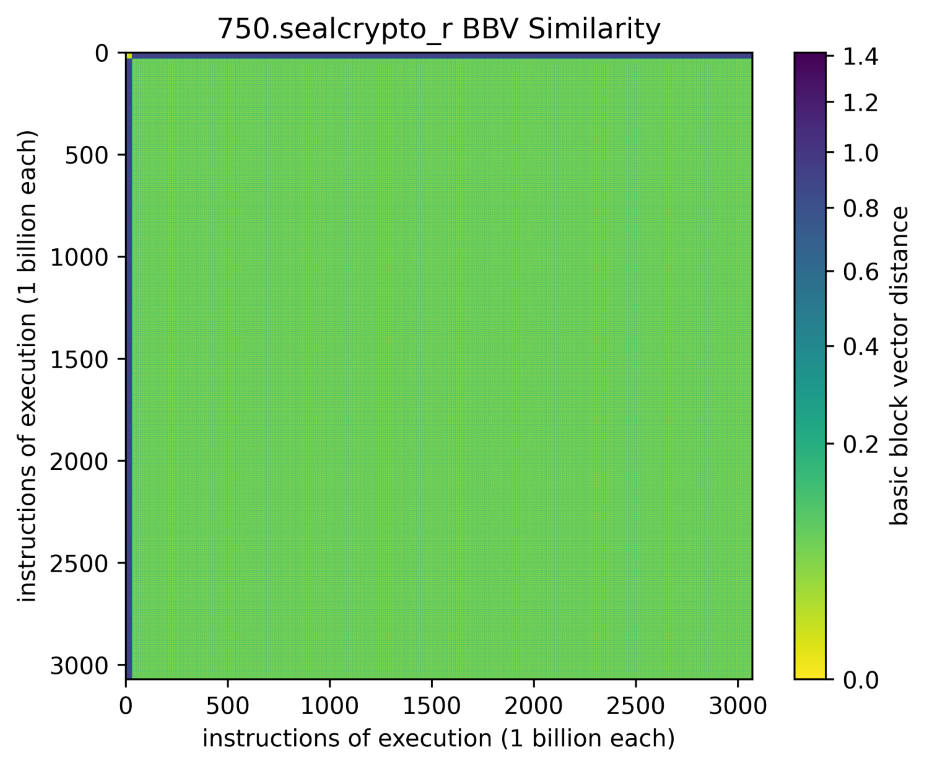}
            \label{fig:bbv_750}
        \end{subfigure}
        \\[-26pt] 

        \begin{subfigure}[b]{\linewidth}
            \centering
            \hspace*{-9pt} 
            \includegraphics[width=0.82\linewidth]{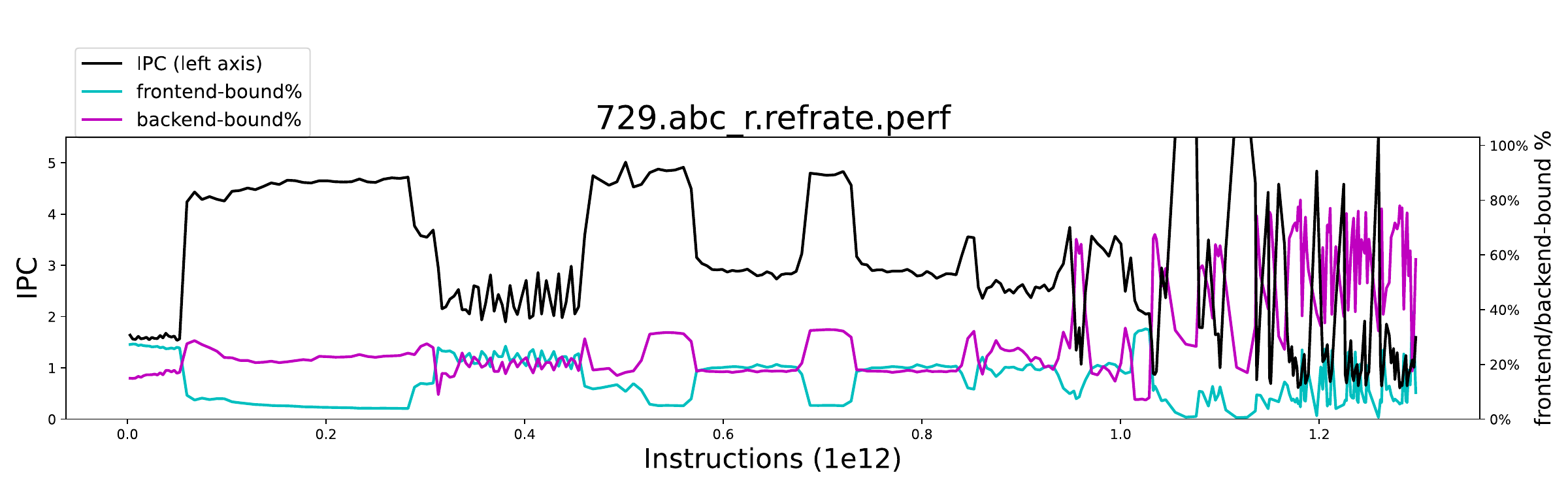}
        \end{subfigure}
        &
        \begin{subfigure}[b]{\linewidth}
            \centering
            \hspace*{-9pt} 
            \includegraphics[width=0.82\linewidth]{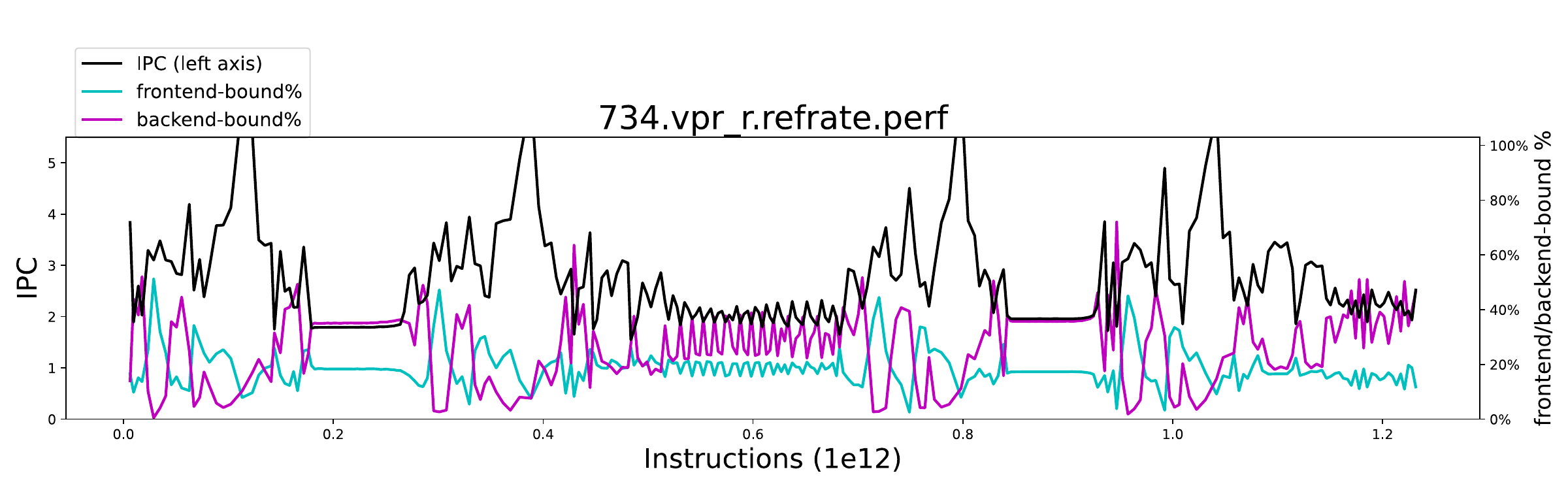}
        \end{subfigure}
        &
        \begin{subfigure}[b]{\linewidth}
            \centering
            \hspace*{-9pt} 
            \includegraphics[width=0.82\linewidth]{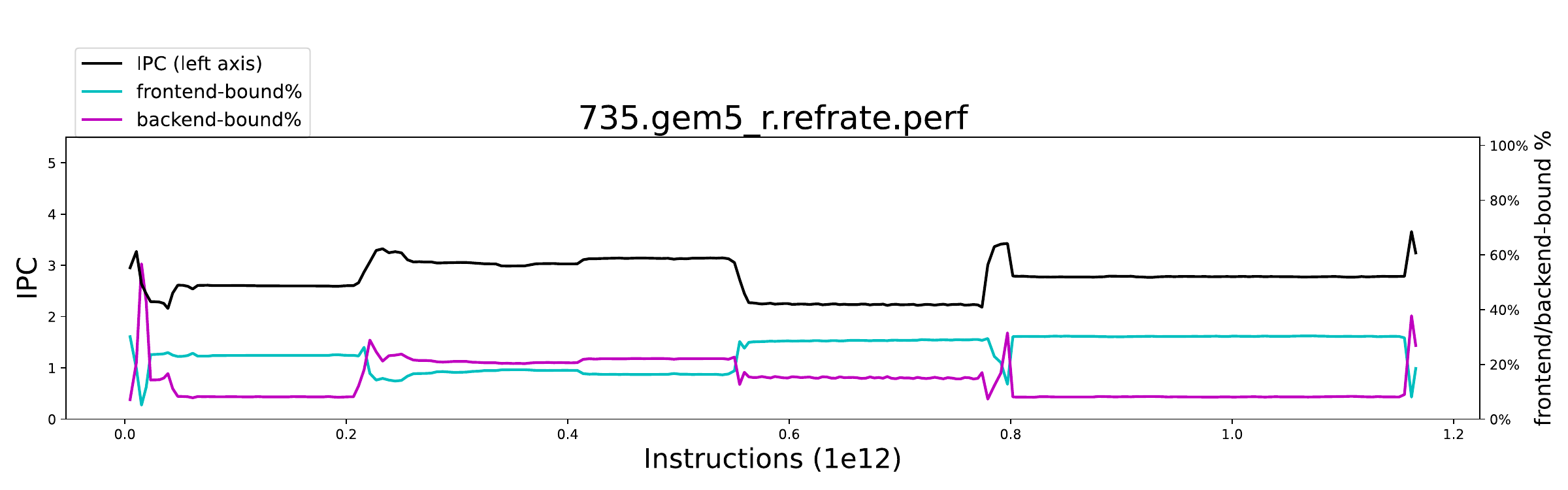}
        \end{subfigure}
        &
        \begin{subfigure}[b]{\linewidth}
            \centering
            \hspace*{-9pt} 
            \includegraphics[width=0.82\linewidth]{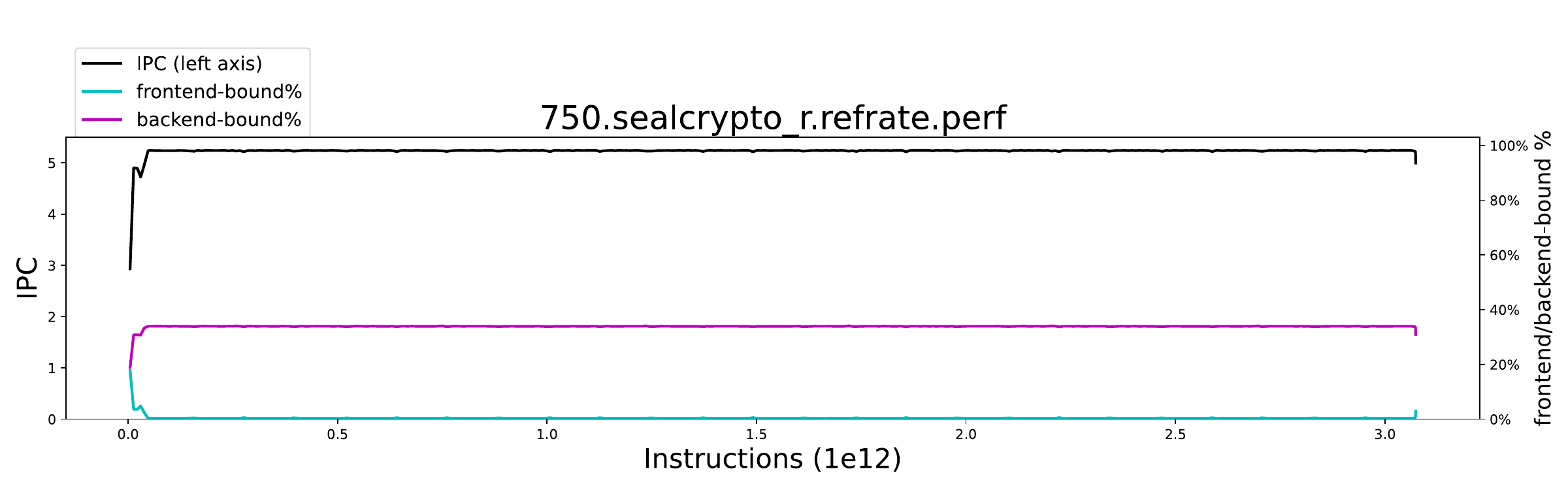}
            
        \end{subfigure}
        \\[+4pt]

        \begin{subfigure}[b]{\linewidth}
            \includegraphics[width=\linewidth]{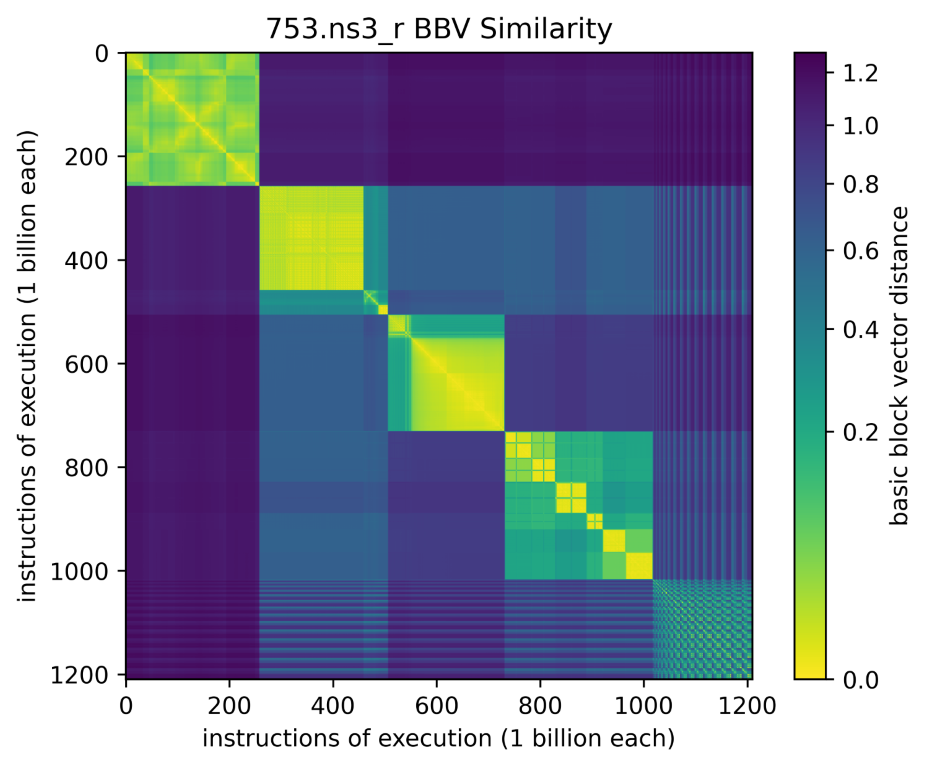}
            \label{fig:bbv_753}
        \end{subfigure}
        &
        \begin{subfigure}[b]{\linewidth}
            \includegraphics[width=\linewidth]{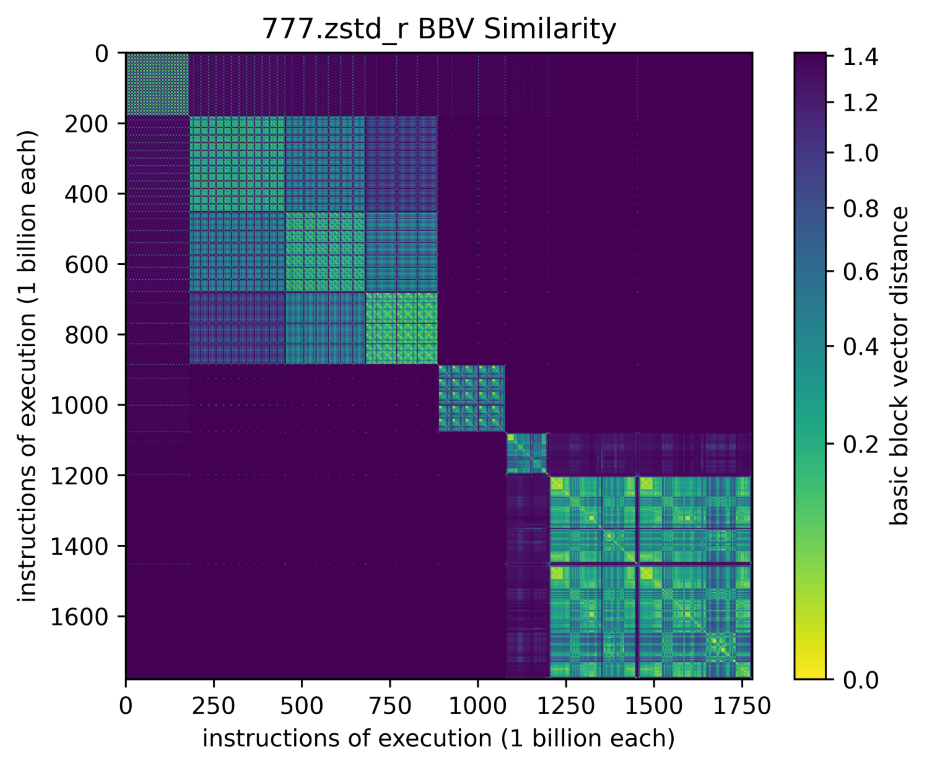}
            \label{fig:bbv_777}
        \end{subfigure}
        &
        & 
        \begin{subfigure}[b]{\linewidth}
            \includegraphics[width=\linewidth]{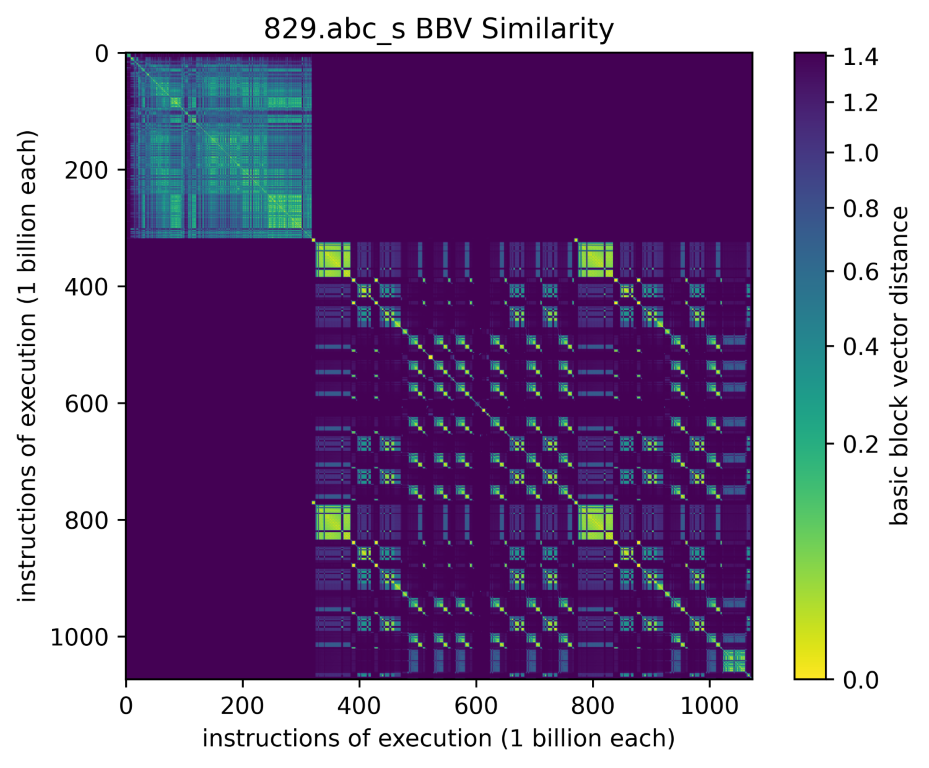}
            \label{fig:bbv_829}
        \end{subfigure}
        \\[-26pt] 

        \begin{subfigure}[b]{\linewidth}
            \centering
            \hspace*{-9pt} 
            \includegraphics[width=0.82\linewidth]{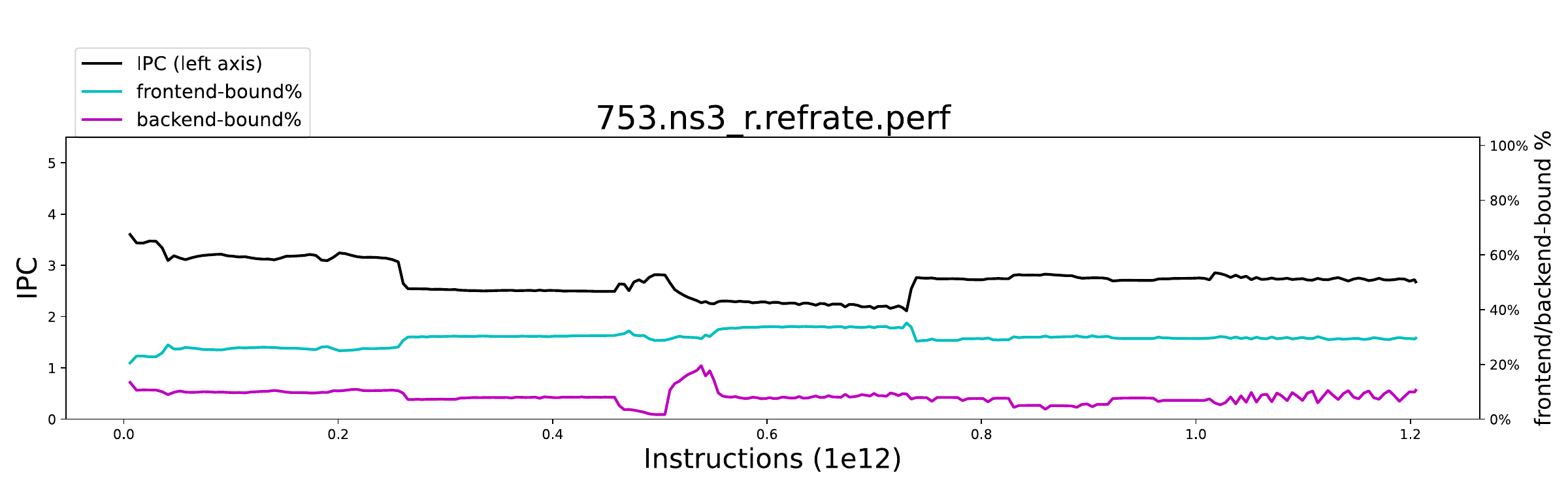}
        \end{subfigure}
        &
        \begin{subfigure}[b]{\linewidth}
            \centering
            \hspace*{-9pt} 
            \includegraphics[width=0.82\linewidth]{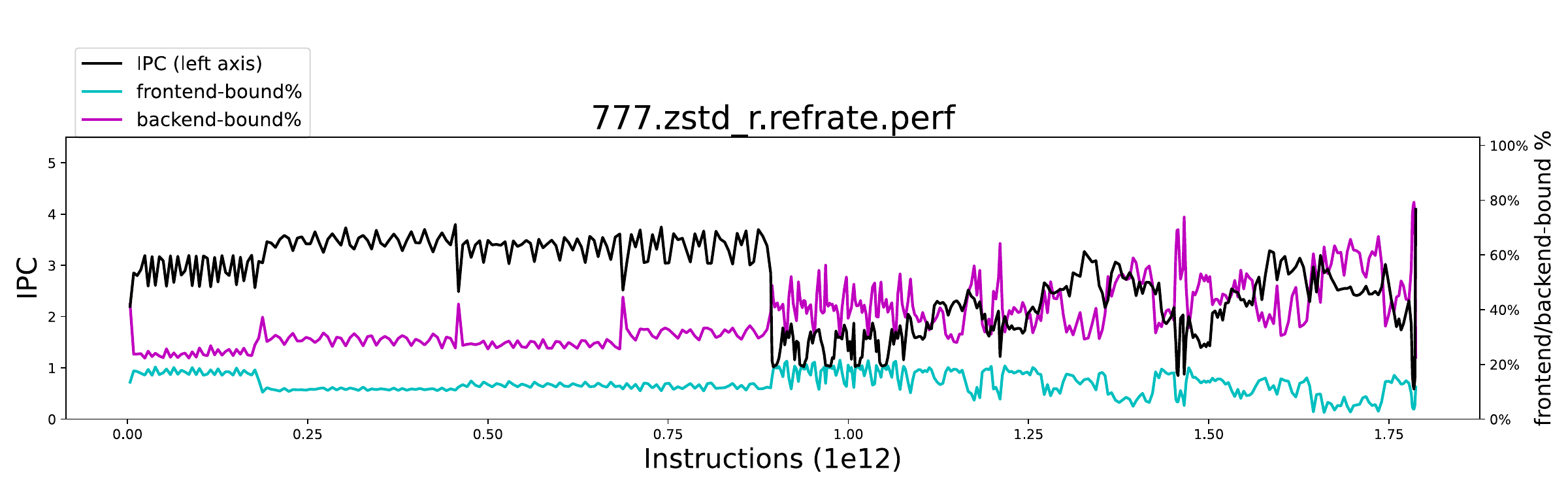}
        \end{subfigure}
        &
        &
        \begin{subfigure}[b]{\linewidth}
            \centering
            \hspace*{-9pt} 
            \includegraphics[width=0.82\linewidth]{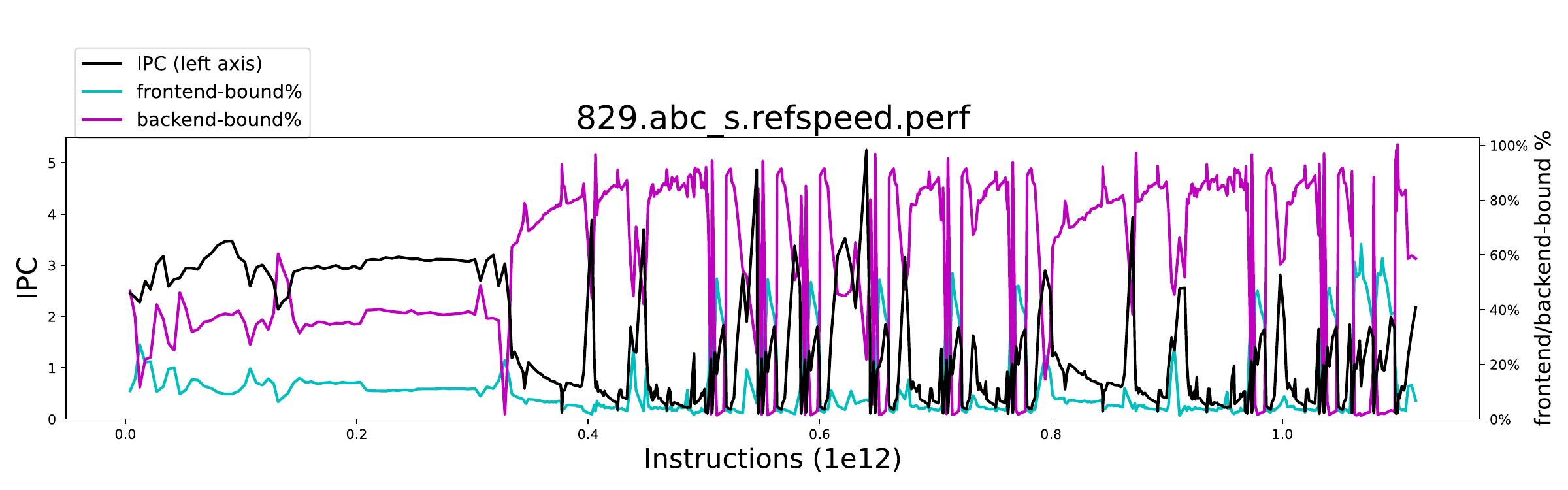}
        \end{subfigure}
        \\[+4pt]

        \begin{subfigure}[b]{\linewidth}
            \includegraphics[width=\linewidth]{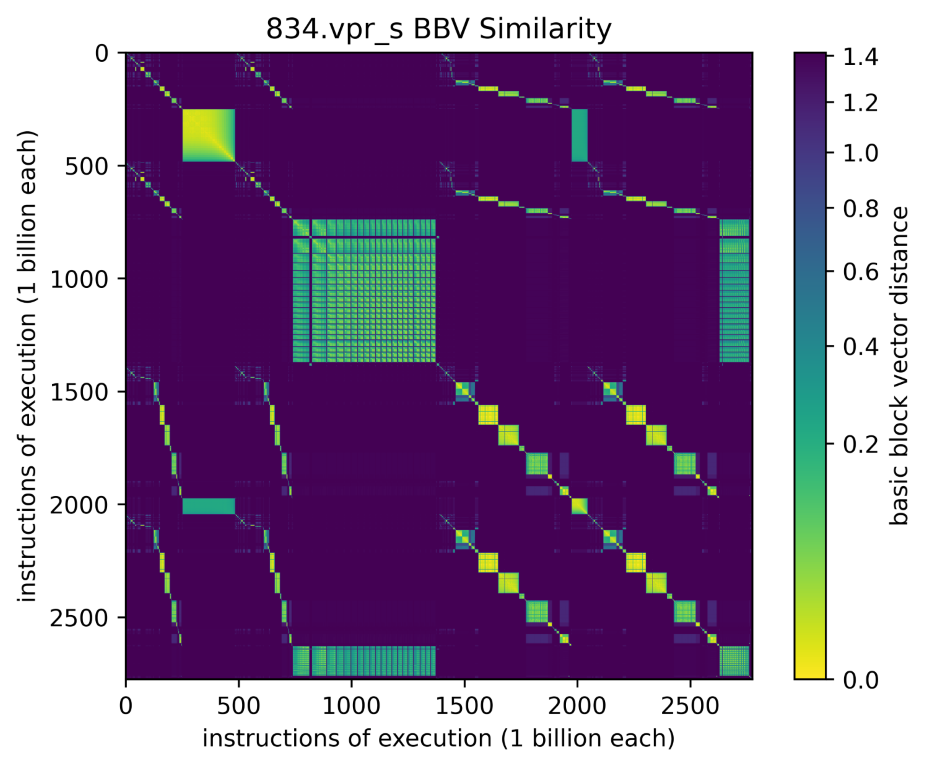}
            \label{fig:bbv_834}
        \end{subfigure}
        &
        \begin{subfigure}[b]{\linewidth}
            \includegraphics[width=\linewidth]{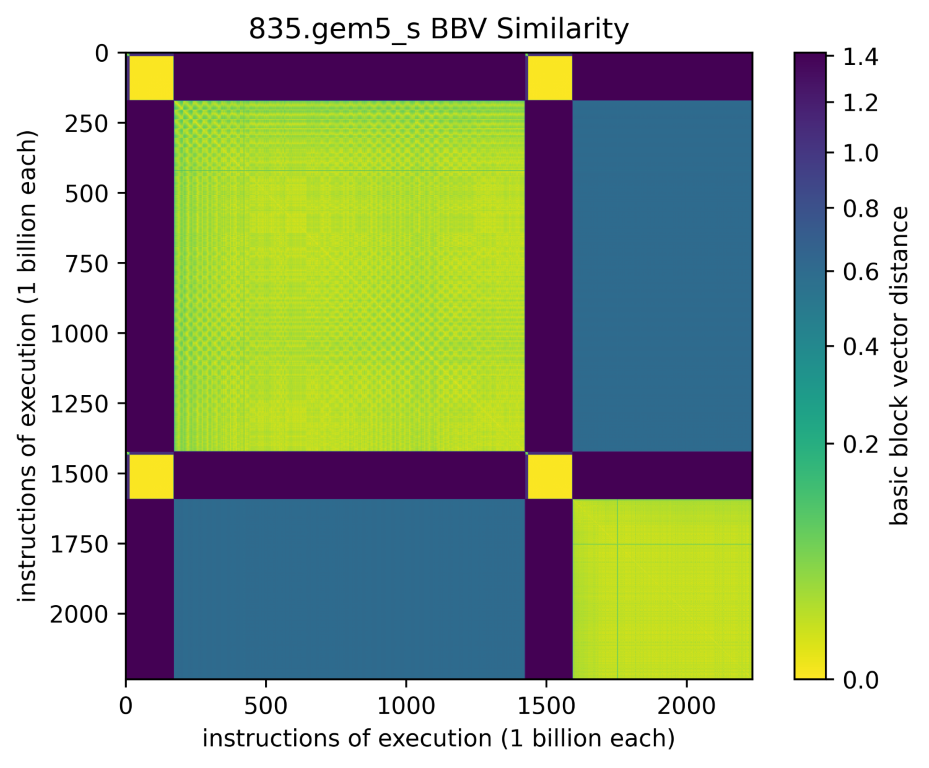}
            \label{fig:bbv_835}
        \end{subfigure}
        &
        \begin{subfigure}[b]{\linewidth}
            \includegraphics[width=\linewidth]{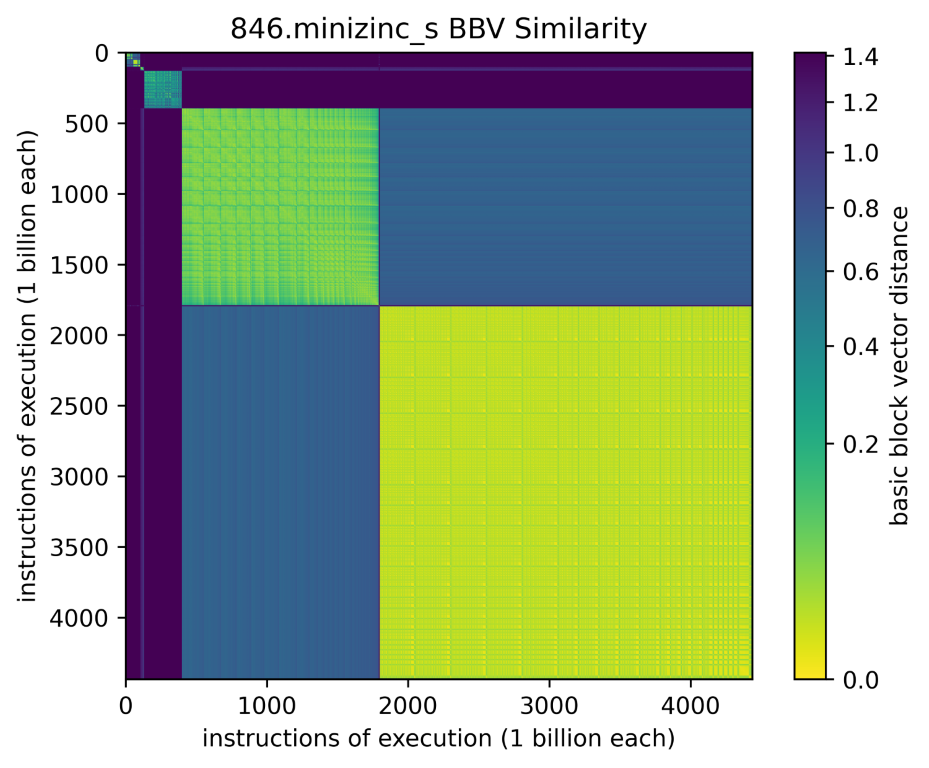}
            \label{fig:bbv_846}
        \end{subfigure}
        &
        \begin{subfigure}[b]{\linewidth}
            \includegraphics[width=\linewidth]{figure/853.ns3_s.png}
            \label{fig:bbv_853}
        \end{subfigure}
        \\[-26pt] 

        \begin{subfigure}[b]{\linewidth}
            \centering
            \hspace*{-9pt} 
            \includegraphics[width=0.82\linewidth]{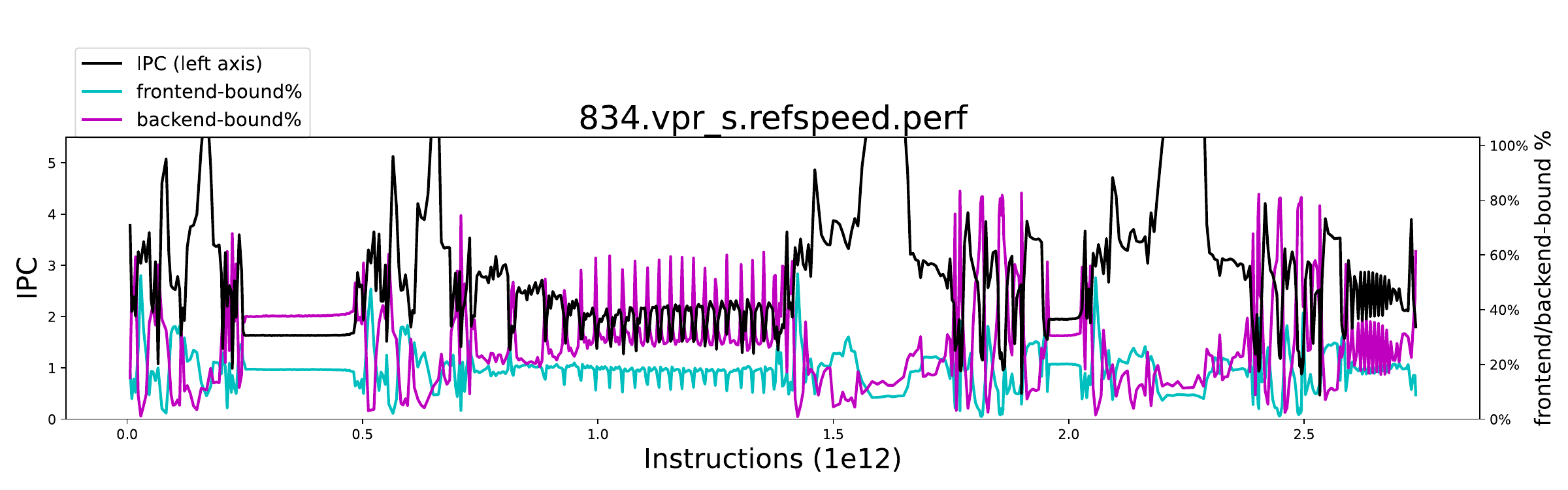}
        \end{subfigure}
        &
        \begin{subfigure}[b]{\linewidth}
            \centering
            \hspace*{-9pt} 
            \includegraphics[width=0.82\linewidth]{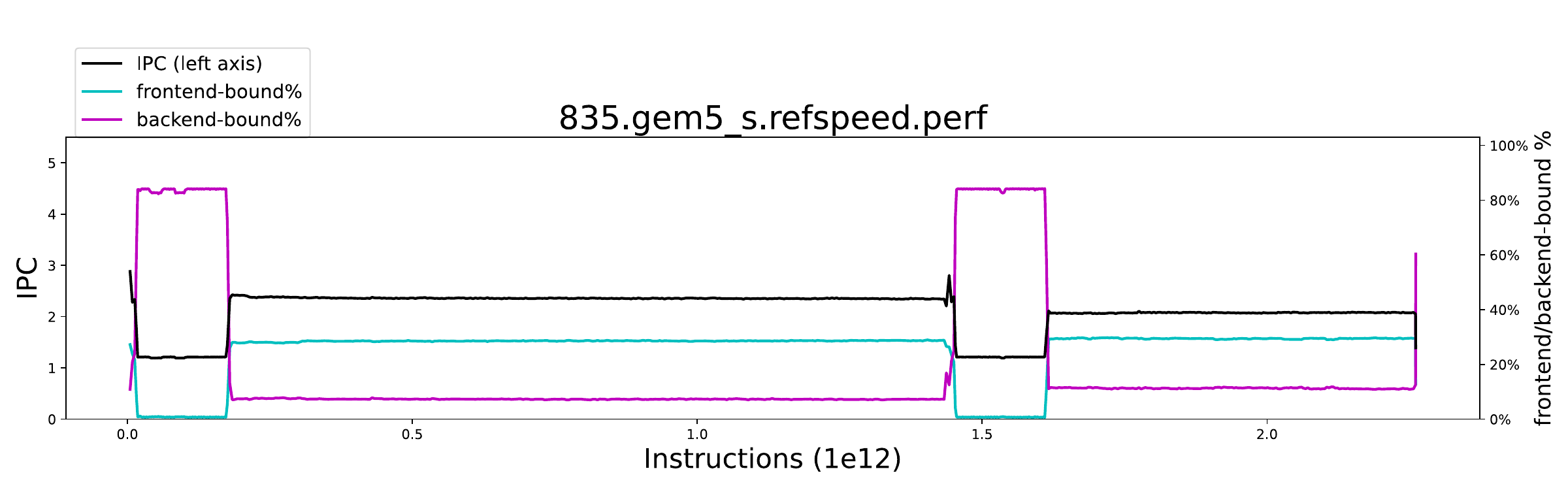}
        \end{subfigure}
        &
        \begin{subfigure}[b]{\linewidth}
            \centering
            \hspace*{-9pt} 
            \includegraphics[width=0.82\linewidth]{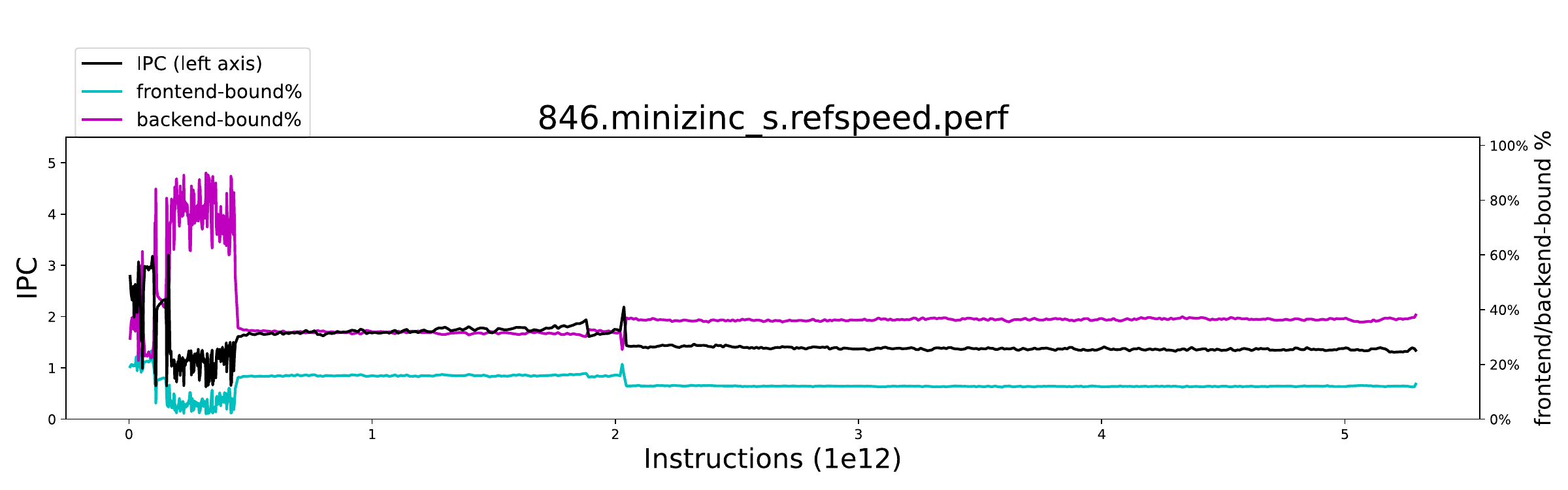}
        \end{subfigure}
        &
        \begin{subfigure}[b]{\linewidth}
            \centering
            \hspace*{-9pt} 
            \includegraphics[width=0.82\linewidth]{figure/853.ns3_s.refspeed.perf.pdf}
                       
        \end{subfigure}
        \\[+4pt]
        
    \end{tabularx} 
    
    \caption{BBV Recurrence and Perf Plots: Integer Rate, and single-threaded Integer Speed}
    \label{fig:bbv_int_rate}
\end{figure*}

%% file: 99.2-bbv_fp_perf_plot.tex
\newcolumntype{C}{>{\centering\arraybackslash}X} 

\begin{figure*}[t]
    \centering 

    \begin{tabularx}{\textwidth}{@{} *{3}{C} @{}}

        \begin{subfigure}[b]{\linewidth}
            \includegraphics[width=\linewidth]{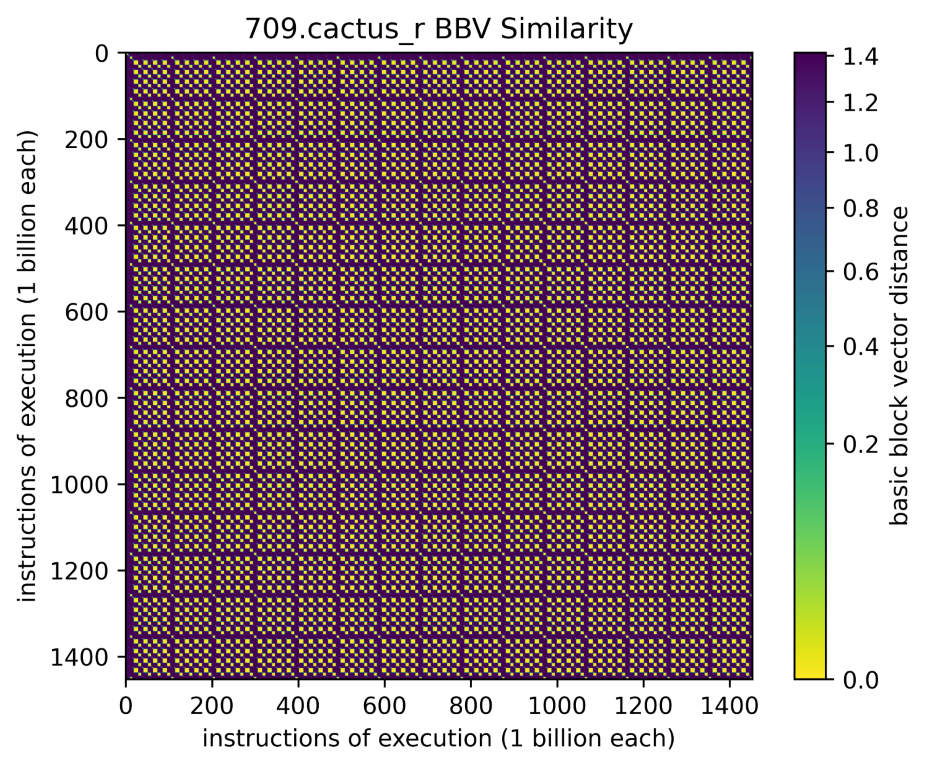}
            \label{fig:bbv_709}
        \end{subfigure}
        &
        \begin{subfigure}[b]{\linewidth}
            \includegraphics[width=\linewidth]{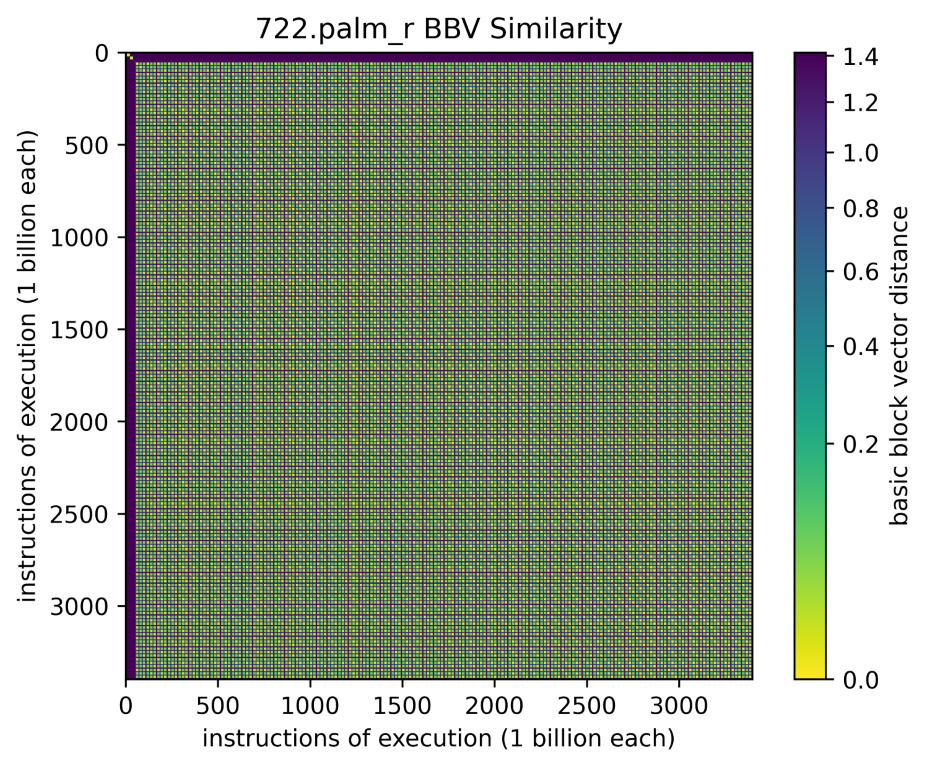}
            \label{fig:bbv_722}
        \end{subfigure}
        &
        \begin{subfigure}[b]{\linewidth}
            \includegraphics[width=\linewidth]{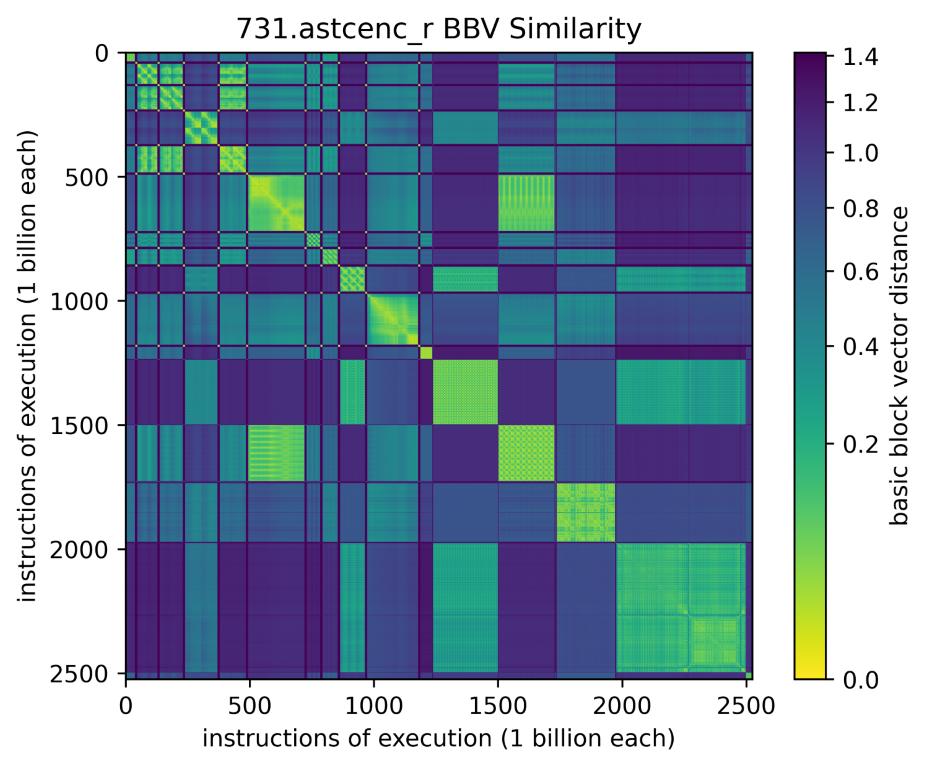}
            \label{fig:bbv_731}
        \end{subfigure}
        \\[-30pt]

        \begin{subfigure}[b]{\linewidth}
            \centering
            \hspace*{-11pt} 
            \includegraphics[width=0.82\linewidth]{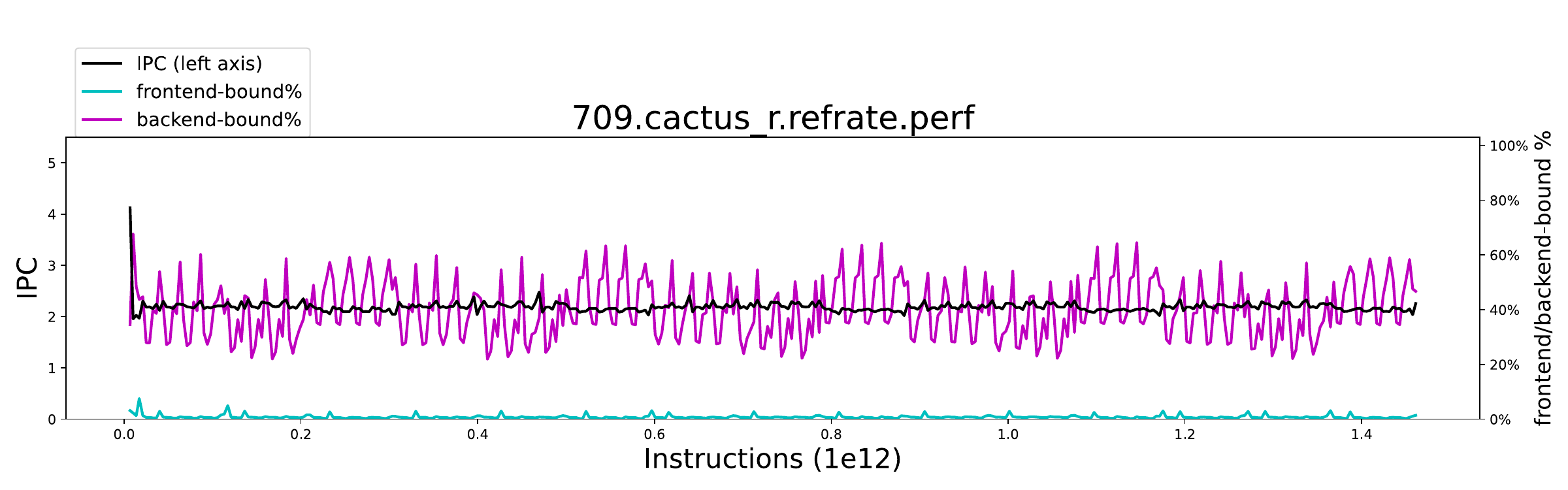}
        \end{subfigure}
        &
        \begin{subfigure}[b]{\linewidth}
            \centering
            \hspace*{-11pt} 
            \includegraphics[width=0.82\linewidth]{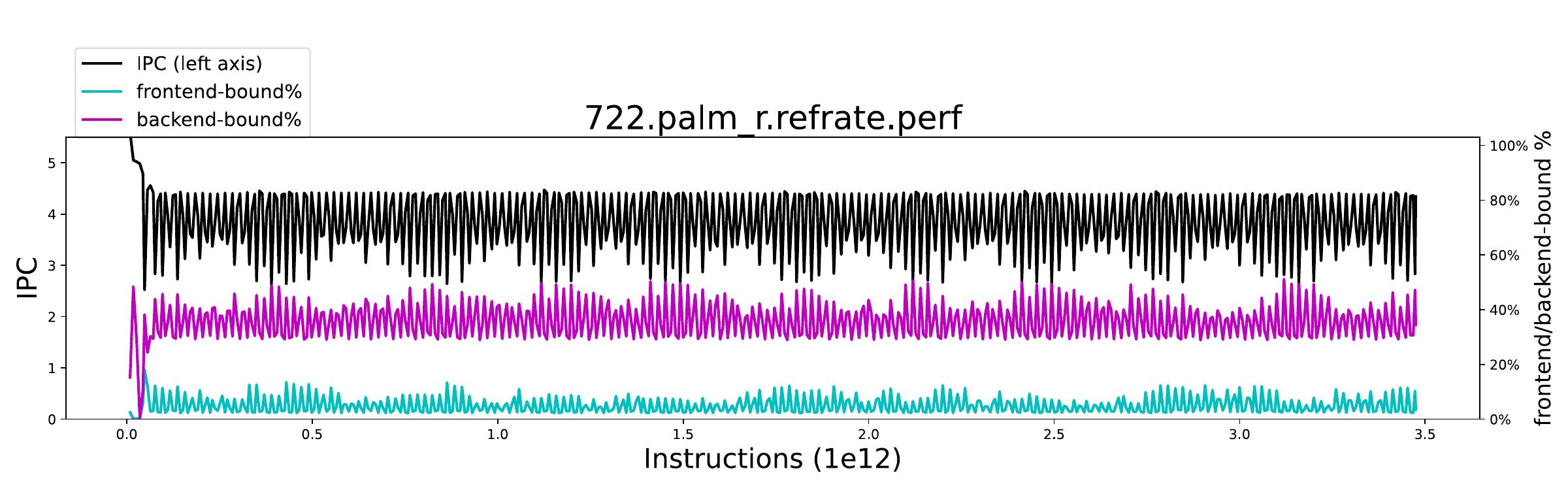}
        \end{subfigure}
        &
        \begin{subfigure}[b]{\linewidth}
            \centering
            \hspace*{-11pt} 
            \includegraphics[width=0.82\linewidth]{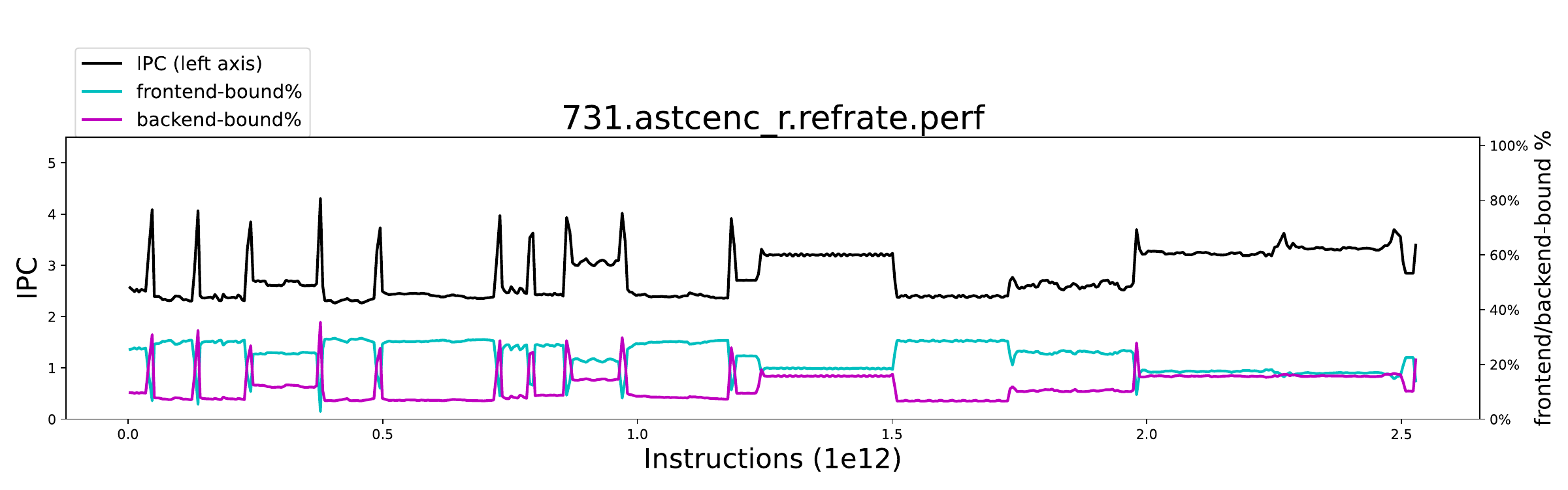}
            
        \end{subfigure}
        \\[+2pt]

        \begin{subfigure}[b]{\linewidth}
            \includegraphics[width=\linewidth]{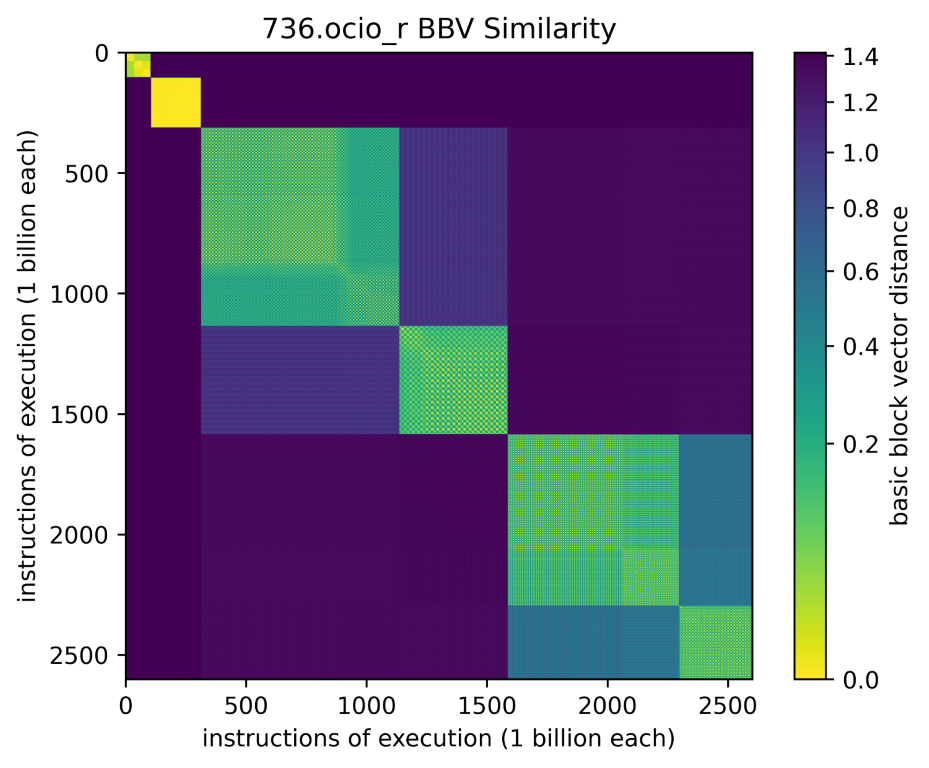}
            \label{fig:bbv_736}
        \end{subfigure}
        &
        \begin{subfigure}[b]{\linewidth}
            \includegraphics[width=\linewidth]{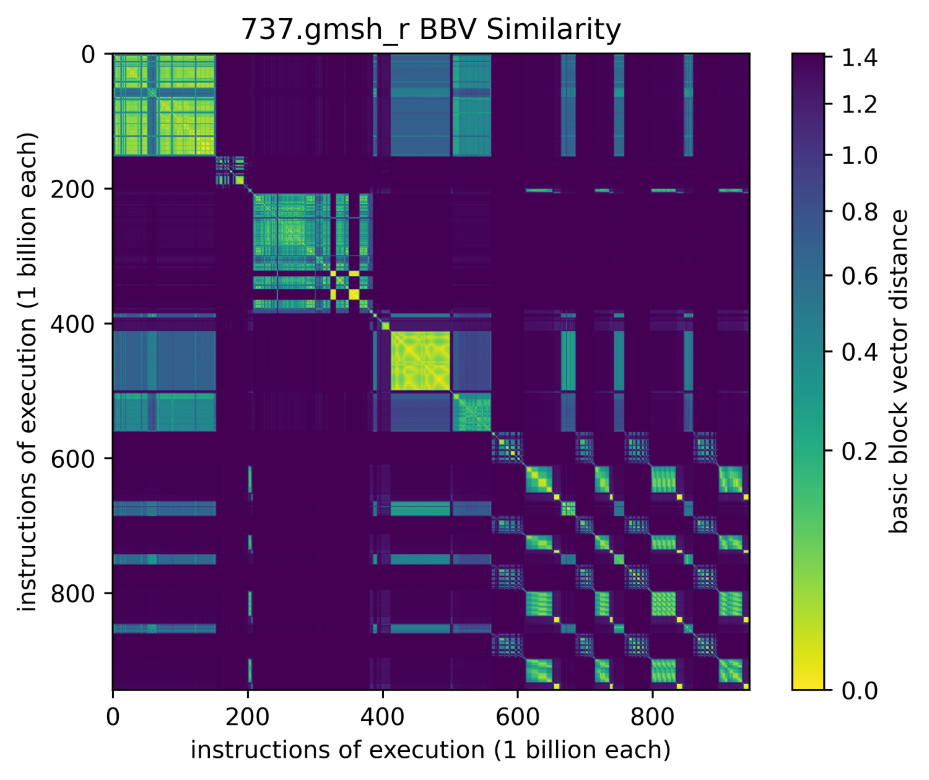}
            \label{fig:bbv_737}
        \end{subfigure}
        &
        \begin{subfigure}[b]{\linewidth}
            \includegraphics[width=\linewidth]{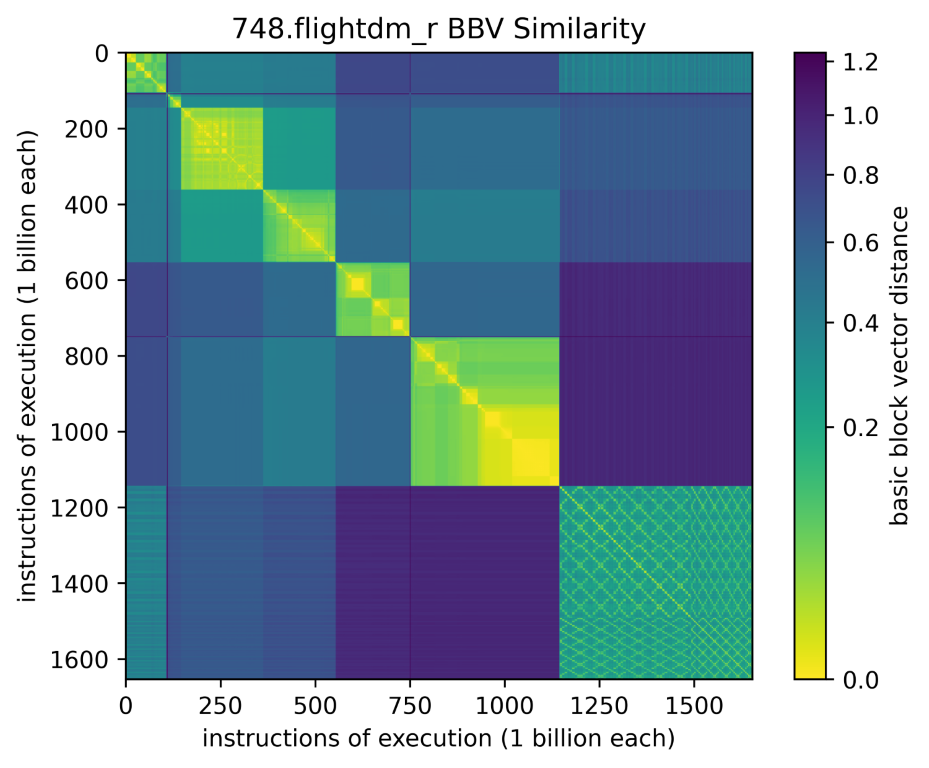}
            \label{fig:bbv_748}
        \end{subfigure}
        \\[-30pt]
        
        \begin{subfigure}[b]{\linewidth}
            \centering
            \hspace*{-11pt} 
            \includegraphics[width=0.82\linewidth]{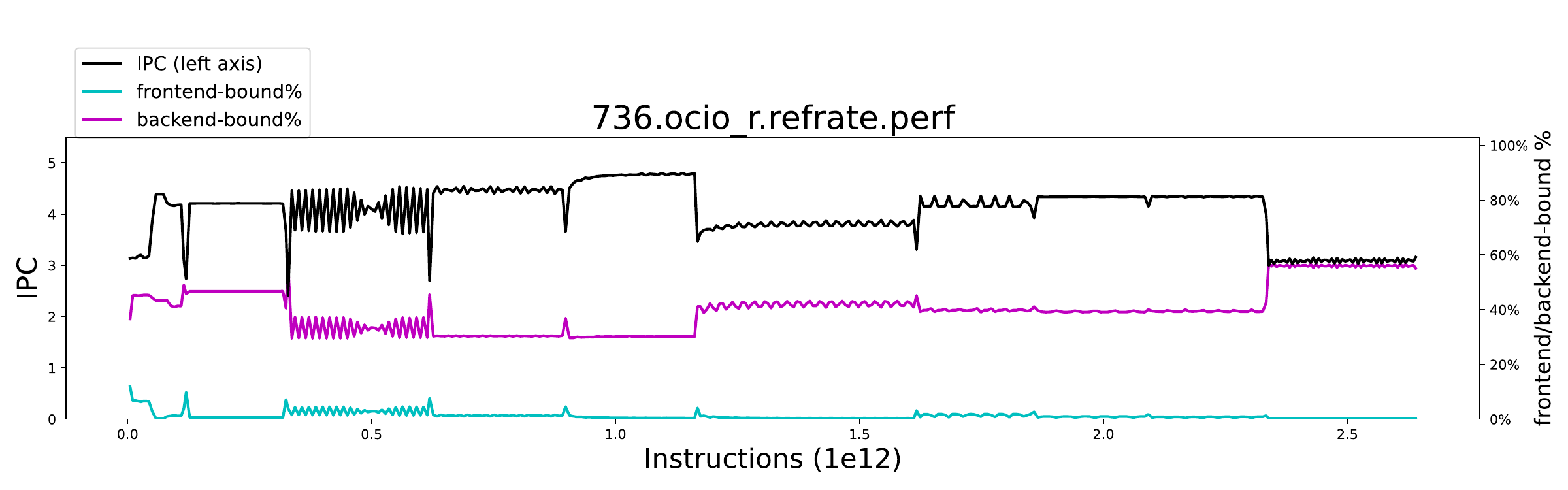}
        \end{subfigure}
        &
        
        \begin{subfigure}[b]{\linewidth}
            \centering
            \hspace*{-12pt} 
            \includegraphics[width=0.83\linewidth]{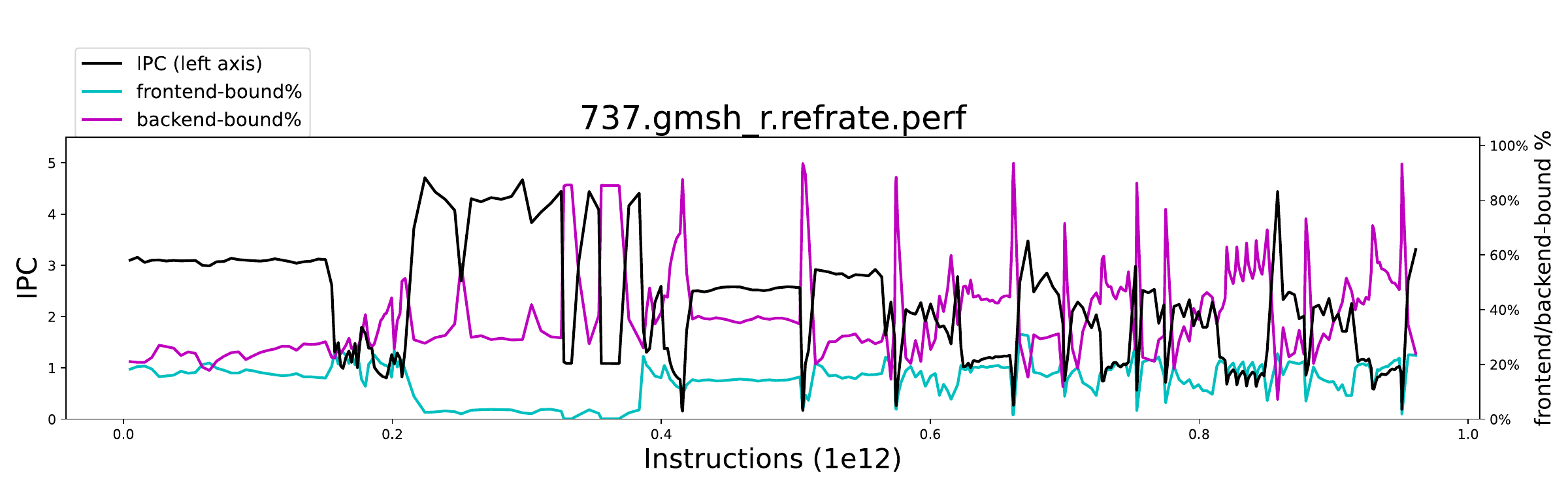}
        \end{subfigure}
        &
        
        \begin{subfigure}[b]{\linewidth}
            \centering
            \hspace*{-11pt} 
            \includegraphics[width=0.82\linewidth]{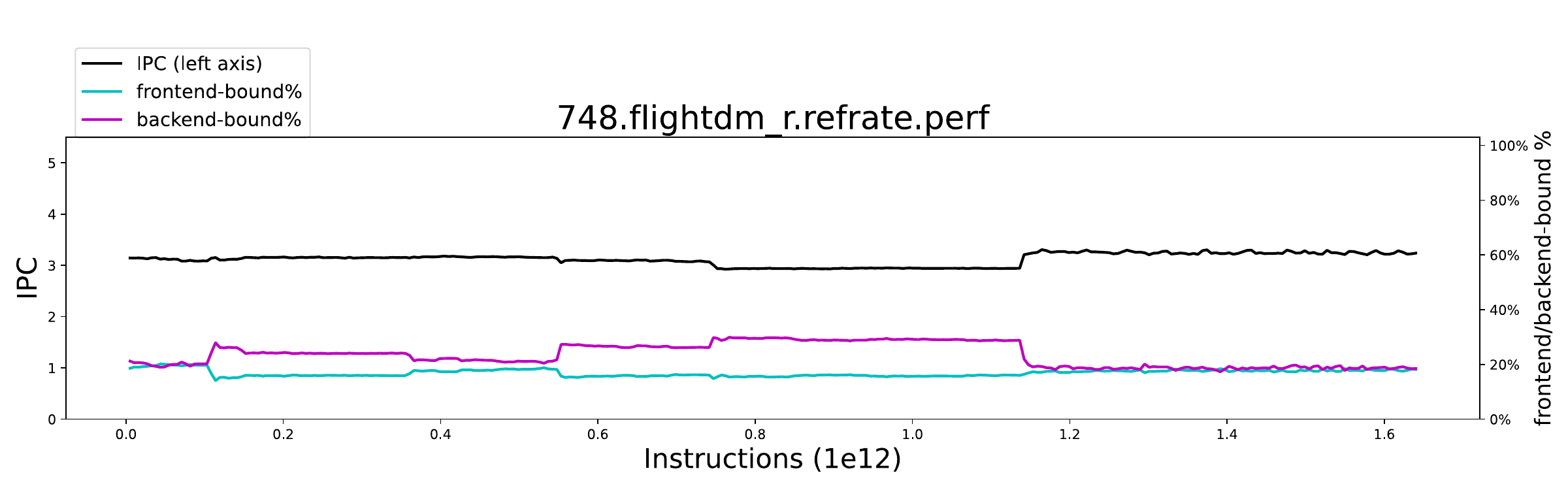}
        \end{subfigure}
        \\[+2pt]

        \begin{subfigure}[b]{\linewidth}
            \includegraphics[width=\linewidth]{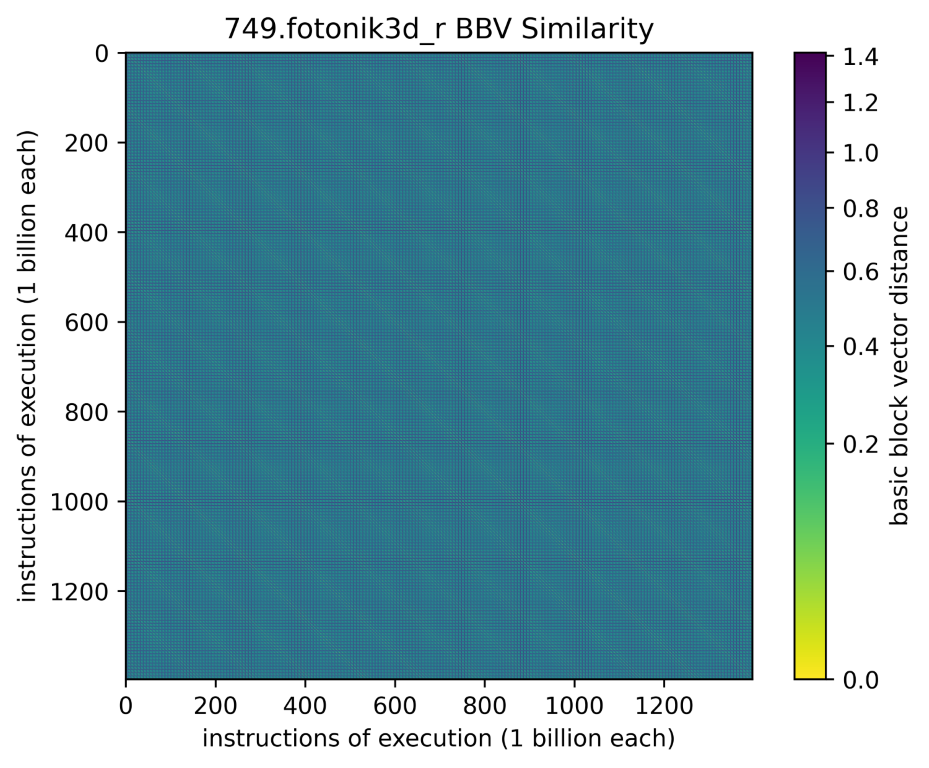}
            \label{fig:bbv_749}
        \end{subfigure}
        &
        \begin{subfigure}[b]{\linewidth}
            \includegraphics[width=\linewidth]{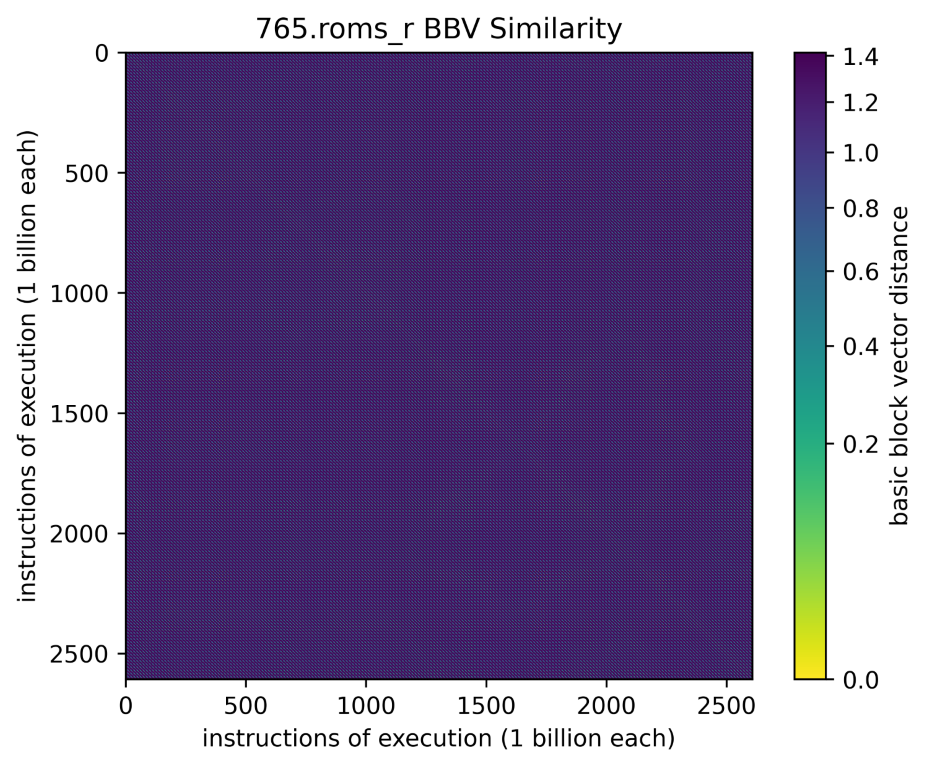}
            \label{fig:bbv_765}
        \end{subfigure}
        &
        \begin{subfigure}[b]{\linewidth}
            \includegraphics[width=\linewidth]{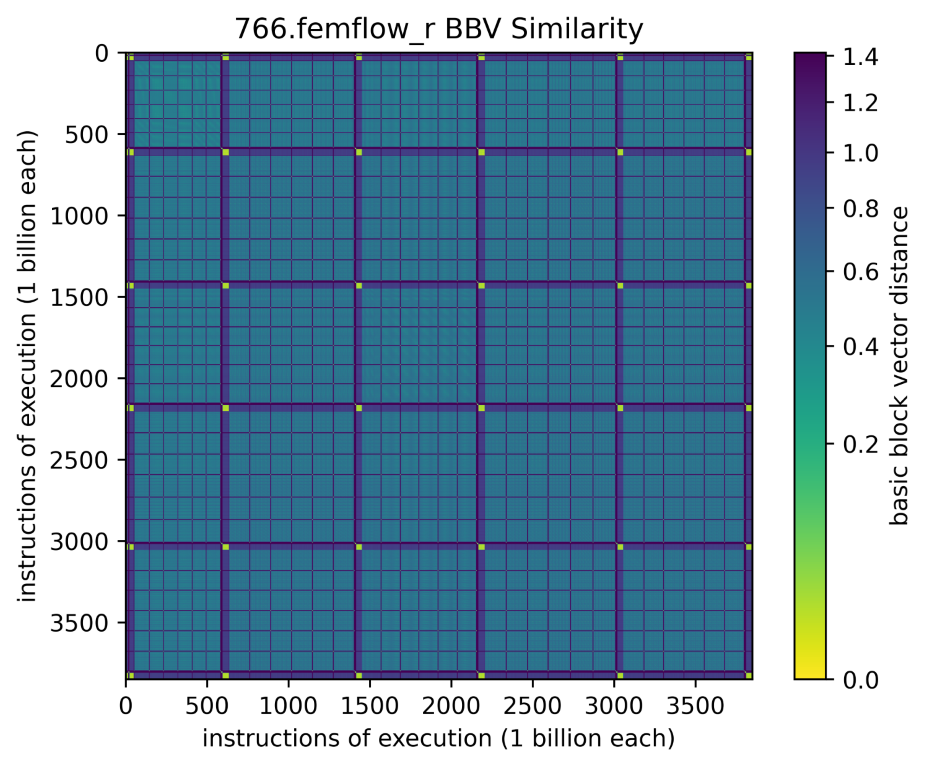}
            \label{fig:bbv_766}
        \end{subfigure}
        \\[-30pt]
        
        \begin{subfigure}[b]{\linewidth}
            \centering
            \hspace*{-11pt} 
            \includegraphics[width=0.82\linewidth]{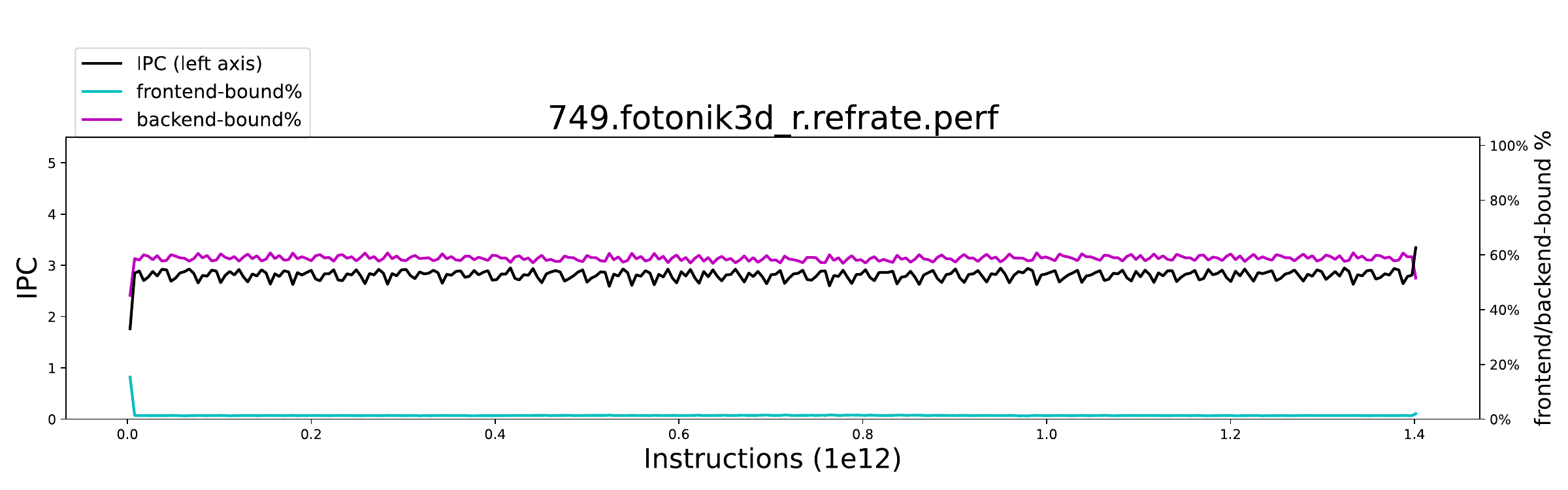}
        \end{subfigure}
        &
        \begin{subfigure}[b]{\linewidth}
            \centering
            \hspace*{-11pt} 
            \includegraphics[width=0.82\linewidth]{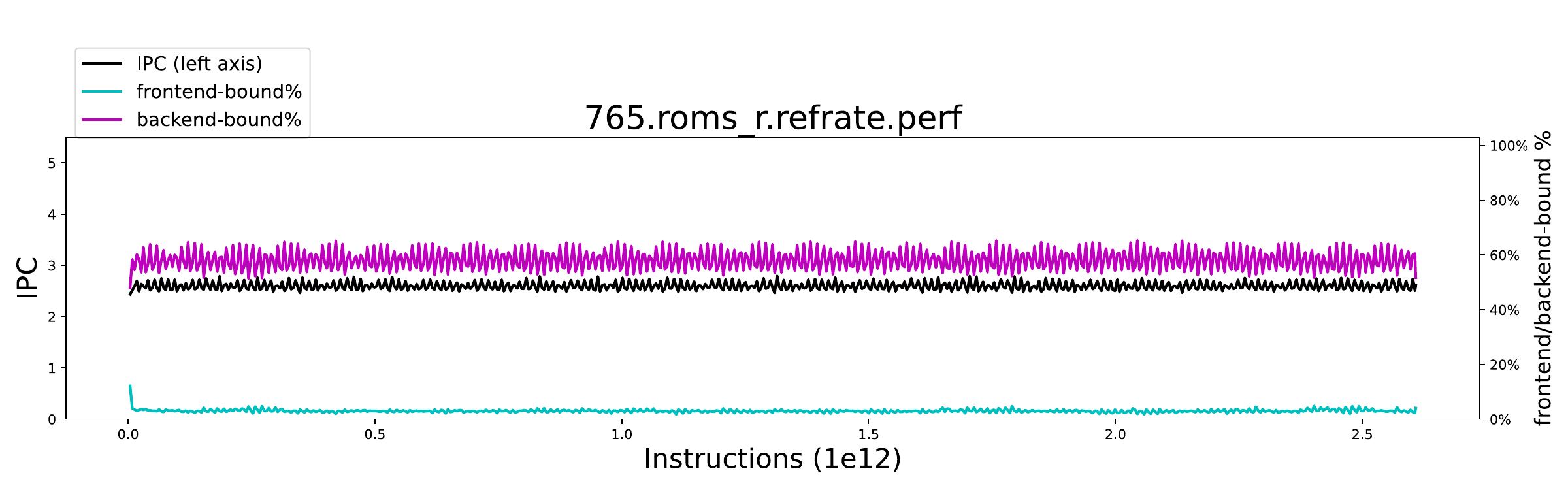}
        \end{subfigure}
        &
        \begin{subfigure}[b]{\linewidth}
            \centering
            \hspace*{-11pt} 
            \includegraphics[width=0.82\linewidth]{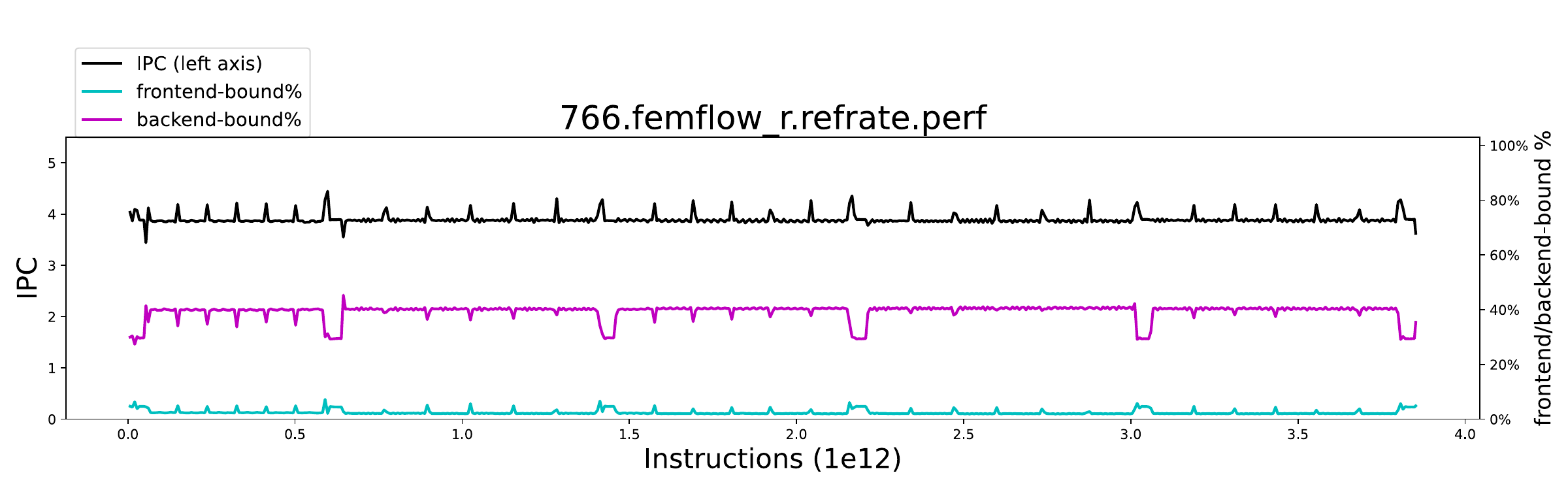}
        \end{subfigure}
        \\[+2pt]

        \begin{subfigure}[b]{\linewidth}
            \includegraphics[width=\linewidth]{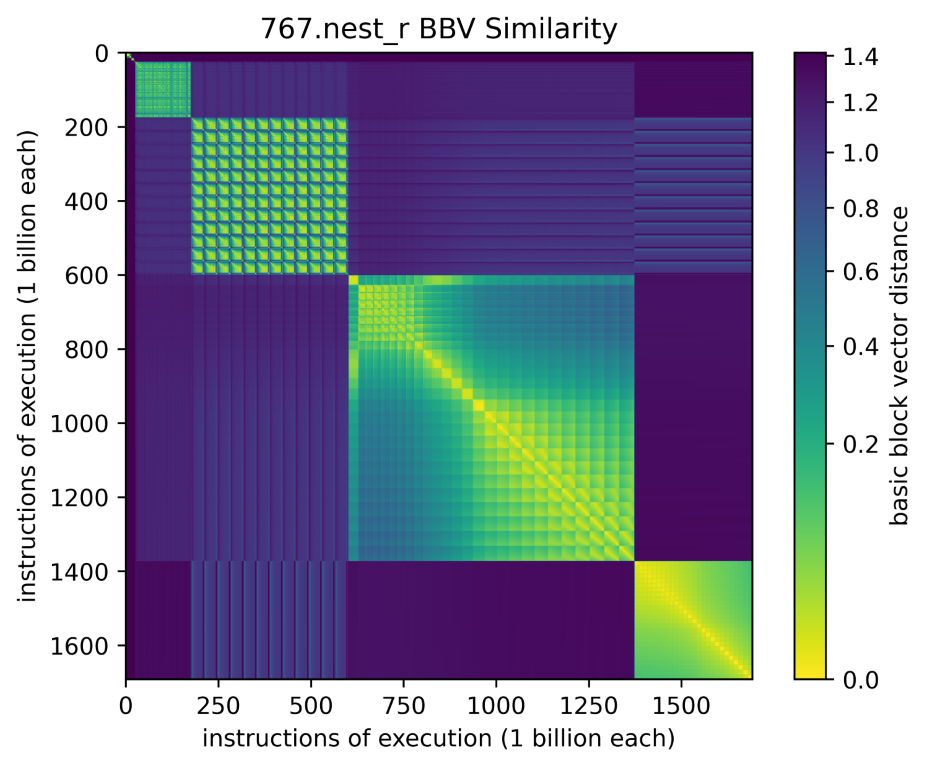}
            \label{fig:bbv_767}
        \end{subfigure}
        &
        \begin{subfigure}[b]{\linewidth}
            \includegraphics[width=\linewidth]{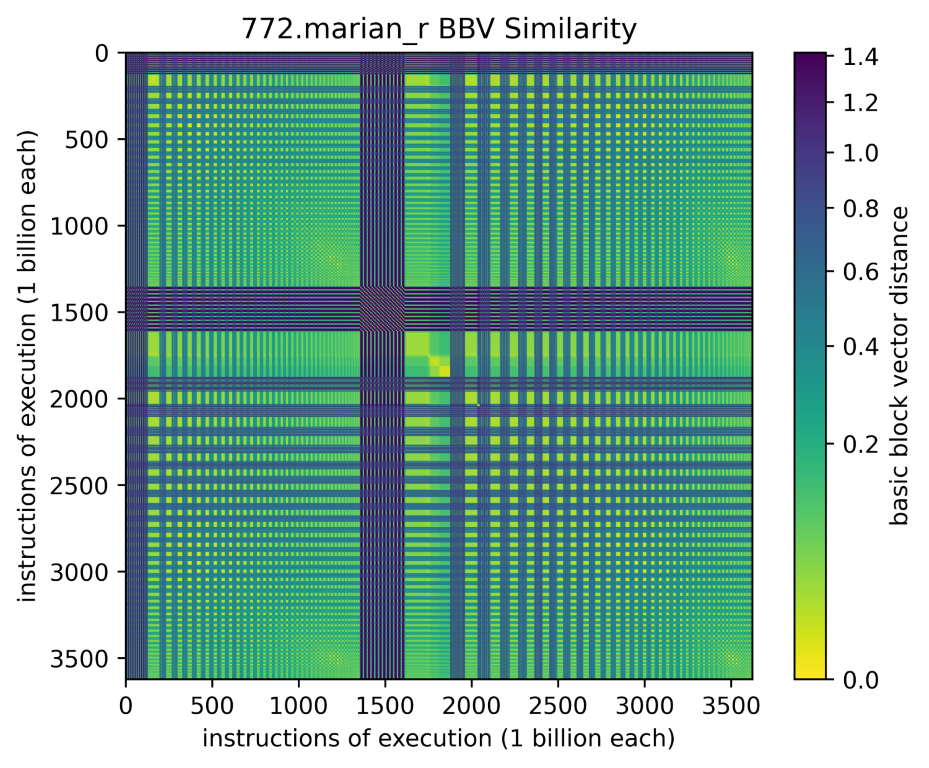}
            \label{fig:bbv_772}
        \end{subfigure}
        &
        \begin{subfigure}[b]{\linewidth}
            \includegraphics[width=\linewidth]{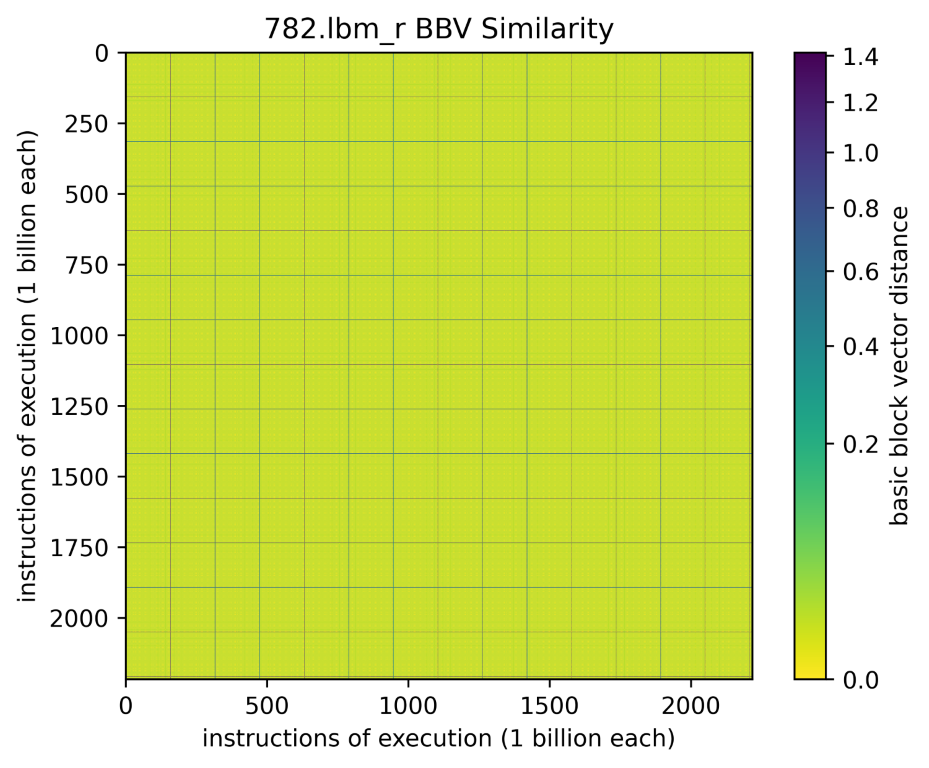}
            \label{fig:bbv_782}
        \end{subfigure}
        \\[-30pt]
        
        \begin{subfigure}[b]{\linewidth}
            \centering
            \hspace*{-11pt} 
            \includegraphics[width=0.82\linewidth]{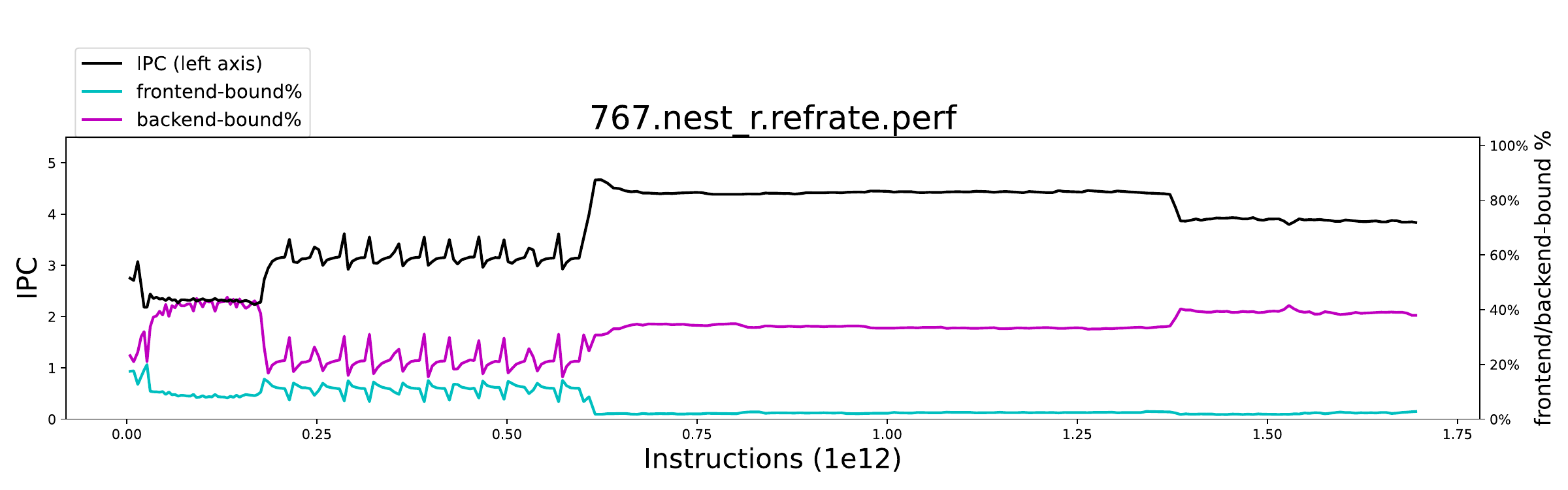}
        \end{subfigure}
        &
        \begin{subfigure}[b]{\linewidth}
            \centering
            \hspace*{-11pt} 
            \includegraphics[width=0.82\linewidth]{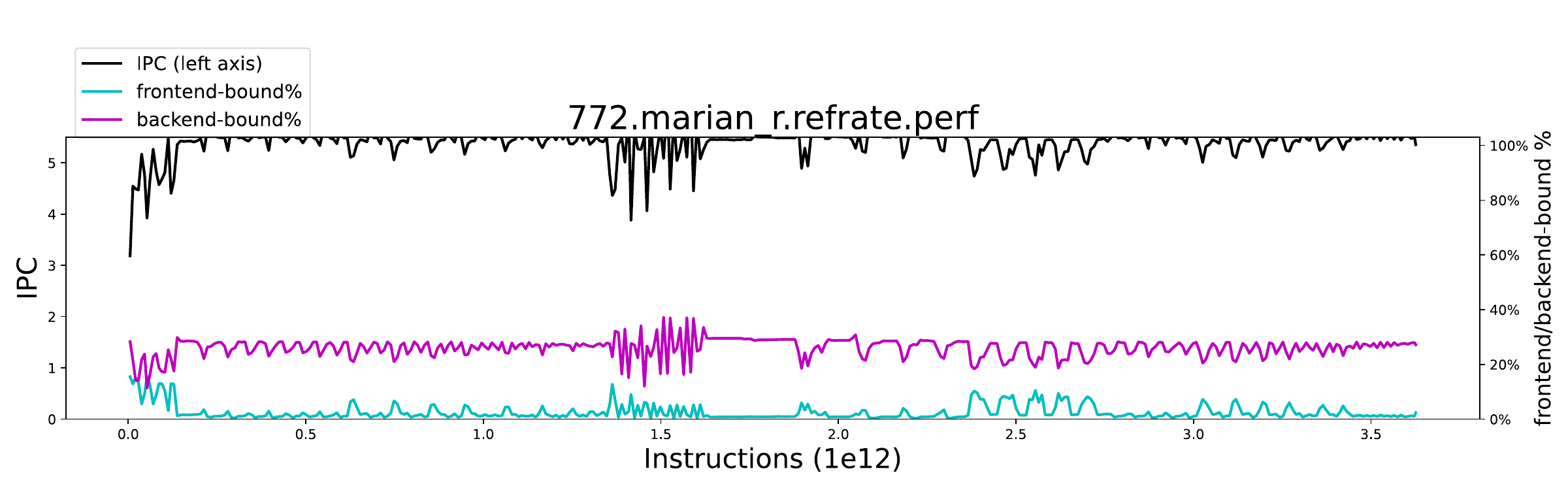}
        \end{subfigure}
        &
        \begin{subfigure}[b]{\linewidth}
            \centering
            \hspace*{-11pt} 
            \includegraphics[width=0.82\linewidth]{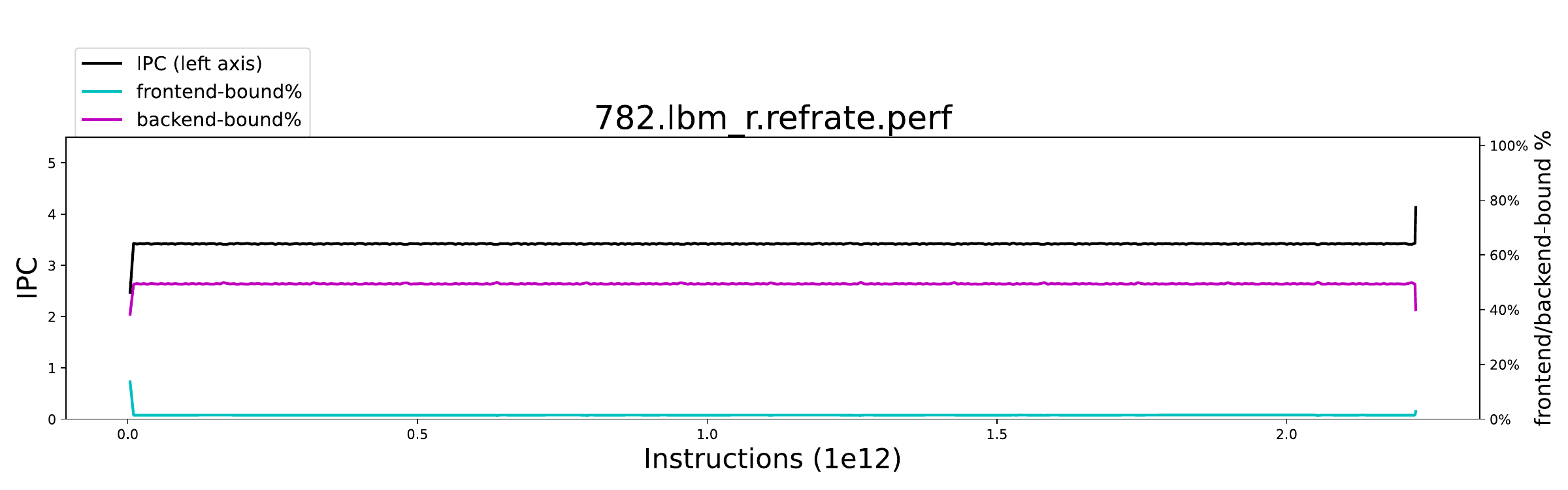}
        \end{subfigure}

    \end{tabularx} 
    
    \caption{BBV Recurrence and Perf Plots: Floating Point Rate}
    \label{fig:bbv_perf_fp}
\end{figure*}

%% file: 99.5-intspeed-2wide.tex
\newcolumntype{C}{>{\centering\arraybackslash}X} 

\begin{figure*}[!ht]
    \centering 

    \begin{tabularx}{\textwidth}{@{} *{2}{C} @{}}

        \begin{subfigure}[b]{\linewidth}
            \centering
            \includegraphics[width=1\linewidth]{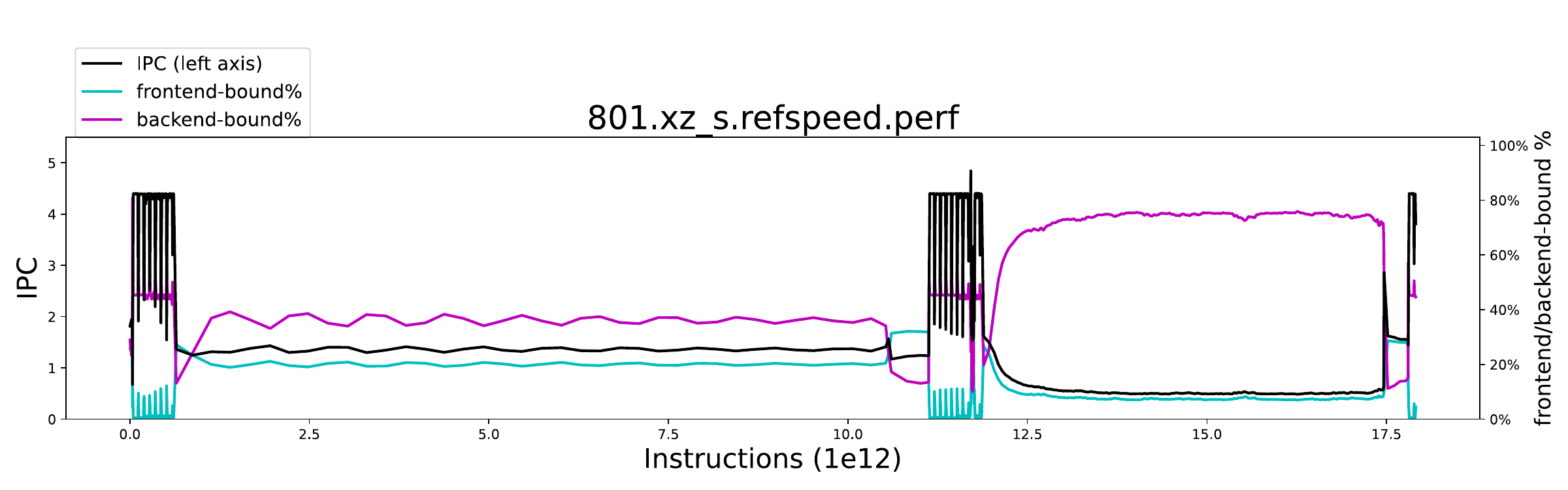}
        \end{subfigure}
        &
        \begin{subfigure}[b]{\linewidth}
            \centering
            \includegraphics[width=1\linewidth]{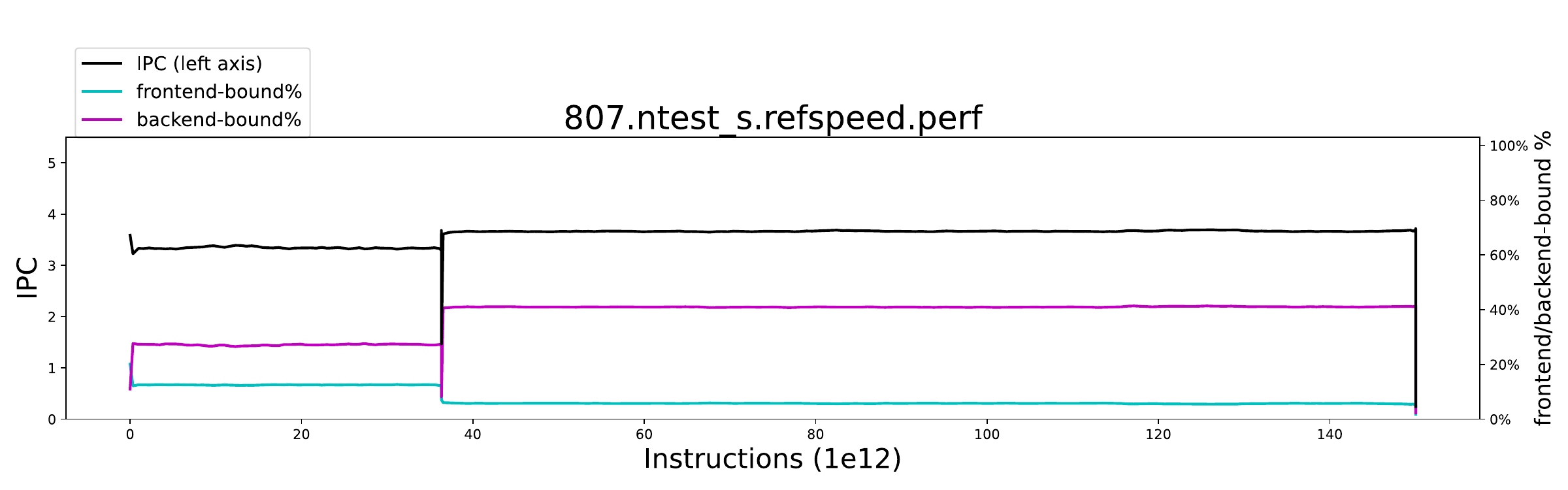}
        \end{subfigure}
        \\[-5pt]
        
        \begin{subfigure}[b]{\linewidth}
            \centering
            \includegraphics[width=1\linewidth]{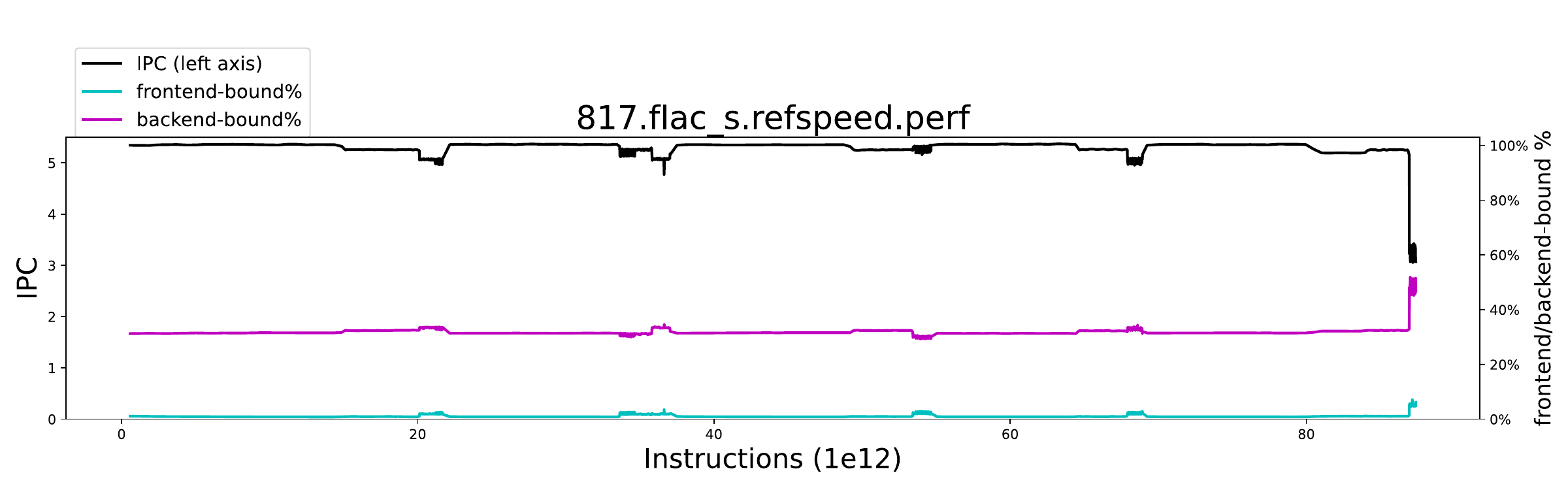}
        \end{subfigure}
        &

        \begin{subfigure}[b]{\linewidth}
            \centering
            \includegraphics[width=1\linewidth]{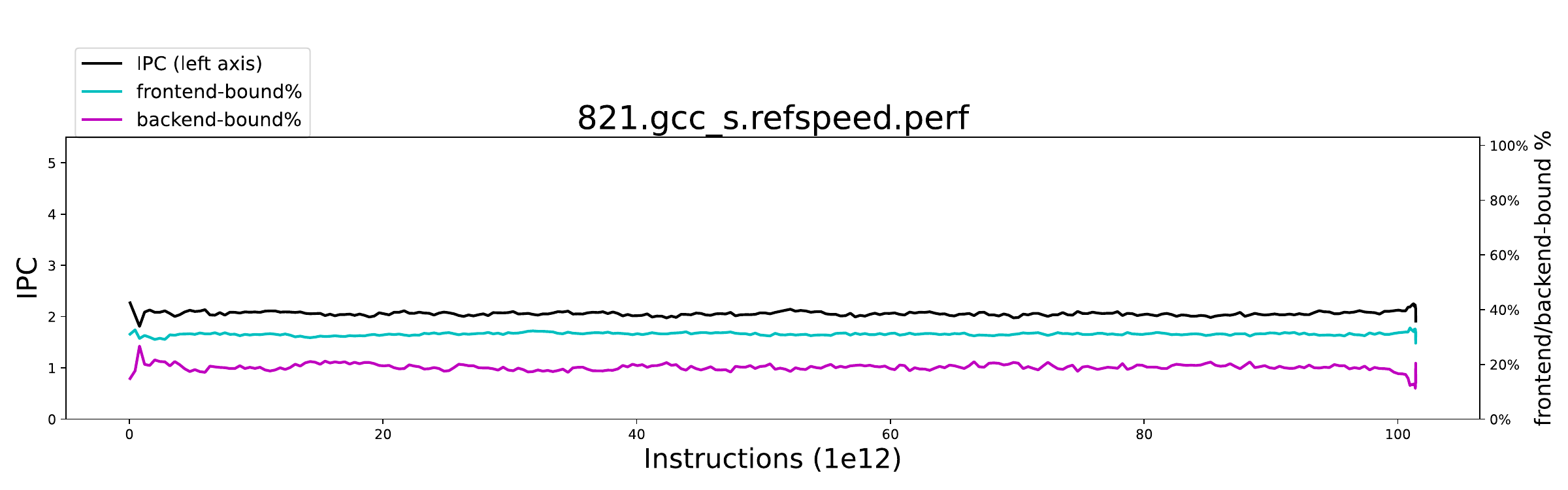}
        \end{subfigure}
        \\[-5pt] 
        \begin{subfigure}[b]{\linewidth}
            \centering
            \includegraphics[width=1\linewidth]{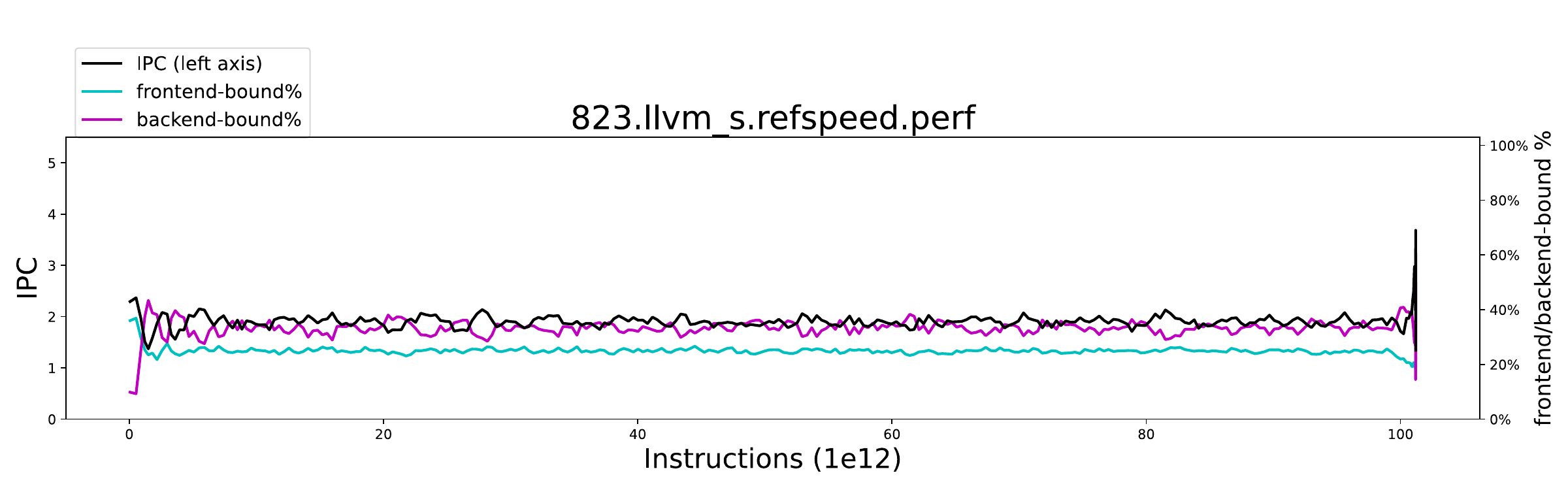}
        \end{subfigure}
        &
        \begin{subfigure}[b]{\linewidth}
            \centering
            \includegraphics[width=1\linewidth]{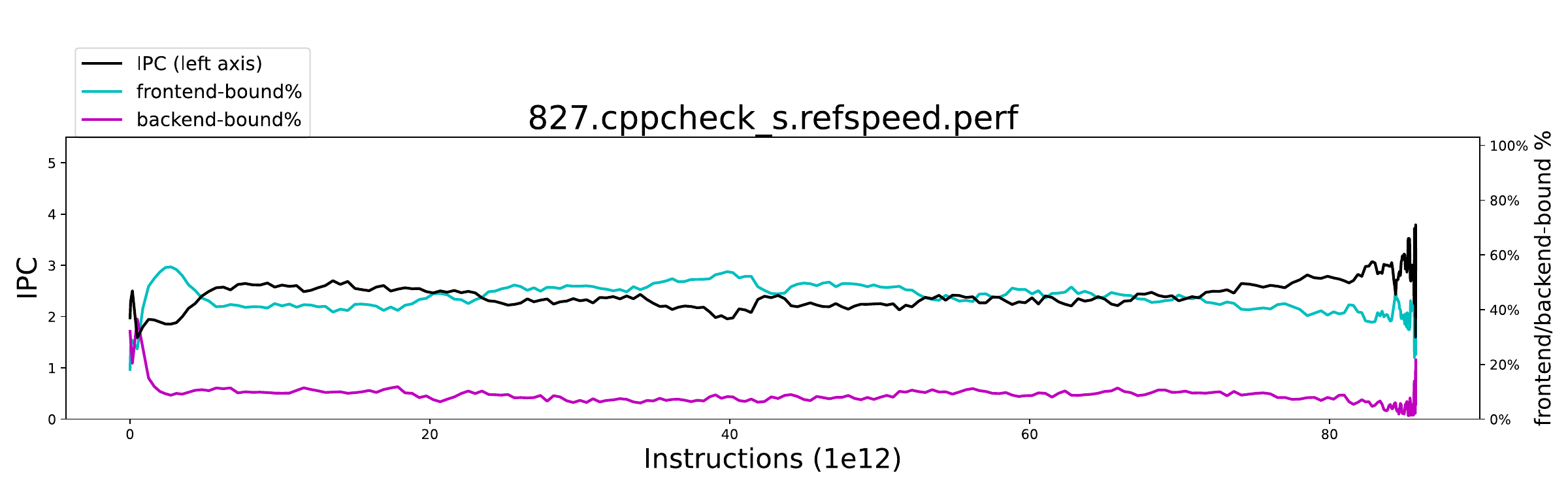}
        \end{subfigure}
        \\[-5pt]

        \begin{subfigure}[b]{\linewidth}
            \centering
            \includegraphics[width=1\linewidth]{figure/829.abc_s.refspeed.perf.pdf}            
        \end{subfigure}
        &
        \begin{subfigure}[b]{\linewidth}
            \centering
            \includegraphics[width=1\linewidth]{figure/834.vpr_s.refspeed.perf.pdf}
        \end{subfigure}
        \\[-5pt] 
        \begin{subfigure}[b]{\linewidth}
            \centering
            \includegraphics[width=1\linewidth]{figure/835.gem5_s.refspeed.perf.pdf}
            \end{subfigure}
        &

        \begin{subfigure}[b]{\linewidth}
            \centering
            \includegraphics[width=1\linewidth]{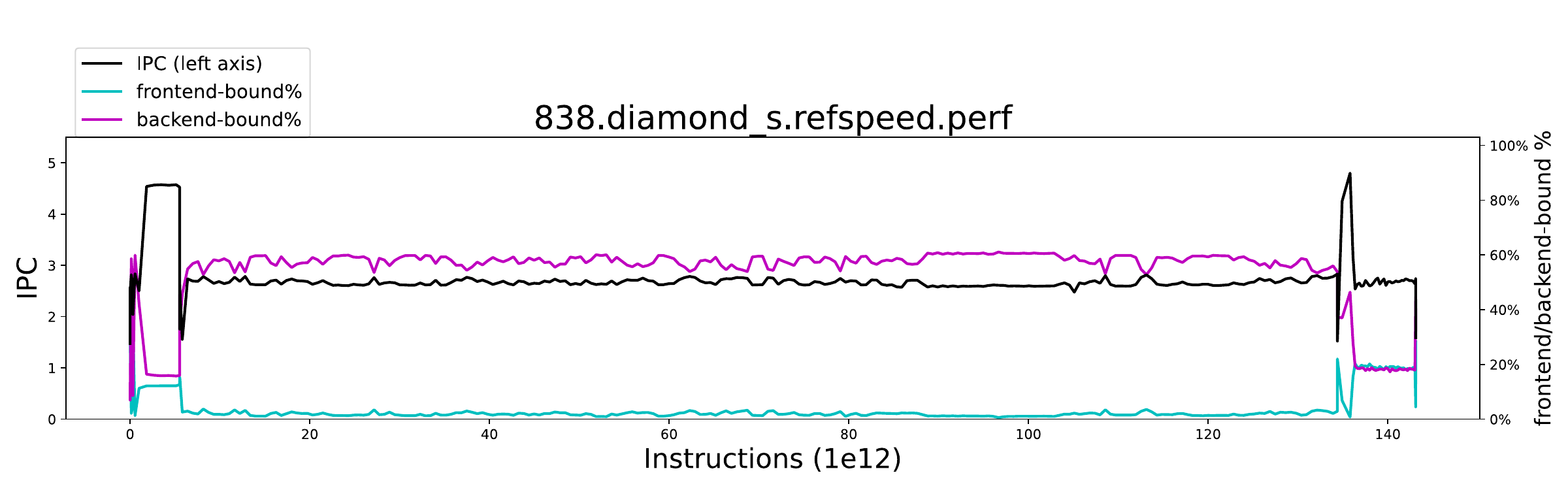}
        \end{subfigure}
        \\[-5pt] 
        \begin{subfigure}[b]{\linewidth}
            \centering
            \includegraphics[width=1\linewidth]{figure/846.minizinc_s.refspeed.perf.pdf}
        \end{subfigure}
        &
        \begin{subfigure}[b]{\linewidth}
            \centering
            \includegraphics[width=1\linewidth]{figure/853.ns3_s.refspeed.perf.pdf}
        \end{subfigure}
        \\[-5pt] 

        \begin{subfigure}[b]{\linewidth}
            \centering
            \includegraphics[width=1\linewidth]{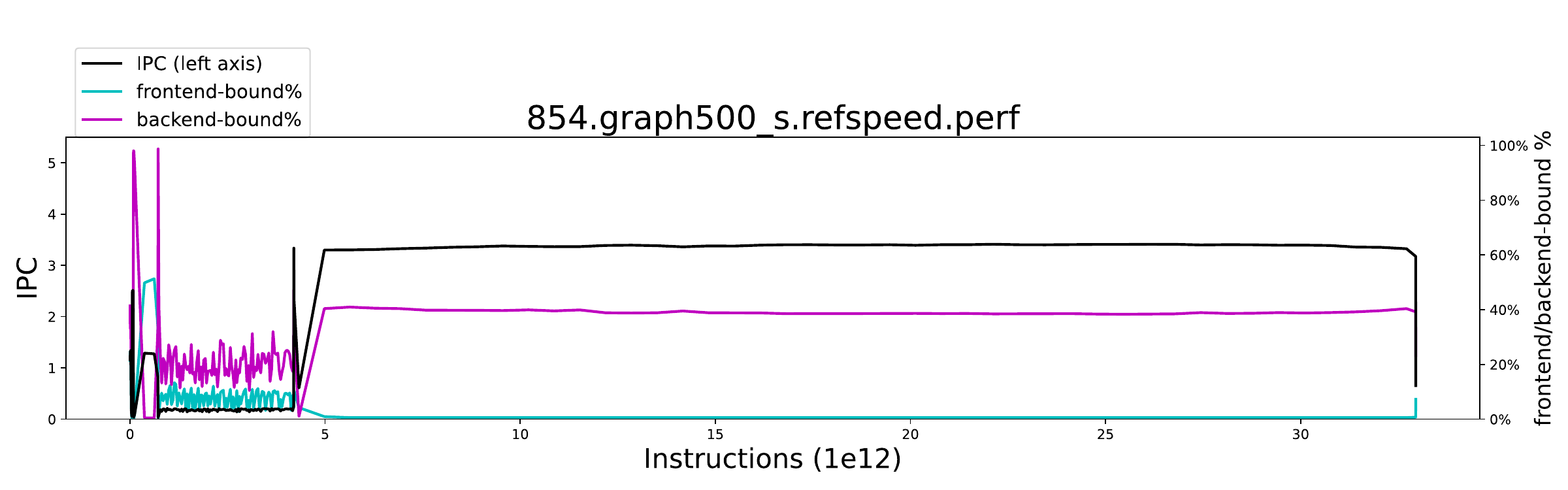}
        \end{subfigure}
        &

    \end{tabularx} 
    
    \caption{Perf Plots: Integer Speed}
    \label{fig:perf_refspeed_int}
\end{figure*}

%% file: 99.6-fpspeed-2wide.tex
\newcolumntype{C}{>{\centering\arraybackslash}X} 

\begin{figure*}[!ht]
    \centering 

    \begin{tabularx}{\textwidth}{@{} *{3}{C} @{}}

        \begin{subfigure}[b]{\linewidth}
            \centering
            \includegraphics[width=1\linewidth]{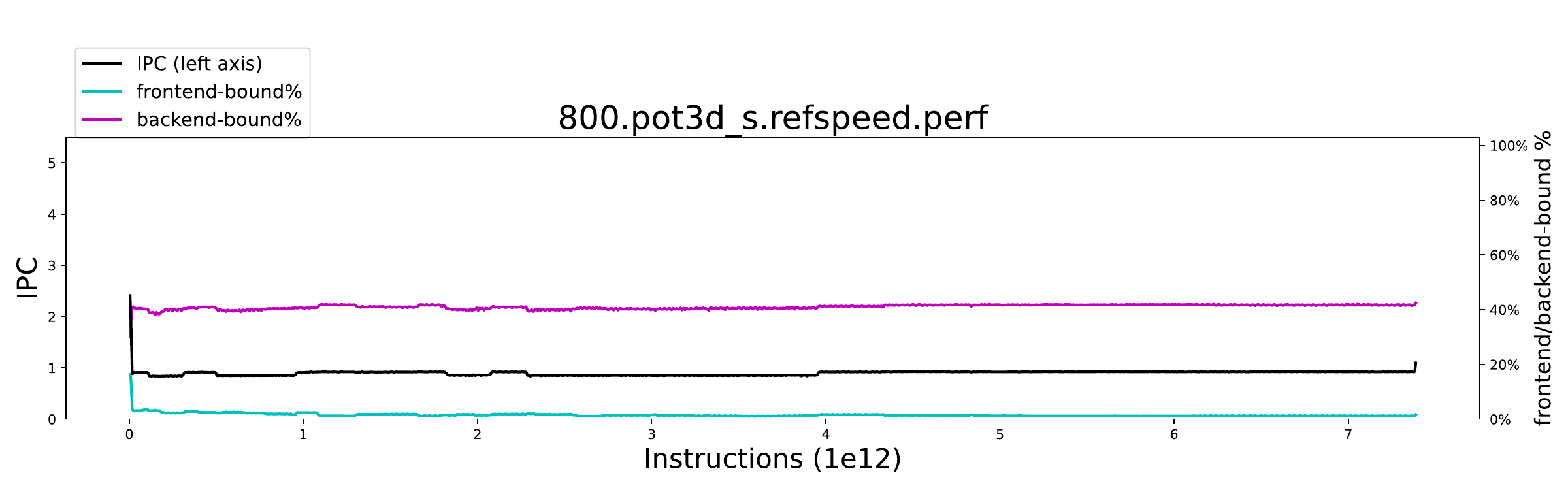}
        \end{subfigure}
        &
        \begin{subfigure}[b]{\linewidth}
            \centering
            \includegraphics[width=1\linewidth]{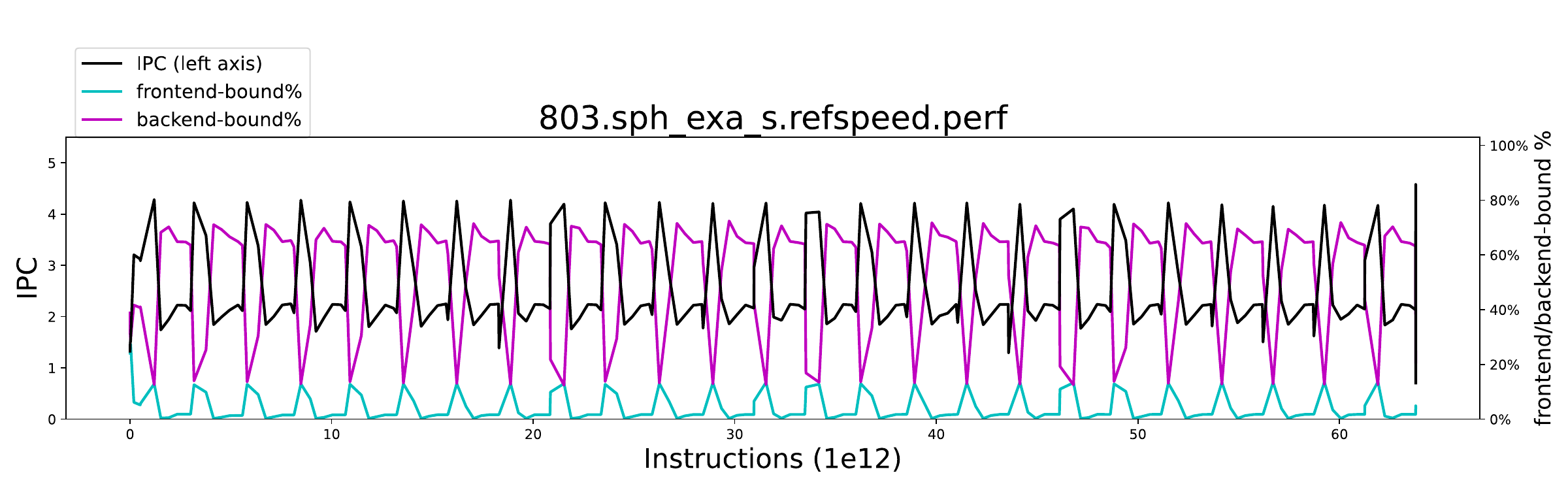}
        \end{subfigure}
        \\[-5pt]
        \begin{subfigure}[b]{\linewidth}
            \centering
            \includegraphics[width=1\linewidth]{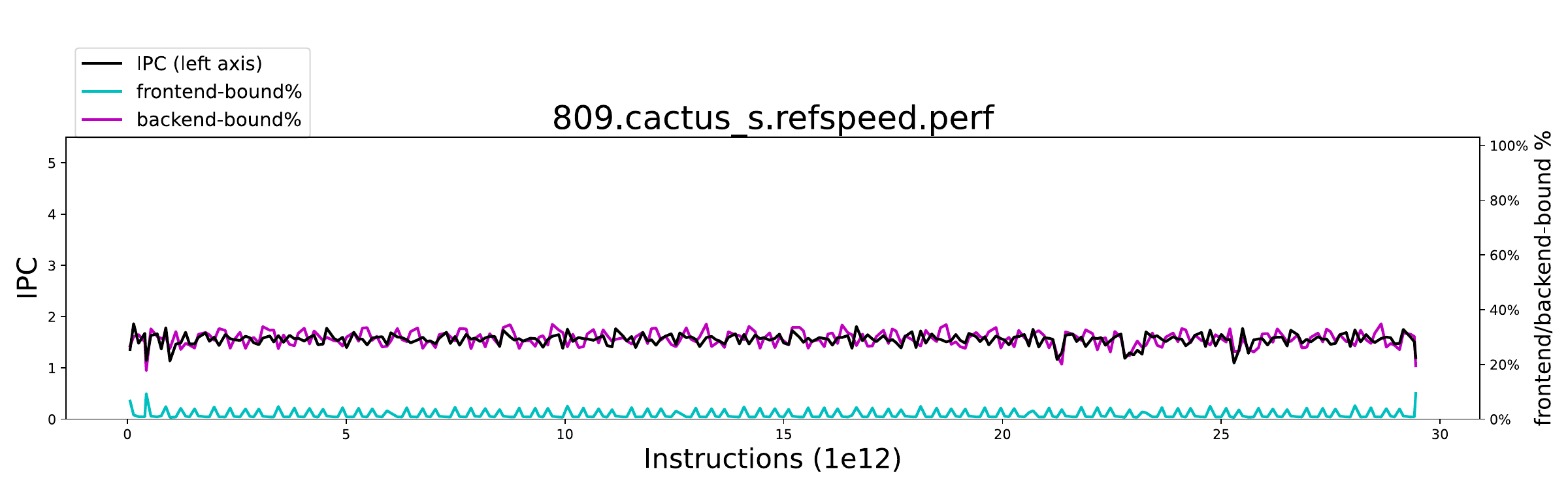}
        \end{subfigure}
        &

        \begin{subfigure}[b]{\linewidth}
            \centering
            \includegraphics[width=1\linewidth]{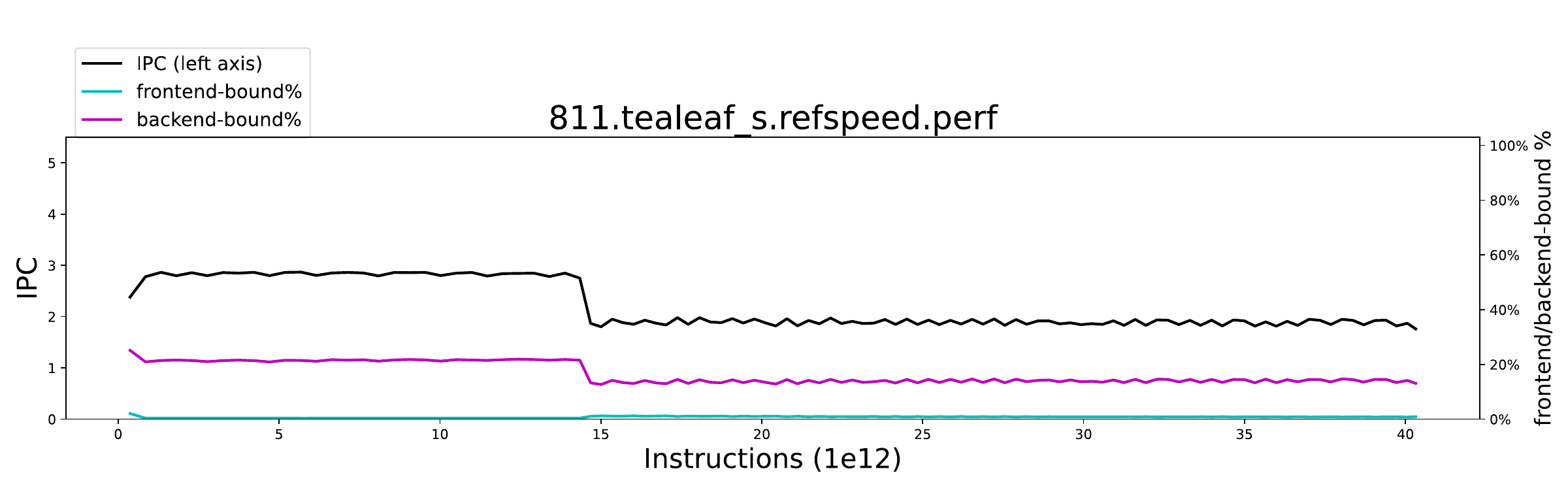}
        \end{subfigure}
        \\[-5pt] 
        \begin{subfigure}[b]{\linewidth}
            \centering
            \includegraphics[width=1\linewidth]{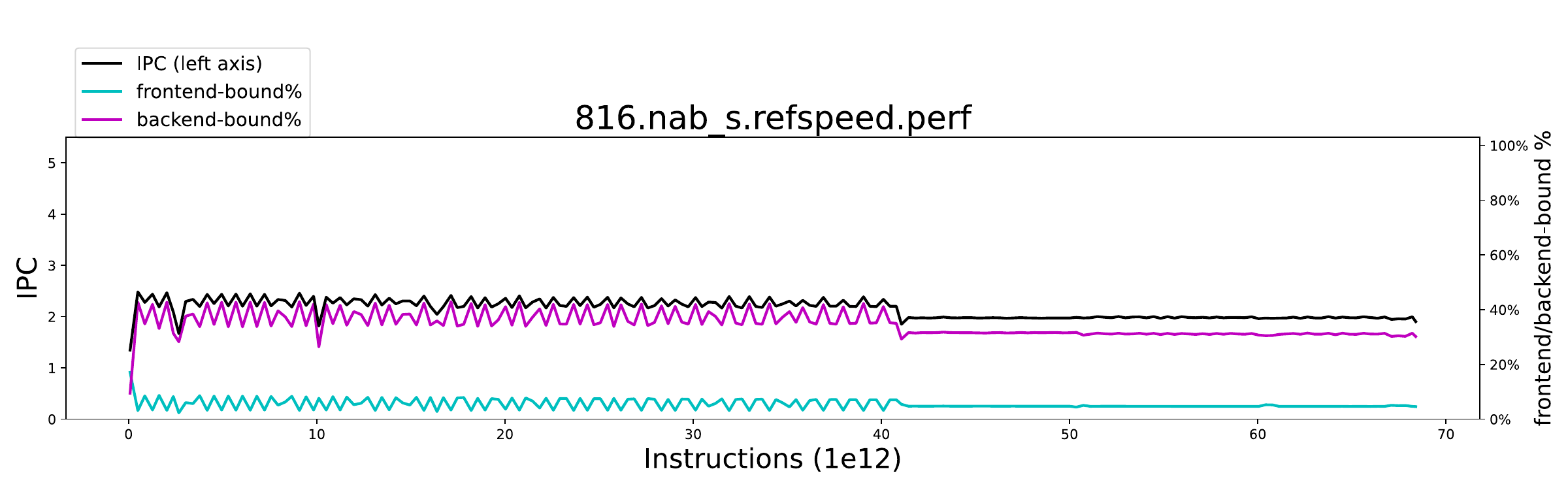}
        \end{subfigure}
        &
        \begin{subfigure}[b]{\linewidth}
            \centering
            \includegraphics[width=1\linewidth]{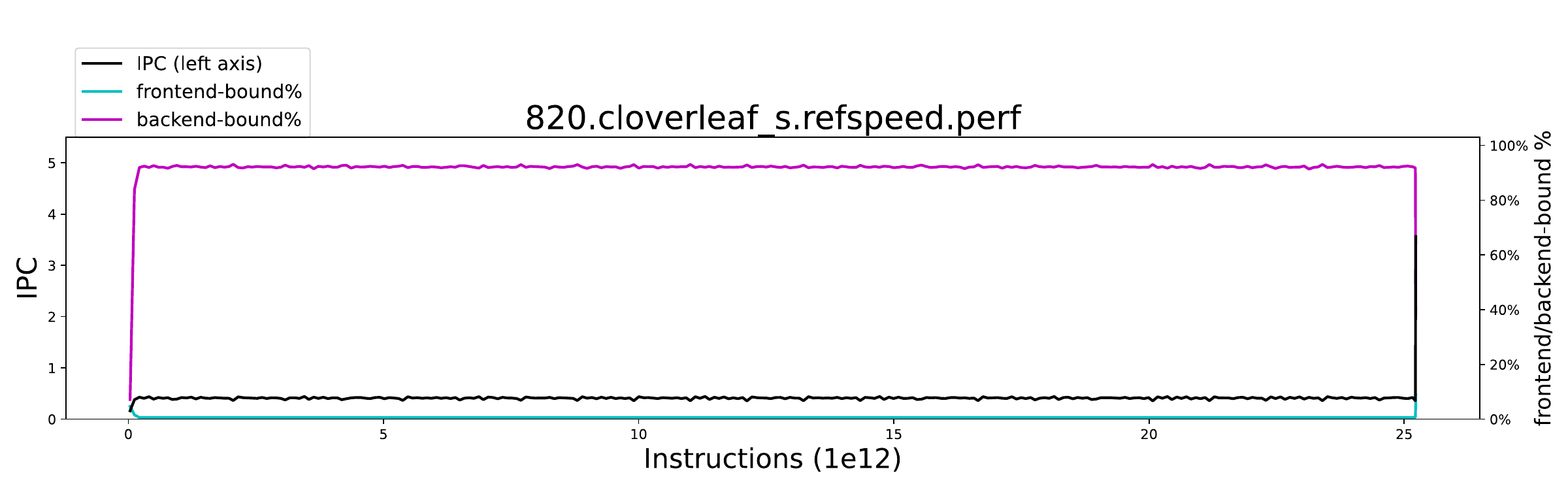}
        \end{subfigure}
        \\[-5pt] 

        \begin{subfigure}[b]{\linewidth}
            \centering
            \includegraphics[width=1\linewidth]{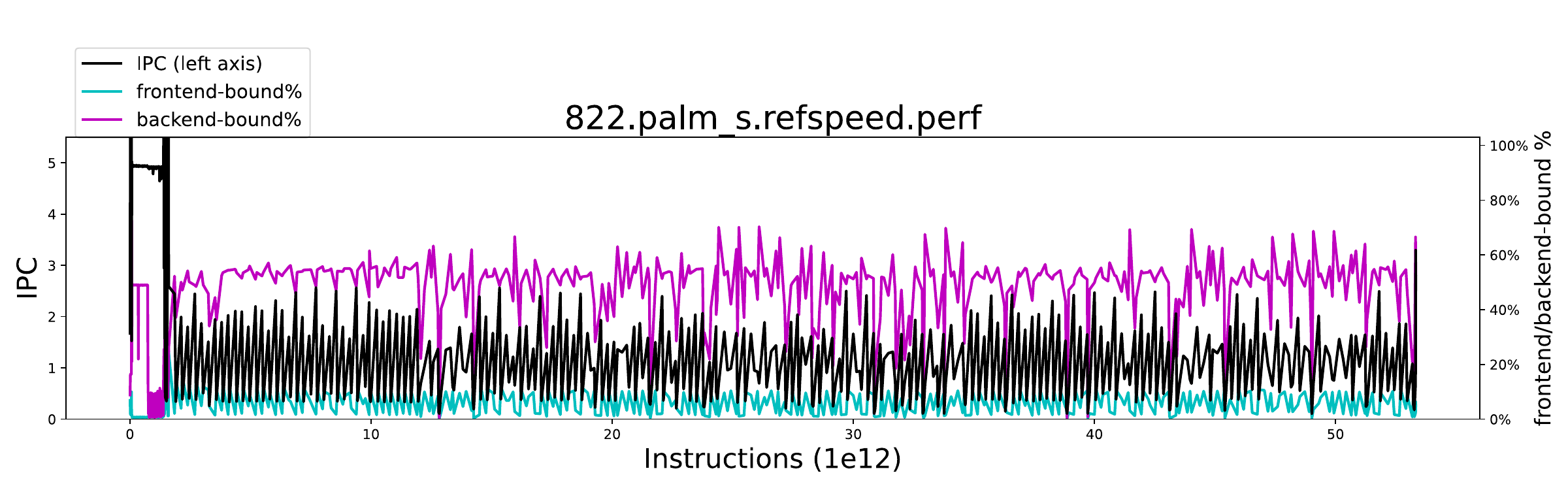}            
        \end{subfigure}
        &
        \begin{subfigure}[b]{\linewidth}
            \centering
            \includegraphics[width=1\linewidth]{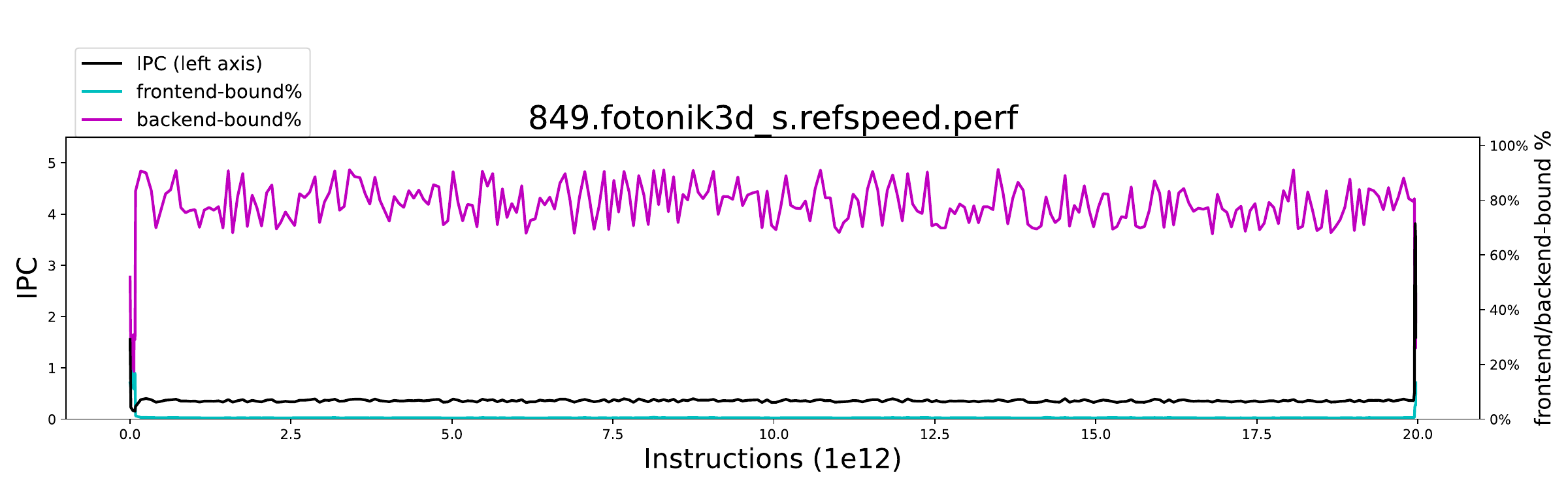}
        \end{subfigure}
        \\[-5pt] 
        \begin{subfigure}[b]{\linewidth}
            \centering
            \includegraphics[width=1\linewidth]{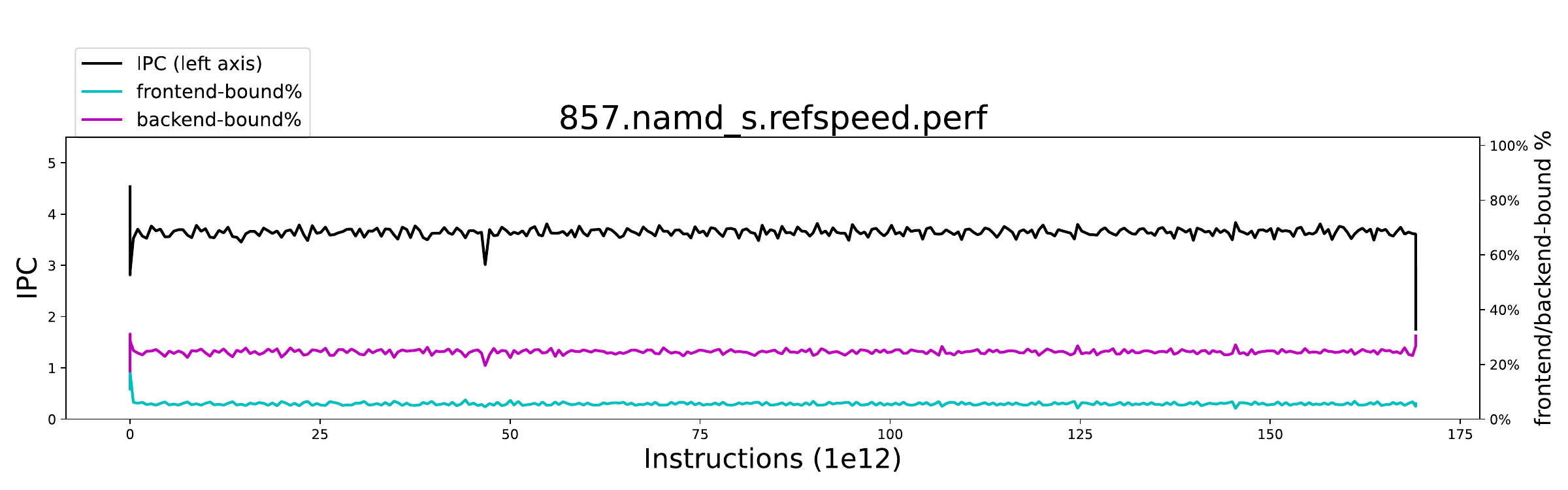}
            \end{subfigure}
        &

        \begin{subfigure}[b]{\linewidth}
            \centering
            \includegraphics[width=1\linewidth]{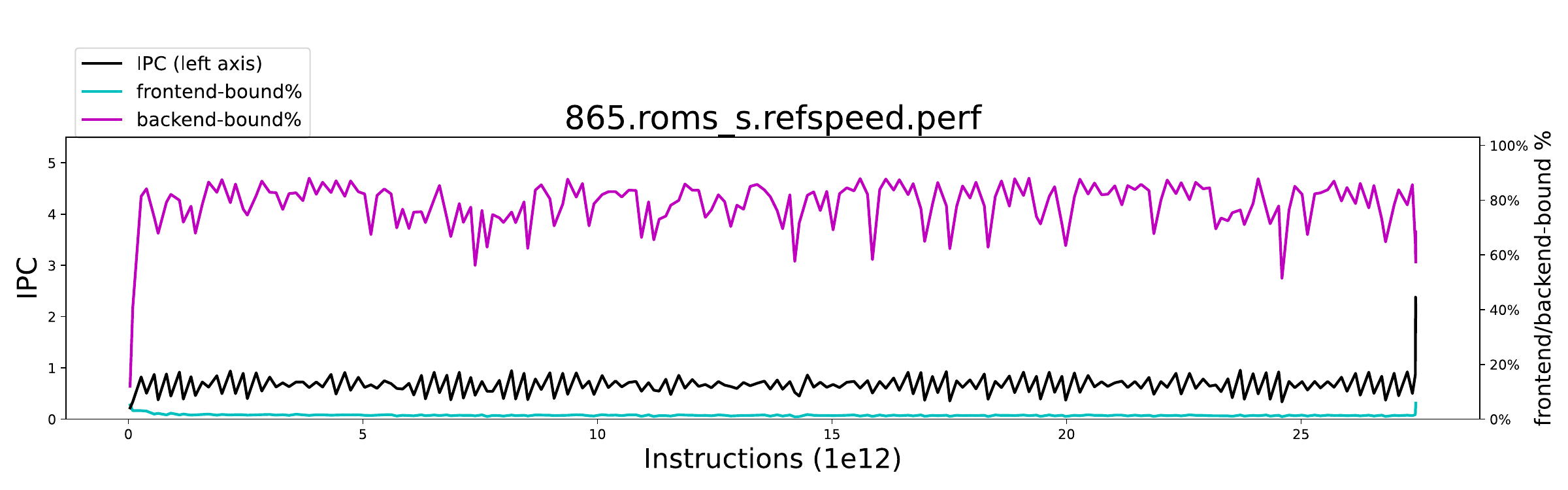}
        \end{subfigure}
        \\[-5pt] 
        \begin{subfigure}[b]{\linewidth}
            \centering
            \includegraphics[width=1\linewidth]{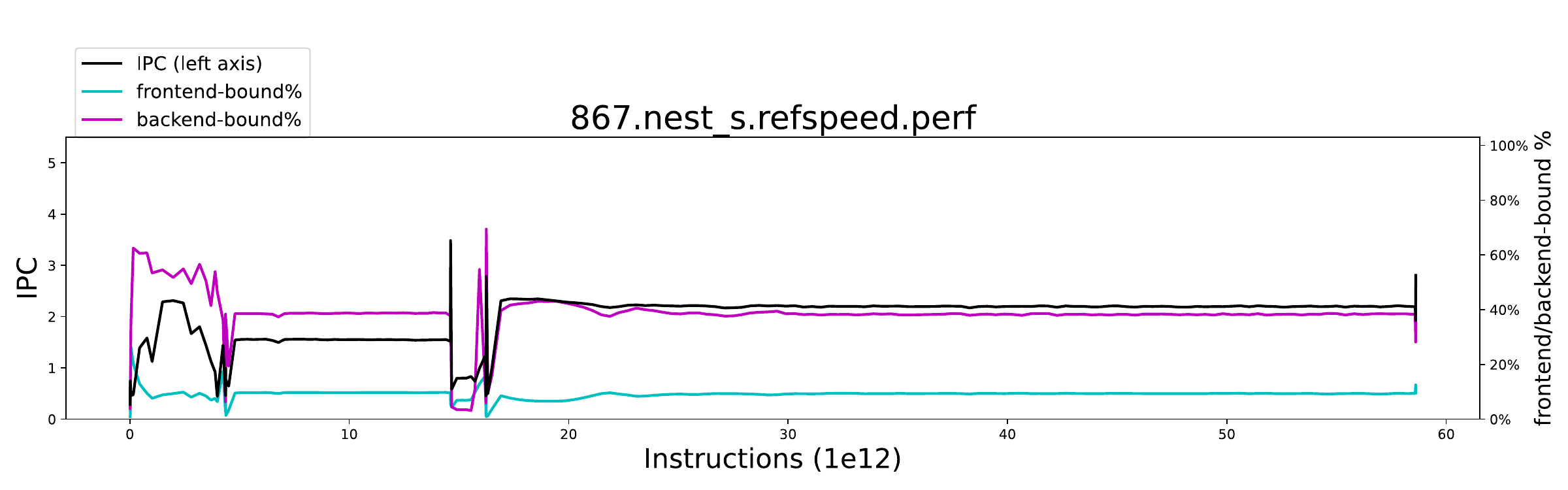}
        \end{subfigure}
        &
        \begin{subfigure}[b]{\linewidth}
            \centering
            \includegraphics[width=1\linewidth]{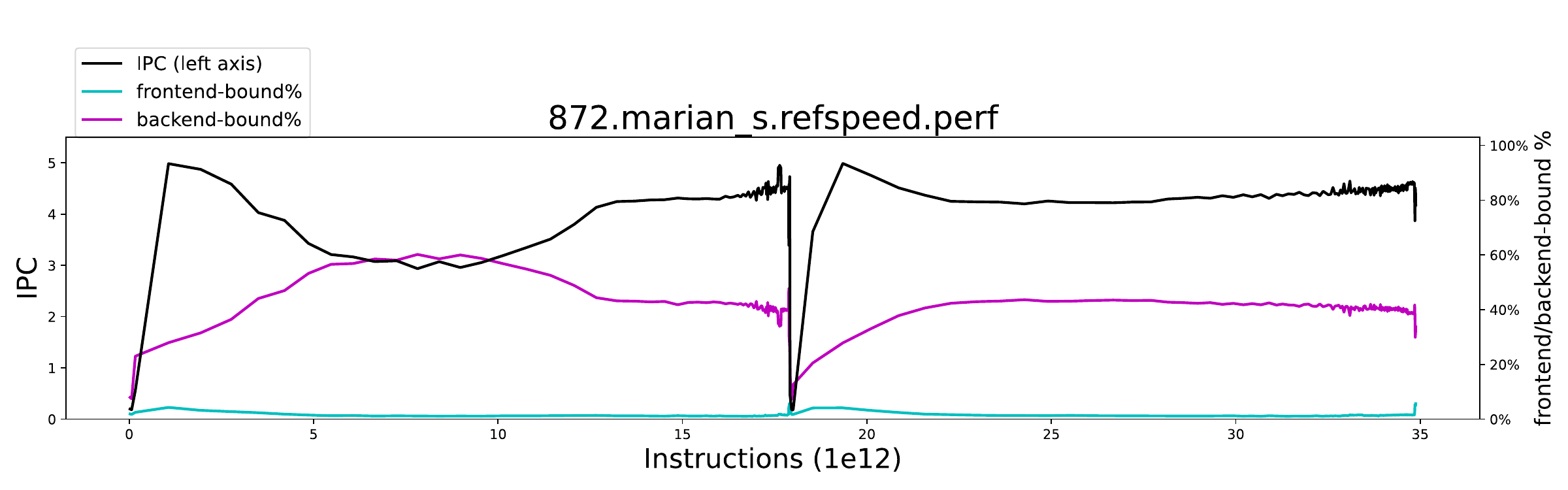}
        \end{subfigure}
        \\[-5pt]

        \begin{subfigure}[b]{\linewidth}
            \centering
            \includegraphics[width=1\linewidth]{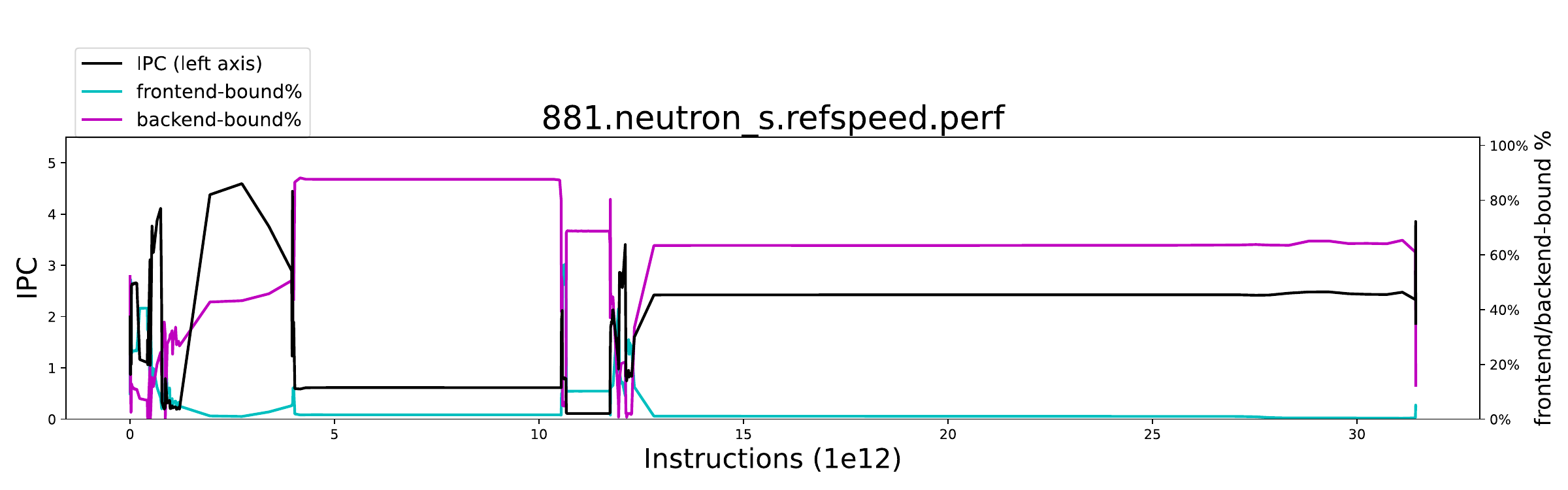}
        \end{subfigure}
        &

    \end{tabularx} 
    
    \caption{Perf Plots: Floating Point Speed}
    \label{fig:perf_refspeed_fp}
\end{figure*}